\documentclass[fleqn,usenatbib]{mnras}

\usepackage{newtxtext,newtxmath}

\usepackage[T1]{fontenc}

\DeclareRobustCommand{\VAN}[3]{#2}
\let\VANthebibliography\thebibliography
\def\thebibliography{\DeclareRobustCommand{\VAN}[3]{##3}\VANthebibliography}

\usepackage{graphicx}	
\usepackage{amsmath}	
\usepackage{threeparttable,booktabs}
\usepackage{float}
\usepackage{orcidlink}

\newcommand{\nwprp}{\bar{n}_2w_{{\rm{p}}}(r_{\rm{p}})}
\newcommand{\nwp}{\bar{n}_2w_{{\rm{p}}}}

\newcommand{\Wp}{w_{{\rm{p}}}}
\newcommand{\ms}{{\rm M}_{\sun}}
\newcommand{\msh}{h^{-1}{\rm M}_{\sun}}

\defcitealias{2022ApJ...925...31X}{Paper I}
\defcitealias{2022ApJ...939..104X}{Paper III}
\defcitealias{2023ApJ...944..200X}{Paper IV}

\title[DESI Y1 GSMF]{PAC in DESI. I. Galaxy Stellar Mass Function into the $10^{6}{\rm M}_{\odot}$ Frontier}

\author[Kun Xu et al.]{
\parbox{\textwidth}{
\Large
Kun Xu\orcidlink{0000-0002-7697-3306},$^{1,2,3}$\thanks{E-mail: kunxu@sas.upenn.edu}
Y.~P.~Jing\orcidlink{0000-0002-4534-3125},$^{3}$
S.~Cole\orcidlink{0000-0002-5954-7903},$^{2}$
C.~S.~Frenk\orcidlink{0000-0002-2338-716X},$^{2}$
S.~Bose\orcidlink{0000-0002-0974-5266},$^{2}$
W.~Elbers\orcidlink{0000-0002-2207-6108},$^{2}$
W.~Wang,$^{4}$
Yirong Wang\orcidlink{0000-0003-3203-3299},$^{4}$
S.~Moore,$^{2}$
J.~Aguilar,$^{5}$
S.~Ahlen\orcidlink{0000-0001-6098-7247},$^{6}$
D.~Bianchi\orcidlink{0000-0001-9712-0006},$^{7,8}$
D.~Brooks,$^{9}$
T.~Claybaugh,$^{5}$
A.~de la Macorra\orcidlink{0000-0002-1769-1640},$^{10}$
Arjun~Dey\orcidlink{0000-0002-4928-4003},$^{11}$
J.~E.~Forero-Romero\orcidlink{0000-0002-2890-3725},$^{12,13}$
E.~Gaztañaga,$^{14,15,16}$
S.~Gontcho A Gontcho\orcidlink{0000-0003-3142-233X},$^{5}$
G.~Gutierrez,$^{17}$
K.~Honscheid\orcidlink{0000-0002-6550-2023},$^{18,19,20}$
M.~Ishak\orcidlink{0000-0002-6024-466X},$^{21}$
T.~Kisner\orcidlink{0000-0003-3510-7134},$^{5}$
S.~E.~Koposov\orcidlink{0000-0003-2644-135X},$^{22,23}$
M.~Landriau\orcidlink{0000-0003-1838-8528},$^{5}$
L.~Le~Guillou\orcidlink{0000-0001-7178-8868},$^{24}$
R.~Miquel,$^{25,26}$
J.~Moustakas\orcidlink{0000-0002-2733-4559},$^{27}$
C.~Poppett,$^{5,28,29}$
F.~Prada\orcidlink{0000-0001-7145-8674},$^{30}$
I.~P\'erez-R\`afols\orcidlink{0000-0001-6979-0125},$^{31}$
G.~Rossi,$^{32}$
E.~Sanchez\orcidlink{0000-0002-9646-8198},$^{33}$
D.~Sprayberry,$^{11}$
G.~Tarl\'{e}\orcidlink{0000-0003-1704-0781},$^{34}$
B.~A.~Weaver,$^{11}$
and H.~Zou\orcidlink{0000-0002-6684-3997}$^{35}$}
\vspace{0.4cm}
\\
\parbox{\textwidth}{Affiliations are listed in Appendix~\ref{sec:author}}
}

\date{Accepted XXX. Received YYY; in original form ZZZ}

\pubyear{2025}

\begin{document}
\label{firstpage}
\pagerange{\pageref{firstpage}--\pageref{lastpage}}
\maketitle

\begin{abstract}
The Photometric Objects Around Cosmic Webs (PAC) method integrates cosmological photometric and spectroscopic surveys, offering valuable insights into galaxy formation. PAC measures the excess surface density of photometric objects, $\nwp$, with specific physical properties around spectroscopic tracers. In this study, we improve the PAC method to make it more rigorous and eliminate the need for redshift bins. We apply the enhanced PAC method to the DESI Y1 BGS Bright spectroscopic sample and the deep DECaLS photometric sample, obtaining $\nwp$ measurements across the complete stellar mass range, from \(10^{5.3} \ms\) to \(10^{11.5} \ms\) for blue galaxies, and from \(10^{6.3} \ms\) to \(10^{11.9} \ms\) for red galaxies. We combine $\nwp$ with $\Wp$ measurements from the BGS sample, which is not necessarily complete in stellar mass. Assuming that galaxy bias is primarily determined by stellar mass and colour, we derive the galaxy stellar mass functions (GSMFs) down to \(10^{5.3} \ms\) for blue galaxies and \(10^{6.3} \ms\) for red galaxies, while also setting lower limits for smaller masses. The blue and red GSMFs are well described by single and double Schechter functions, respectively, with low-mass end slopes of $\alpha_{\rm{blue}}=-1.54^{+0.02}_{-0.02}$ and $\alpha_{\rm{red}}=-2.50^{+0.08}_{-0.08}$, resulting in the dominance of red galaxies below \(10^{7.6} \ms\). Stage-IV cosmological photometric surveys, capable of reaching 2-3 magnitudes deeper than DECaLS, present an opportunity to explore the entire galaxy population in the local universe with PAC. This advancement allows us to address critical questions regarding the nature of dark matter, the physics of reionization, and the formation of dwarf galaxies.
\end{abstract}




\section{Introduction}
Over the past two decades, both cosmological spectroscopic surveys \citep{2000AJ....120.1579Y,2001MNRAS.328.1039C,2004MNRAS.355..747J,2005A&A...439..845L,2011MNRAS.413..971D,2013ApJS..208....5N,2014A&A...562A..23G} and photometric surveys  \citep{2000AJ....120.1579Y,2005astro.ph.10346T,2012MNRAS.427..146H,2015A&A...582A..62D,2018PASJ...70S...4A,2019AJ....157..168D} have significantly contributed to shaping our understanding of galaxy formation and cosmology. These two types of surveys possess unique advantages and disadvantages. Spectroscopic surveys can offer precise redshift measurements for studying 3D galaxy clustering and assessing galaxy properties. However, they still focus on bright and selected tracers, leading to complex selection effects. Photometric surveys are significantly deeper and more complete than spectroscopic surveys, and the gap is expected to widen further in the future. Photometric surveys can also provide spatially resolved galaxy properties such as morphology and shape. Nonetheless, obtaining precise redshift measurements from photometric surveys remains a challenging task \citep{2022ARA&A..60..363N}. Moreover, due to their distinct characteristics, spectroscopic and photometric surveys are often aligned with different scientific objectives. For instance, photometric surveys are typically driven by weak lensing measurements \citep{1993ApJ...404..441K,2001PhR...340..291B}, whereas spectroscopic surveys are primarily pursued to study baryon acoustic oscillations \citep[BAO;][]{2003ApJ...598..720S} and redshift space distortions \citep[RSD;][]{1987MNRAS.227....1K}.

If we can integrate these two types of surveys and make full use of their respective strengths, we can generate a wealth of information to advance our understanding of galaxy formation and cosmology. To fully leverage the deep photometric surveys, as both photometric and spectroscopic objects trace the comic web, it is theoretically possible to statistically assign redshifts by cross-correlating them with spectroscopic tracers, eliminating the use of photometric redshift (photoz). However, in practice, a well-thought-out design is essential to achieve this goal. Various existing approaches, such as the clustering-z method in weak lensing \citep{2008ApJ...684...88N,2013arXiv1303.4722M}, have made attempts in this direction, but they still rely on certain assumptions, such as those related to galaxy bias \citep{2013MNRAS.431.3307S,2015MNRAS.447.3500R}. 

To achieve this, building upon the work of \citet{2011ApJ...734...88W}, we introduced the Photometric objects Around Cosmic webs (PAC) method in \citet[][hereafter \citetalias{2022ApJ...925...31X}]{2022ApJ...925...31X}. The PAC method estimates the excess surface density, \(\nwp\), of photometric objects with specific physical properties, such as stellar mass, luminosity, rest-frame colour, and star formation rate, around spectroscopic tracers. Utilizing Stage-III cosmological spectroscopic and photometric surveys such as the Sloan Digital Sky Survey (SDSS) \citep{2000AJ....120.1579Y,2013AJ....145...10D}, the Hyper Suprime-Cam (HSC) SSP Survey \citep{2018PASJ...70S...4A}, and the DECam Legacy Survey \citep[DECaLS,][]{2019AJ....157..168D}, we have conducted a series of investigations using the PAC method. These studies have yielded innovative measurements and results in various aspects of galaxy formation, including precise measurements of the galaxy stellar mass function \citep[GSMF;][hereafter \citetalias{2022ApJ...939..104X}]{2022ApJ...939..104X} and the stellar mass-halo mass relation \citep[SHMR;][hereafter \citetalias{2023ApJ...944..200X}]{2023ApJ...944..200X} down to stellar mass \(10^{8.0}\ms\), as well as their evolution up to redshift 0.7. Additionally, we have investigated galaxy assembly bias \citep{2022ApJ...926..130X}, environmental quenching \citep{2024ApJ...969..129Z,2025arXiv250100986Z}, and the quasar-halo connection \citep{2024ApJ...967...17G}. In \citetalias{2023ApJ...944..200X}, we also developed accurate galaxy-halo connections for the SDSS spectroscopic CMASS and LOWZ samples, which have been applied to study the intrinsic alignment of massive galaxies \citep{2023ApJ...954....2X} and gravitational lensing magnification around them \citep{2024ApJ...973..102X}. Furthermore, many other studies utilized a concept similar to the PAC method, with a focus on studying the distribution of satellite galaxies \citep{2009ApJ...699.1333H,2011MNRAS.417..370G,2012ApJ...760...16J,2012MNRAS.424.2574W,2012MNRAS.427..428G,2014MNRAS.442.1363W,2016MNRAS.459.3998L,2021ApJ...919...25W,2021MNRAS.505.5370T,2025arXiv250303317W}. These studies underscore the well-established status of the PAC method and have already contributed to enhancing our understanding of galaxy formation.

We are now entering the Stage-IV era, with various next-generation cosmological spectroscopic and photometric surveys either commencing or planning to release their data in the next few years, including the Dark Energy Spectroscopic Instrument \citep[DESI;][]{2016arXiv161100036D}, the Subaru Prime Focus Spectrograph \citep[PFS;][]{2014PASJ...66R...1T}, the Legacy Survey of Space and Time \citep[LSST;][]{2019ApJ...873..111I}, Euclid \citep{2011arXiv1110.3193L}, the Chinese Space Station Optical Survey \citep[CSS-OS;][]{2019ApJ...883..203G}, and surveys conducted with the Nancy Grace Roman Space Telescope \citep[Roman;][]{2015arXiv150303757S}. DESI  aims to provide more than 40 million spectroscopic tracers up to $z\sim3.5$, which is approximately 40 times greater than the number available in SDSS. LSST, Euclid, CSS-OS and Roman will provide large area multi-band galaxy images ranging from near-ultraviolet to near-infrared to unparalleled depth, reaching approximately $r\sim27$ or even deeper, which is $2-3$ magnitudes deeper than DECaLS we utilized in previous studies. The Stage-IV surveys present promising opportunities for us to employ the PAC method in exploring fainter sources and earlier cosmic epochs and obtaining more precise measurements. These innovative measurements can contribute to addressing fundamental questions regarding the nature of dark matter and dark energy, the epoch of reionization, structure formation in the universe, and the formation of galaxies.

In this paper, as the first application of the PAC method in the Stage-IV era, we focus on measuring the local GSMF down to much lower mass limits than previously achieved. The GSMF is a crucial measurement for understanding galaxy evolution. The amplitude and shape of the GSMF encode vital information about galaxy star formation and quenching mechanisms. Consequently, it has been utilized to infer and constrain galaxy formation models in numerous studies ranging from simple empirical models to numerical simulations \citep{2010ApJ...721..193P,2013ApJ...770...57B,2015MNRAS.446..521S,2015MNRAS.451.2663H,2018MNRAS.473.4077P,2019MNRAS.486.2827D,2023MNRAS.526.4978S}. In previous works, large flux-limited spectroscopic surveys, particularly the Two Degree Field Galaxy Spectroscopic Survey \citep[2dF-GRS;][]{2001MNRAS.328.1039C}, SDSS, and Galaxy and Mass Assembly \citep[GAMA;][]{2011MNRAS.413..971D}, are commonly employed to investigate the local GSMF \citep{2001MNRAS.326..255C,2008MNRAS.388..945B,2009MNRAS.398.2177L,2022MNRAS.513..439D}. However, due to depth limitations, accurate and reliable measurements of the local GSMF from spectroscopic surveys remain constrained to relatively high masses. At lower masses, measurements are typically refined to the Local Volume, which lies within or around the Local Void \citep{2008ApJ...676..184T,2010Natur.465..565P, 2012AJ....144....4M, 2019Ap.....62..293K,2020A&A...633A..19B}, often leading to underestimation of the GSMF. Using the ELUCID simulation \citep{2014ApJ...794...94W,2016ApJ...831..164W}, which was run based on the reconstructed initial density field of the nearby universe, \citet{2019ApJ...872..180C} quantified the underestimation of the GSMF by galaxy surveys with various magnitude limits. They identified significant underestimation at the low-mass end in current surveys, starting at stellar mass \( 10^{8.0} \ms - 10^{9.0} \ms \). Therefore, measuring the low-mass end of the GSMF using spectroscopic surveys alone remains highly challenging.

To reach lower stellar masses, we employ the PAC method in combination with deeper photometric surveys, rather than relying solely on spectroscopic data, as have been achieved in \citetalias{2022ApJ...939..104X}. Notably, comparing to the stellar mass limits of $10^{8.2}\ms$ reached in \citetalias{2022ApJ...939..104X}, we enhance various aspects of the previous PAC method and take a significant step into the $10^{6.0} \ms$ frontier with the DESI Y1 Bright Galaxy Survey (BGS) Bright sample and DECaLS photometric sample. This advance is facilitated by leveraging numerous low-redshift tracers available in BGS, enabling accurate $\nwp$ measurements at lower redshifts. Furthermore, we derive the GSMF for blue and red galaxy samples separately, offering deeper insights into the galaxy population. Our results mark significant progress in the measurement of the local GSMF and have the potential to impact our understanding of dark matter and reionization. Specifically, the number density of galaxies with stellar masses around $10^{6.0} \ms$ becomes sensitive to dark matter particle mass \citep{2000ApJ...542..622C,2001ApJ...556...93B,2005PhRvD..71f3534V,2012MNRAS.420.2318L,2013MNRAS.428..882M,2016MNRAS.456.4346H,2017MNRAS.464.4520B} and to the epoch and physics of reionization \citep{1992MNRAS.256P..43E,2001ARA&A..39...19L,2008MNRAS.390..920O,2017MNRAS.465.3913B,2018ApJ...863..123B}. Additionally, our findings can also contribute to advancing the understanding of dwarf galaxy formation.

We introduce the DESI and DECaLS data in Section~\ref{sec:2}. Section~\ref{sec:3} provides a detailed description of the PAC method, including various improvements made to enhance it. The measurements and the resulting GSMF are presented in Section~\ref{sec:4}. In Section~\ref{sec:5}, we provide a concise summary. We adopt the cosmology with $\Omega_{\rm m} = 0.268$, $\Omega_{\Lambda} = 0.732$ and $H_0 = 71{\rm \ km/s/Mpc}$ throughout the paper \citep{2013ApJS..208...19H}.

\section{Data} \label{sec:2}

In this section, we briefly introduce the DESI Y1 BGS spectroscopic samples and DECaLS photometric sample. 

\subsection{DESI Y1 BGS}\label{sec:2.1}
DESI is a prominent Stage-IV dark energy survey with the goal of obtaining spectra for approximately 40 million extragalactic objects over a five-year period \citep{2013arXiv1308.0847L, 2016arXiv161100036D, 2016arXiv161100037D, 2022AJ....164..207D,2024AJ....167...62D,2024AJ....168...58D}. Covering more than 14,000 $\mathrm{deg^2}$, the survey is conducted using a multi-object fibre-fed spectrograph mounted on the prime focus panel of the 4-meter Mayall Telescope at Kitt Peak National Observatory \citep{2022AJ....164..207D}. The spectrometer operates in the wavelength range of $3600-9800\,$\r{A} and can allocate fibres to 5,000 objects during a visit \citep{2016arXiv161100037D, 2023AJ....165....9S, 2024AJ....168...95M,2024AJ....168..245P}. Additional pipelines supporting the DESI experiment are detailed in \cite{2023AJ....165..144G, redrock2023, fba, 2023AJ....166..259S, 2023AJ....165...50M}. The parent catalogue used for DESI target selections is constructed from Data Release 9 of the DESI Legacy
Imaging Surveys \citep{2017PASP..129f4101Z,2019AJ....157..168D}. The photometric data contains three optical bands $grz$ from the DECam Legacy Survey \citep[DECaLS;][]{2019AJ....157..168D}, the Dark Energy Survey \citep[DES;][]{2005astro.ph.10346T}, the Beijing-Arizona Sky Survey \citep[BASS;][]{2017PASP..129f4101Z} and the Mayall $z$-band Legacy
Survey (MzLS).

As part of its core observations, DESI is conducting the Bright Galaxy Survey \citep[BGS;][]{2023AJ....165..253H}. BGS spans 14,000 $\mathrm{deg^2}$ footprint and includes low-redshift galaxies with \(z < 0.6\) that can be observed during bright time, when the night sky is approximately \(2.5\times\) brighter than nominal dark conditions. BGS provides two galaxy samples: the BGS Bright sample, a \(r < 19.5\) magnitude-limited sample of approximately 10 million galaxies, and the BGS Faint sample, a fainter \(19.5 < r < 20.175\) sample of approximately 5 million galaxies selected based on surface brightness and colour. The selection and completeness of the BGS samples are detailed in \citet{2023AJ....165..253H}.

We use the DESI Y1 BGS Bright sample in this study, covering 5300 and 2173 deg$^2$ in the Northern and Southern Galactic caps (NGC and SGC), with an average completeness of 0.656. Our analysis focuses on the BGS Bright sample overlapped with the DECaLS footprint, restricted to $\rm{decl}.<32^{\circ}$, leading to a reduced total area of 5349 deg$^2$. Each BGS source is assigned a weight
\begin{equation}
    w_{\rm{tot}} = w_{\rm{comp}} w_{\rm{zfail}}\,,
\end{equation}
where $w_{\rm{comp}}$ corrects for fibre-assignment incompleteness, and $w_{\rm zfail}$ adjusts for changes in the relative redshift success rates \citep{2025JCAP...01..125R}.

\begin{figure}
    \centering
    \includegraphics[width=\columnwidth]{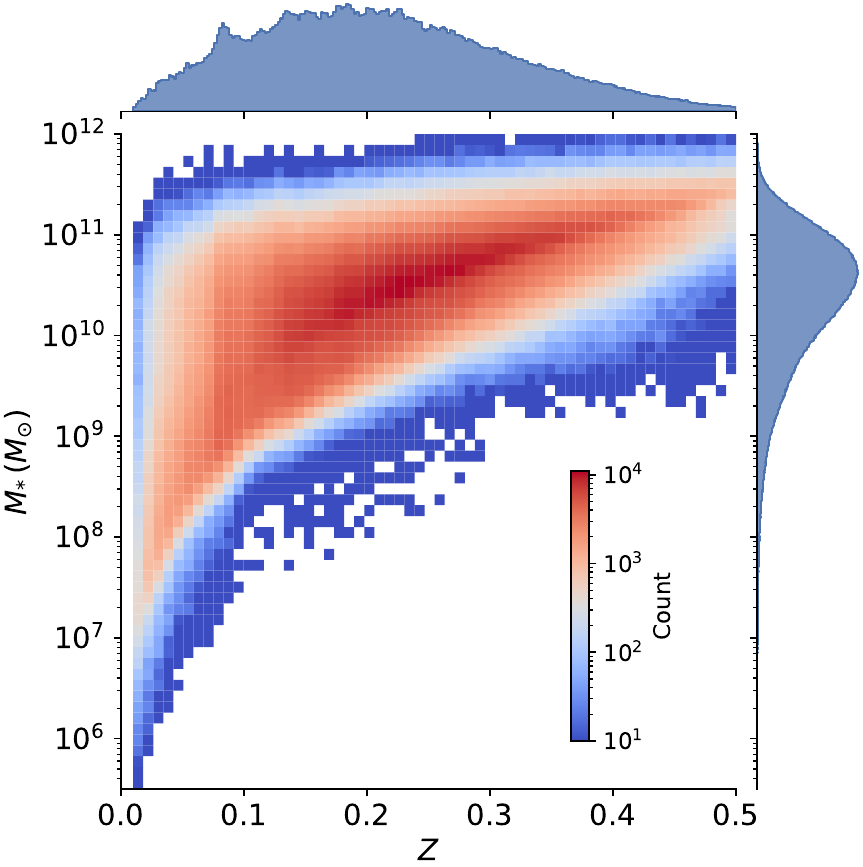}
    \caption{Distributions of redshift and stellar mass for the DESI Y1 BGS Bright sample that overlaps with the DECaLS footprint.}
    \label{fig:bgs_distribution}
\end{figure}

In Figure~\ref{fig:bgs_distribution}, we present the redshift and stellar mass distributions of the BGS Bright sample within the DECaLS region. We compute their physical properties by applying the SED fitting code {\tt{CIGALE}} \citep{2019A&A...622A.103B}, utilizing the $grz$ band fluxes from DECaLS and redshifts from DESI spectra. The calculations are based on the stellar population synthesis models by \citet{2003MNRAS.344.1000B}, which employ a \citet{2003PASP..115..763C} initial mass function and a star formation history characterized by a delayed exponential function $\phi(t)\approx t\exp{(-t/\tau)}$. We consider three metallicities, specifically $Z/Z_{\odot}=0.4,\ 1,\ \text{and}\ 2.5$, where $Z_{\odot}$ represents the metallicity of the Sun. Additionally, we incorporate the \citet{2000ApJ...533..682C} extinction law to account for dust reddening, within the range of $0<E(B-V)<0.5$. 

\begin{figure}
    \centering
    \includegraphics[width=\columnwidth]{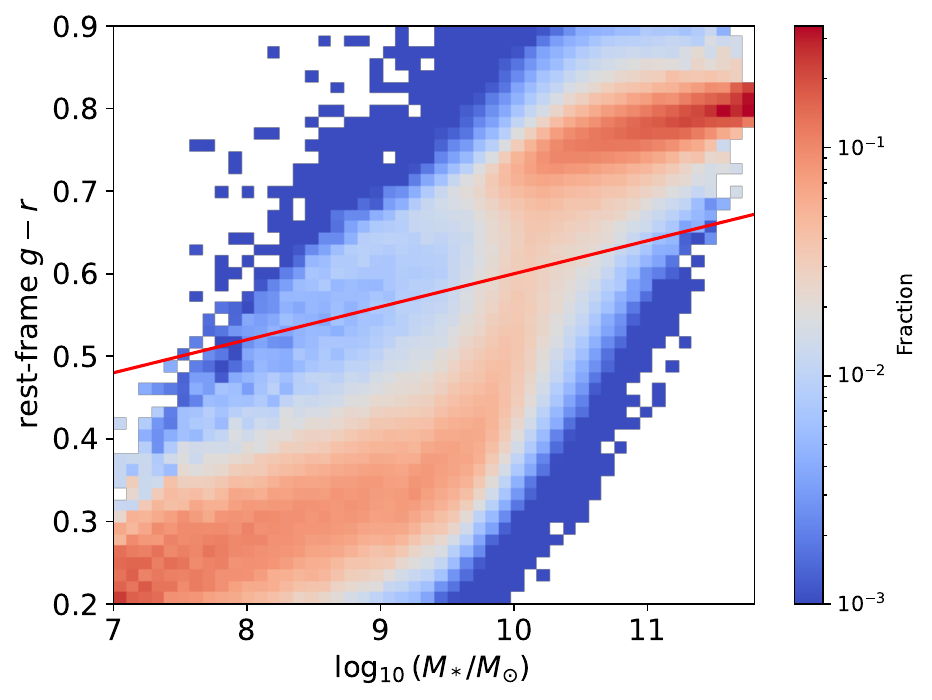}
    \caption{$V_{\rm{max}}$-weighted rest-frame $g-r$ colour vs. stellar mass distribution for BGS with $z<0.2$. The distribution is further normalized within each stellar mass bin for clearer visualization. The line $g-r=0.04\log_{10}(M_*/\ms)+0.2$ is used to separate the blue and red populations.}
    \label{fig:colour_distribution}
\end{figure}

Since we aim to study the blue and red populations separately, Figure~\ref{fig:colour_distribution} presents the rest-frame $g-r$ colour vs. stellar mass distributions, weighted by 1/$V_{\rm{max}}$, where $V_{\rm{max}}$ corresponds to $z_{\rm{max}}$, the maximum redshift at which a galaxy can still meet our sample selection criteria. The $V_{\rm{max}}$ values are taken from \citet{2024ApJ...971..119W}. We apply the colour cut $g-r=0.04\log_{10}(M_*/\ms)+0.2$ to separate the blue and red populations. This cut is chosen to closely match the one used in \citet{2024ApJ...969..129Z}, where they studied environmental quenching using PAC and SDSS data. Their colour cut, based on $u-r$, could not be directly used because DECaLS does not include $u$-band measurements. Instead, we used $g-r$ and determined the cut based on the relationship between the $u$ and $g$ bands from SDSS data.

To calculate the correlation functions, we use a random catalogue provided by DESI \citep{2025JCAP...01..125R}, which matches the footprint and redshift distribution of the BGS sample used in our analysis.

\subsection{DECaLS}
We utilize the photometric catalogue from DECaLS in Data Release 9 of the DESI Legacy Imaging Surveys \citep{2019AJ....157..168D}. This catalogue covers approximately 9000 deg$^2$ in the Northern and Southern Galactic caps (NGC and SGC) with $\rm{decl}.<32^{\circ}$. Observations were conducted in the $g$, $r$, and $z$ bands, with median 5$\sigma$ point source depths of 24.9, 24.2, and 23.3, respectively. DECaLS also includes data from DES, which covers an additional 5000 deg$^2$ in the SGC. 

Image processing was performed using {\texttt{Tractor}} \citep{2016ascl.soft04008L} to extract sources. These sources are modelled using parametric profiles convolved with specific point spread functions (PSFs), including a delta function for point sources, an exponential profile, a de Vaucouleurs profile, and a S\'ersic profile. The best-fit model magnitudes are used as the default photometry, with correction for Galactic extinction \citep{1998ApJ...500..525S}. 

We restrict our analysis to the footprint that has been observed at least once in all three bands. We also apply bright star masks and bad pixel masks using the MASKBITS\footnote{https://www.legacysurvey.org/dr9/bitmasks/} provided by the Legacy Survey. Additional masks are applied to match the final DESI footprint using the {\texttt{in\_desi}} flag, resulting in a final coverage of 10324 deg$^2$. We exclude all the sources with PSF morphology to reject stars, and adopt a conservative magnitude cut $r<23.5$ to save computation time. 
This results in a DECaLS photometric sample of 0.56 billion sources. While we calculate the physical properties for the entire sample to prepare for future studies, our current focus in this study is exclusively on the DESI Y1 BGS Bright sample. Therefore, the footprint is further restricted to 5349 deg$^2$ during the measurement of the excess surface density $\nwp$.

To calculate the correlation functions, we use a random catalogue provided by  DESI Legacy Imaging Surveys, which matches the footprint of the DECaLS sample used in our analysis.

\begin{figure}
	\includegraphics[width=\columnwidth]{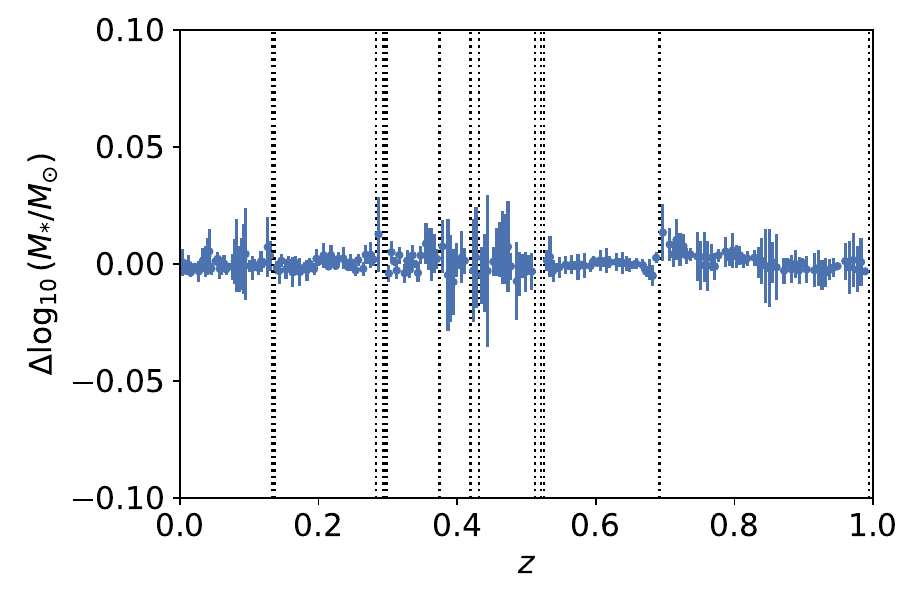}
    \caption{Mean errors and 1 $\sigma$ scatter in interpolated stellar mass $\Delta\log_{10}M_{*}$ at 241 redshifts interpolated from the values at 44 reference redshifts within the range of $0.001<z<1$. For these reference redshifts, 31 of them are chosen with an equal interval of $\Delta\log_{10}(1+z)=0.01$, while the other 13 indicated by dotted lines are included to account for the rapid changes in stellar mass with a given redshift due to specific structures in the galaxy spectrum.}
    \label{fig:test_intepolation}
\end{figure}

\begin{figure*}
	\includegraphics[width=2\columnwidth]{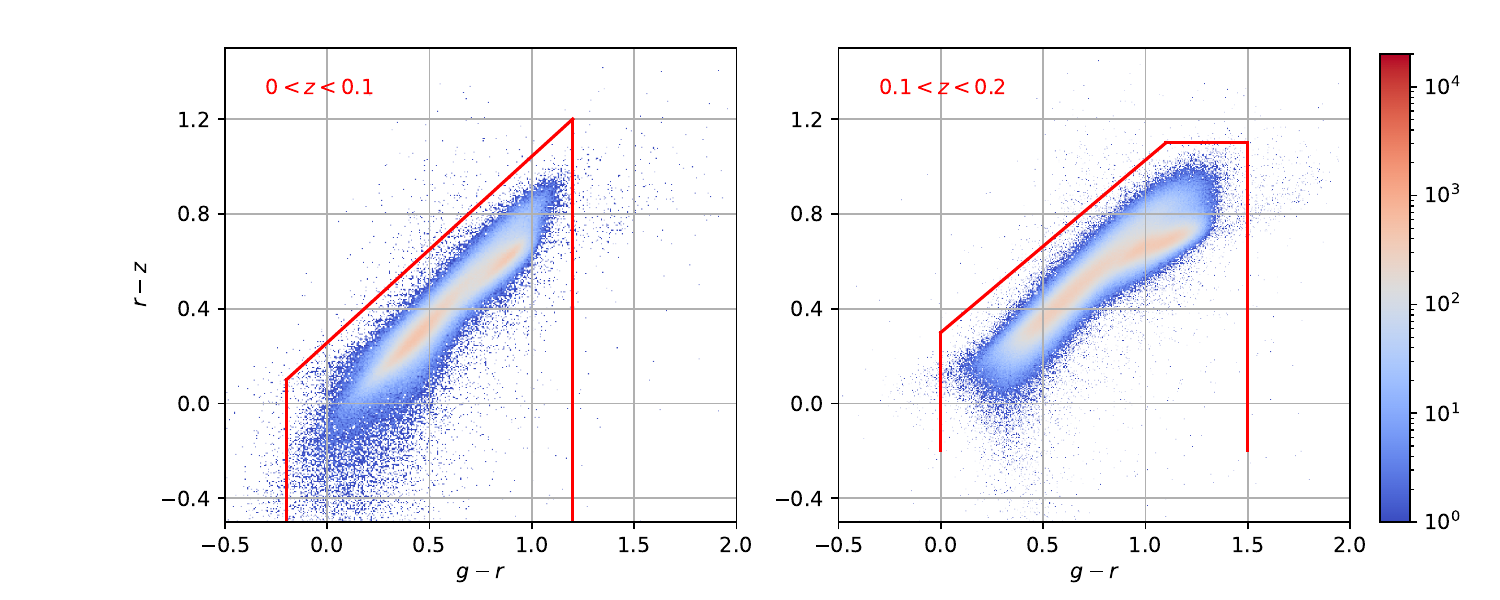}
    \caption{The observed $g-r$ vs. $r-z$ colour diagrams for DESI BGS for two redshift bins. The red solid lines represent the colour cuts used to exclude background galaxies in DECaLS for PAC measurements at each redshift bin.}
    \label{fig:colour_cut}
\end{figure*}

\section{Improving the PAC method} \label{sec:3}

In this section, we introduce improvements in the PAC method, including improvements to the calculation of physical properties, a new correlation function estimator, colour cuts to exclude background sources, and a new method to determine at what stellar mass limits DECaLS is complete.
We also validate the approximations used in the calculation. 

\subsection{Overview of the previous PAC method}
Suppose we aim to investigate two distinct populations of galaxies: one, denoted as ${\rm{pop}}_1$, is drawn from a spectroscopic catalogue, while the other, referred to as ${\rm{pop}}_2$, is sourced from a photometric catalogue. These populations are characterized within a relatively narrow redshift range. In our prior work \citetalias{2022ApJ...925...31X}, we introduced a method called PAC, which enables the accurate measurement of the excess surface density, $\nwprp$, of objects with selected physical properties from  ${\rm{pop}}_2$ around objects from ${\rm{pop}}_1$. It is based on
\begin{equation}
    \bar{n}_2w_{\rm{p}}(r_{\rm{p}}) \approx \frac{\bar{S}_2}{r_1^2}\omega_{12,\rm{w}}(\theta)\,,\label{eq:1}
\end{equation}
where $\bar{n}_2$ and $\bar{S}_2$ represent the mean number density and mean angular surface density of ${\rm{pop}}_2$, respectively. The parameter $r_1$ denotes the comoving distance to ${\rm{pop}}_1$ and $\Wp(r_{\rm{p}})$ and $\omega_{12,\rm{w}}(\theta)$ are, respectively, the projected cross-correlation function (PCCF) and the weighted angular cross-correlation function (ACCF) between ${\rm{pop}}_1$ and ${\rm{pop}}_2$, where $r_{\rm{p}}=r_1\sin{\theta}$. We weight $\omega_{12}(\theta)$ by $1/r^2_1$ at fixed $\theta$, to account for the fact that, at a fixed $\theta$, $r_{\rm{p}}$ scales with $r_1$. Using PAC, we can statistically determine the rest-frame physical properties of ${\rm{pop}}_2$ without requiring redshift information for this population; this enables us to make full use of the deep photometric surveys. The key steps involved in the previous PAC method are outlined below:
\begin{enumerate}[(i)]
    \item Divide ${\rm{pop}}_1$ into narrower redshift bins to limit the range of $r_1$ in each bin.
    \item Under the false assumption that all galaxies in ${\rm{pop}}_2$ share the same redshift as the mean redshift of a particular redshift bin, calculate the physical properties of ${\rm{pop}}_2$ using techniques such as spectral energy distribution (SED) fitting. In this way, for each redshift bin in ${\rm{pop}}_1$, a catalogue of physical properties for ${\rm{pop}}_2$ is created.
    \item Within each redshift bin, select ${\rm{pop}}_2$ objects with specific physical properties and compute $\nwprp$ using Equation~\ref{eq:1}. The foreground and background objects with incorrect physical properties are uncorrelated with ${\rm{pop}}_1$ and so do not contribute to the ACCF. Thus the measured ACCF depends only on the objects at the same redshift as  ${\rm{pop}}_1$ and these
    have the correct physical properties.
    \item Combine the results from different redshift bins by averaging with appropriate weights.
\end{enumerate}
For more details, see \citetalias{2022ApJ...925...31X}. 
\smallskip

With this method, $\nwp$ is calculated in each redshift bin and then combined, potentially introducing two sources of error. First, the properties of ${\rm{pop}}_2$ are all computed using the mean redshift of the bin, while they are correlated with ${\rm{pop}}_1$ objects spanning the full width of the redshift bin. This can lead to inaccuracies in the properties due to differences in luminosity distances and the rest-frame wavelength covered by the photometric bands. Moreover, the measured $\omega_{12,\rm{w}}(\theta)$ is multiplied by the same $\bar{S}_2/r_1^2$ factor, even though it should vary for each source in ${\rm{pop}}_1$. Although, as demonstrated in \citetalias{2022ApJ...925...31X},  these approximations are found to be acceptable at high redshift, they become less accurate at lower redshifts ($z<0.1$) due to the much more significant fractional variation in $r_1$. Hence, we now present an improved PAC method that operates without the need for redshift bins.

We improve various aspects of the PAC method. To obtain accurate physical properties for ${\rm{pop}}_2$, we present a method to infer the physical properties for the entire photometric sample across the entire redshift range in Section~\ref{sec:3.2}. To eliminate the need for redshift bins and account for the variation in $\bar{S}_2/r_1^2$, we introduce a new estimator for the PAC method in Section~\ref{sec:3.3}. In Section~\ref{sec:3.4}, we employ colour cuts to exclude foreground and background galaxies from the photometric sample, enhancing the accuracy of our measurements. We use a new method in Section~\ref{sec:3.5} to determine the stellar mass limits achievable with DECaLS, allowing us to fully exploit the available data. In Section~\ref{sec:3.6}, we assess the accuracy of the improved PAC method.

\subsection{Properties of photometric objects across the entire redshift range}\label{sec:3.2}

To solve the first source of error, it is necessary to determine the properties of each source in ${\rm{pop}}_2$ at all assumed redshifts. 
This can be achieved by defining a set of reference redshifts, determining the physical properties of each object assuming it is at each reference redshift and then interpolating these properties to any required intermediate redshift. Given the 0.56 billion sources, it is crucial to minimize the number of reference redshifts. To assess the accuracy of the interpolation and establish the required reference redshifts, we randomly select 1 per cent of the sources from the entire DECaLS sample. We compute their physical properties using the SED code {\tt{CIGALE}} and the same template described in Section~\ref{sec:2.1} for each reference redshift.

We perform SED fitting to the 1 per cent sub-sample at 241 redshifts equally spaced in $\Delta\log_{10}(1+z)$ within the redshift range of $0.001<z<1$. As demonstrated in Figure~\ref{fig:test_intepolation}, our results indicate that employing 31 reference redshifts with spacing $\Delta\log_{10}(1+z)=0.01$, in combination with an extra 13 redshifts linked to specific features in galaxy spectra, enables accurate interpolation of the stellar masses for all sources throughout the entire redshift range. The need for the 13 extra redshifts can be illustrated by considering
the case of $z\sim0.13$ where the D4000 feature begins to enter the $g$ band. Here, many galaxies that at a lower assumed redshift would be classified as red are suddenly classified as blue.
This results in the inferred stellar mass varying rapidly with redshift, necessitating the inclusion of additional reference redshifts to maintain the same level of accuracy at these specific redshifts. Using these 44 reference redshifts, we perform the interpolation under the assumption that \(\log_{10}(M_*)\) varies linearly with \(\log_{10}(D_{\rm{L}})\), where \(D_{\rm{L}}\) denotes the luminosity distance. This approach achieves an accuracy of \(\Delta\log_{10}M_{*} < 0.02\), validated by comparing interpolated \(M_*\) values with those directly calculated on a finer grid, as shown in Figure~\ref{fig:test_intepolation}.

\begin{figure}
    \centering
    \includegraphics[width=\columnwidth]{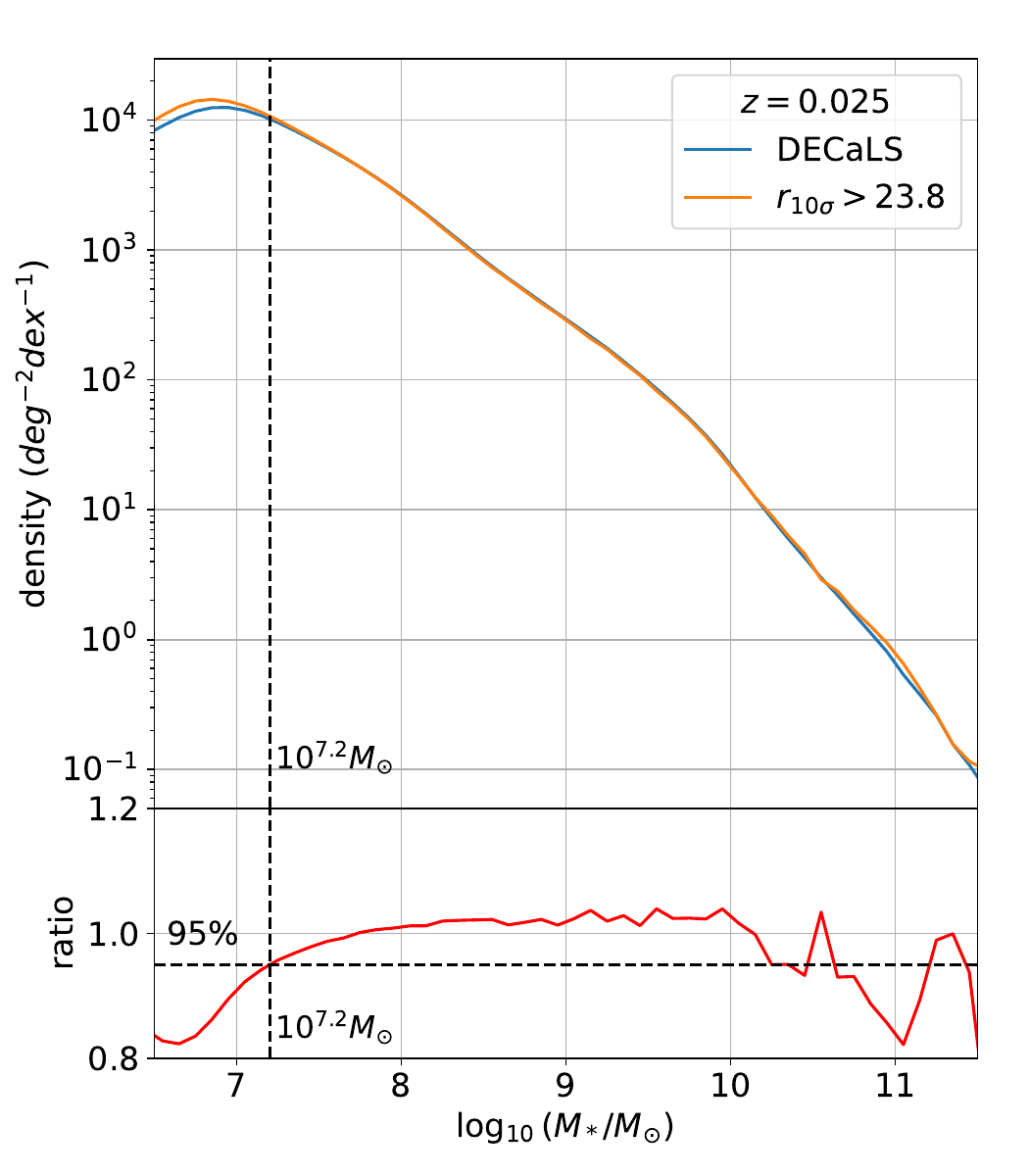}
    \caption{Comparing the angular surface density of galaxies as a function of stellar mass, calculated at an assumed redshift of $z=0.025$, between the entire DECaLS sample and for deep regions with an $r$-band $10\sigma$ PSF depth greater than 23.8. The stellar mass limit is defined as the stellar mass at which the angular surface density of DECaLS decreases to 95 per cent of the deeper sample, which in this case occurs at $10^{7.2}{\rm M}_{\odot}$ for $z=0.025$.}
    \label{fig:compare_SM}
\end{figure}

\begin{figure}
    \centering
    \includegraphics[width=\columnwidth]{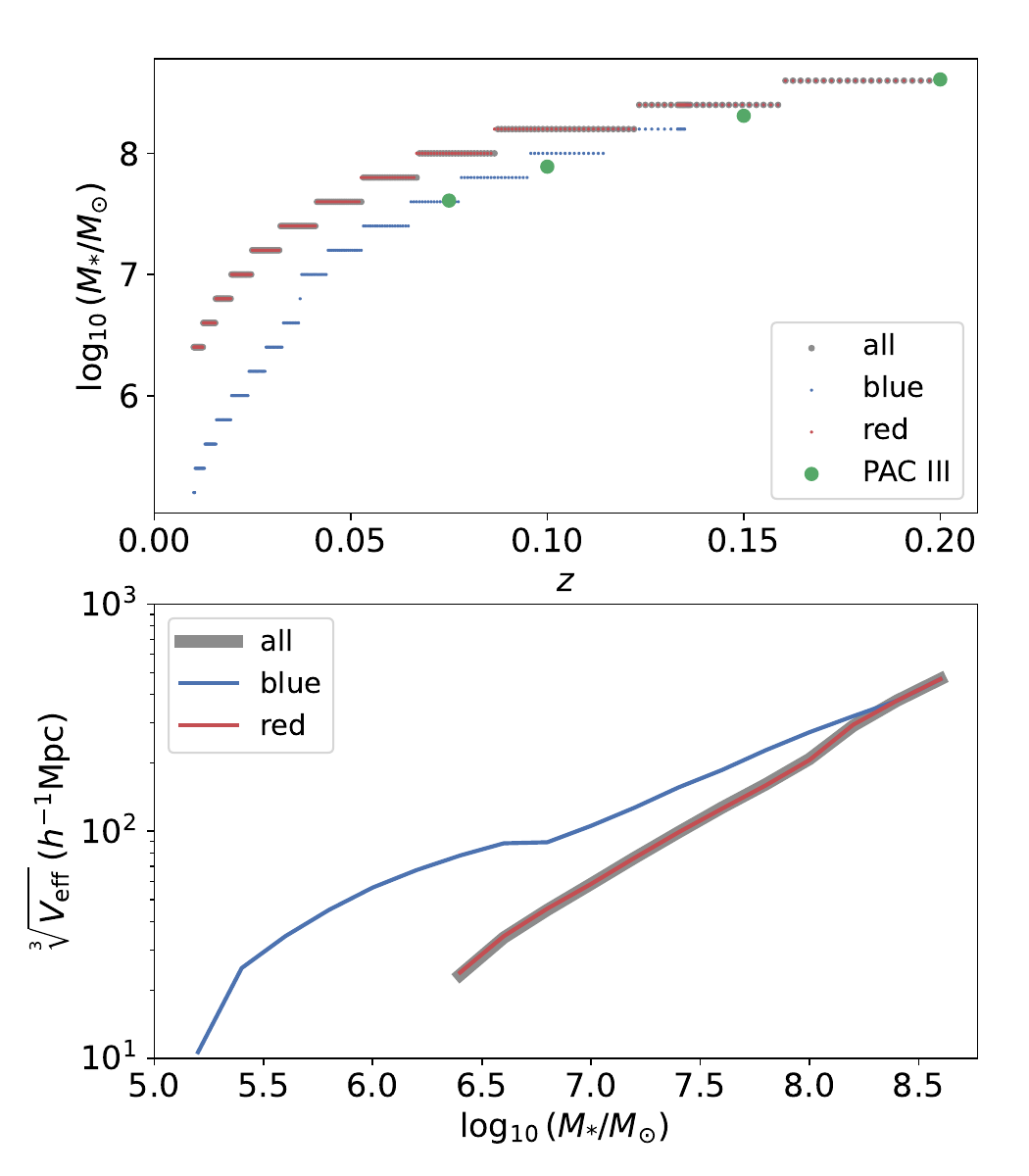}
    \caption{Top: Stellar mass limits obtained using the method demonstrated in Figure~\ref{fig:compare_SM} out to $z<0.2$. Separate limits are shown for the complete population and for red and blue subsets. These are compared to the limits for the complete population previously used in Paper III that combined DECaLS and GAMA.
Bottom: The effective volume as a function of stellar mass, calculated as the volume between a lower limit of z=0.01 and the redshift to which the sample is complete. The area is set to 5349 deg$^2$, corresponding to the overlap between the DECaLS and DESI Y1 BGS footprints.}
    \label{fig:SM_limits}
\end{figure}

\begin{figure*}
    \centering
    \includegraphics[width=0.9\textwidth]{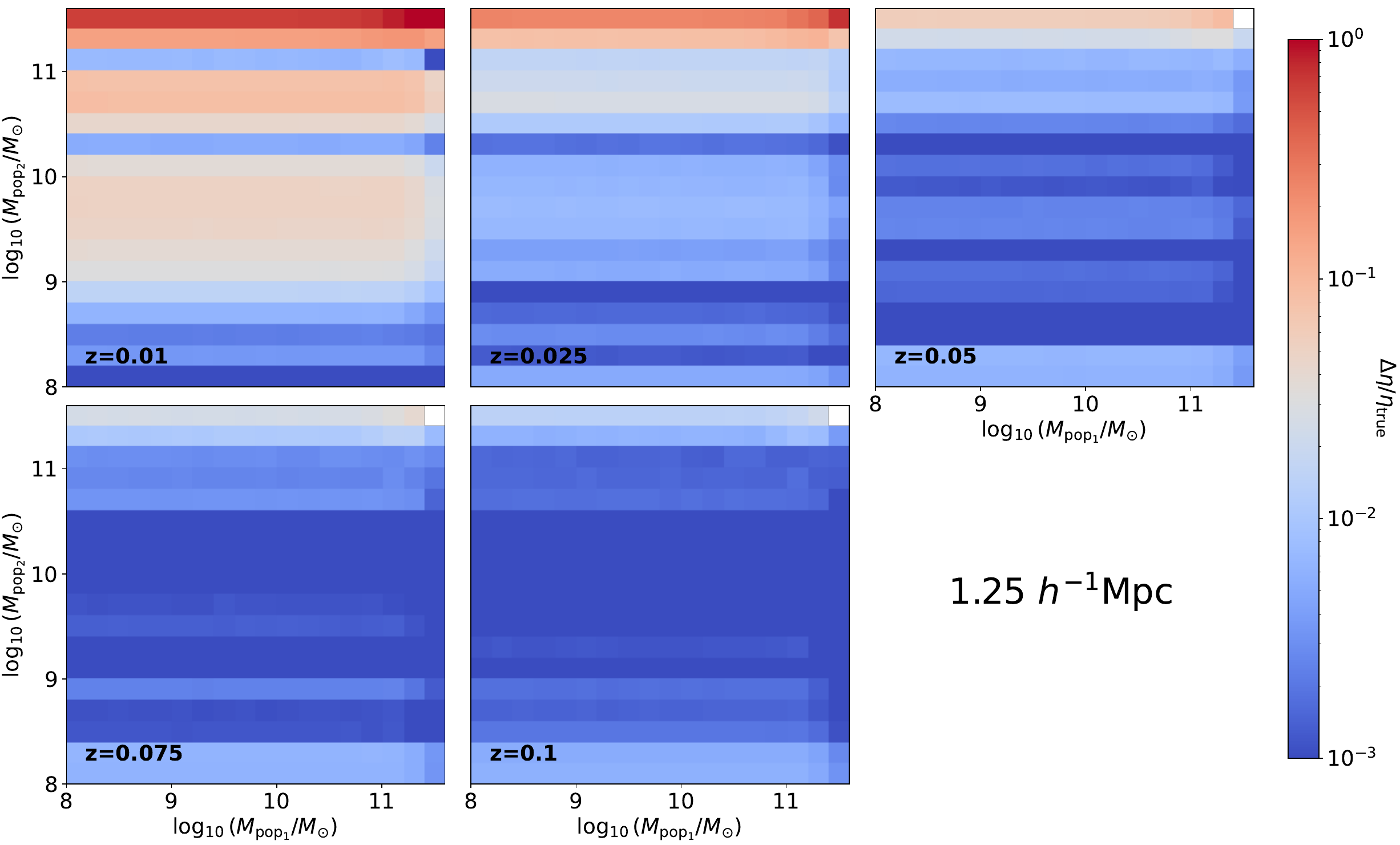}
    \includegraphics[width=0.9\textwidth]{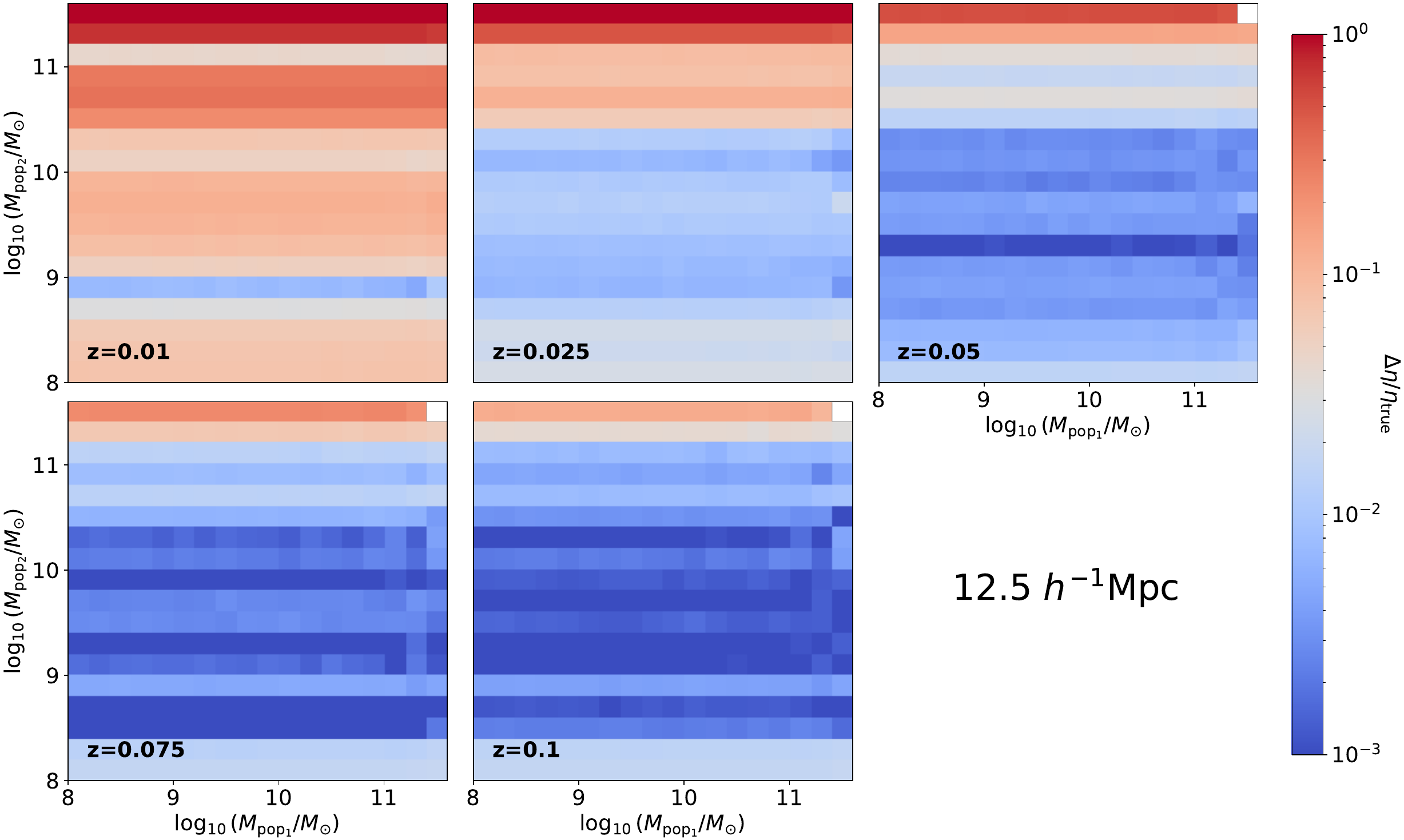}
    \caption{The accuracy $\Delta\eta/\eta_{\rm{true}}$ of the PAC method for various ${\rm{pop}}_1$ and ${\rm{pop}}_2$ stellar mass bins and different redshifts at two typical scales, $r_{\rm{p}}=1.25h^{-1}\rm{Mpc}$ and $12.5h^{-1}\rm{Mpc}$.}
    \label{fig:test_PAC}
\end{figure*}

Subsequently, we perform SED fitting for the entire DECaLS sample at the 44 reference redshifts. While it is possible to obtain the properties at the redshift of each ${\rm{pop}}_1$ source, such an approach is unrealistic and unnecessary. Instead, we opt for interpolating and saving the stellar masses at 634 redshifts within the range of $0.001<z<1$. The mean change in stellar mass between adjacent redshifts in this interpolated dataset is smaller than $\Delta\log_{10}M_{*}=0.01$, which provides a level of accuracy that is sufficient for our analysis. For example, if 2 adjacent reference redshifts among the 44 identified above show a mean $\Delta\log_{10}M_{*}=0.1$, we insert 9 additional redshifts in between, ensuring a consistent spacing of $\Delta\log_{10}M_{*}$ values through interpolation. For each ${\rm{pop}}_1$ source, we utilize the physical properties of ${\rm{pop}}_2$ at the nearest available redshift.

Although the above approach is illustrated using stellar mass as an example, it can be applied to any other physical properties under investigation. We have tested the precision of the interpolation for rest-frame $g$- and $r$-band luminosities and found similar accuracy. Consequently, we expect that the precision for the rest-frame $g-r$ colour is also sufficient for our analysis.

\subsection{A new estimator for the PAC method}\label{sec:3.3}

To address the second source of error and enable flexible data combinations
 without the necessity of redshift bins, we introduce a new estimator for the PAC method. 

In the previous PAC method, for each redshift bin we first compute $\omega_{12,\rm{w}}$ using the Landy–Szalay estimator \citep{1993ApJ...412...64L} with two corresponding random samples
\begin{equation}
    \omega_{12,\rm{w}}(\theta)=\frac{D_1D_2(\theta)-D_1R_2(\theta)-R_1D_2(\theta)+R_1R_2(\theta)}{R_1R_2(\theta)}\,,
\end{equation}
where $DD$, $DR$ and $RR$ are the data-data, data-random and random-random pair counts. The angle $\theta$ varies for different pairs of $D_1$ and $R_1$ and is defined as $\theta=\arcsin{(r_{\rm{p}}/r_1)}$. Subsequently, $\nwprp$ is obtained using Equation~\ref{eq:1} by multiplying it with $\bar{S}_2/r_1^2$, which is calculated using the mean redshift of each bin. 

Instead, we now incorporate $\bar{S}_2$ and $r_1^2$ as weights of ${\rm{pop}}_2$ and ${\rm{pop}}_1$ sources. First, we calculate $\bar{S}_2$ for galaxies with varying properties at the saved redshifts. For instance, if we want to calculate $\nwp$ for ${\rm{pop}}_2$ within $N$ stellar mass bins  at $z<0.2$ across 478 saved redshifts. This results in a total of $478\times N$ $\bar{S}_2$ measurements. Then, we compute $r_1^2$ for each source and random point of ${\rm{pop}}_1$. The weighted pairs are defined as:
\begin{align}
    &D_{1,\rm{w}}D_{2,\rm{w}} = D_2\bar{S}_2D_1/r^2_1\,, \notag\\
    &D_{1,\rm{w}}R_{2,\rm{w}} = R_2\bar{S}_2D_1/r^2_1\,, \notag\\
    &R_{1,\rm{w}}D_{2,\rm{w}} = D_2\bar{S}_2R_1/r^2_1\,, \notag\\
    &R_{1,\rm{w}}R_{2,\rm{w}} = R_2\bar{S}_2R_1/r^2_1\,, \notag
\end{align}
where $\bar{S}_2$ is chosen as the value corresponding to ${\rm{pop}}_2$ with the relevant properties of interest at the nearest redshift of each ${\rm{pop}}_1$ source or random point. Furthermore, the above formula does not include the weights from the spectroscopic catalogues as outlined in Section~\ref{sec:2.1}. These can be incorporated by simply multiplying them with $D_{1,\rm{w}}$, $R_{1,\rm{w}}$ and $R_1$. Finally, we  introduce the new estimator
\begin{equation}
    \nwprp = \frac{D_{1,\rm{w}}D_{2,\rm{w}}-D_{1,\rm{w}}R_{2,\rm{w}}-R_{1,\rm{w}}D_{2,\rm{w}}+R_{1,\rm{w}}R_{2,\rm{w}}}{R_1R_2}\,,\label{eq:3}
\end{equation}
where the pairs are counted in the bin around $\theta=\arcsin{(r_{\rm{p}}/r_1)}$. With this estimator, we can flexibly combine the measurements for ${\rm{pop}}_1$ as long as the evolution of $\nwp$ is sufficiently small to allow for statistical combination.

\subsection{Excluding foreground and background galaxies with colour cuts}\label{sec:3.4}
Deep magnitude-limited photometric samples typically encompass sources spanning a broad range of redshifts, with only a small fraction of them being correlated with ${\rm{pop}}_1$. Therefore, we implement conservative colour cuts to exclude as many foreground and background galaxies as possible, while retaining all galaxies within the redshift ranges of interest. This strategy helps minimize noise and enhances the precision of our measurements.

As an example, in this paper, we focus on the redshift range of interest at $z<0.2$. In Figure~\ref{fig:colour_cut}, we plot the DESI Y1 BGS galaxies in two narrower redshift bins $z<0.1$ and $0.1<z<0.2$ on the $g-r$ vs. $r-z$ colour-colour diagram. According to their colour distribution, we derived colour cuts for these two redshift ranges. For $z<0.1$,
\begin{align}
\left\{
\begin{array}{ll}
    -0.2 < g-r < 1.2, \\
    r-z < \frac{11}{14}(g-r) + \frac{9}{35}.
\end{array}
\right.
\end{align}
For $0.1<z<0.2$,
\begin{align}
\left\{
\begin{array}{ll}
    0<g-r<1.5\,, \notag \\
    r-z<1.1 \,, \notag \\
    r-z<\frac{8}{11}(g-r)+\frac{3}{10}\,.
\end{array}
\right.
\end{align}

The $g-r$ color cut range we used is sufficiently wide compared to the color distribution of BGS in Figure \ref{fig:colour_distribution}, ensuring that all low-redshift sources are retained. These colour cuts result in a reduction of more than half the galaxies in DECaLS. Additionally, it is important to note that $\bar{S}_2$ should also be calculated using the sample after applying the colour cut, rather than the entire sample. This approach can be extended to higher redshifts with an appropriate reference sample, which will be explored in future studies.

\subsection{Determining stellar mass limits}\label{sec:3.5}
While it is evident that DECaLS can extend to fainter magnitude and lower mass limits, determining these values is essential for fully exploiting the dataset. However, this task is not straightforward as we are reaching unexplored territory, with limited knowledge about these faint galaxies. In our previous studies \citepalias{2022ApJ...939..104X,2023ApJ...944..200X}, we first determined the $r$ or $z$ band galaxy depths of DECaLS. Then, we created apparent magnitude vs. stellar mass diagrams using GAMA DR4 data \citep{2022MNRAS.513..439D} with $V_{\rm{max}}$ correction for galaxies at $z<0.2$ and using DES deep field data \citep{2022MNRAS.509.3547H} with photoz for galaxies at $0.2<z<0.7$. Subsequently, we determined the stellar mass for which DECaLS achieves 95 per cent completeness at each redshift with the galaxy depths we obtained. These results are sensitive to the determination of galaxy depths, as well as to corrections for incompleteness, cosmic variance in GAMA, and photo-z errors in DES. Additionally, both GAMA and DESI struggle to measure enough galaxies with $M_{*}<10^{8.0}\ms$, making this method impractical for the lower mass end.

To address this issue, we present a straightforward method based only on DECaLS itself to establish the stellar mass limit in the PAC method. We show an example for $z=0.025$ in Figure~\ref{fig:compare_SM}. We select sources from the deepest regions where the $r$-band $10\sigma$ PSF depth exceeds 23.8 and compare the angular surface density of galaxies as a function of stellar mass, assuming all galaxies are at $z=0.025$, between these deep regions and the entire DECaLS sample. According to the cosmological principle, the angular surface density of DECaLS should be identical to that of the deeper regions for the complete mass range and should begin to drop once it reaches the stellar mass limit. We define the stellar mass at which the angular surface density of DECaLS decreases to 95 per cent of that in the deeper sample as the stellar mass limit. As an example, we present the results for the entire population at $z=0.025$ in Figure~\ref{fig:compare_SM}, where we find that the stellar mass limit in DECaLS is $10^{7.2}\ms$.

We repeat the process at various redshifts for $z<0.2$ for the entire, blue, and red populations. We present the results in the top panel of Figure~\ref{fig:SM_limits} and also plot the results for the entire population from \citetalias{2022ApJ...939..104X} based on GAMA data for comparison. The results are discrete because we measure the angular surface density in stellar mass bins with an interval of $\Delta\log_{10}(M_{*}/\ms)=0.2$. The stellar mass limits are nearly identical for the red and entire populations but are lower for blue galaxies. This is because red galaxies have higher mass-to-light ratios than blue ones, making them fainter at fixed stellar masses. As a result, the stellar mass limits for the entire population are primarily determined by the red galaxies. Compared to \citetalias{2022ApJ...939..104X}, we find that the stellar mass limits derived from GAMA are overly optimistic. This may suggest that GAMA does not fully sample the entire colour space of low-mass galaxies, potentially missing red ones. In the bottom panel, we further display the effective volume calculated using the maximum complete redshift, as well as with a redshift cutoff of $z>0.01$, for various stellar masses. Our results indicate the potential to investigate galaxies with stellar masses of $10^{6}\ms$ within a sufficiently large volume using the PAC method with BGS and DECaLS.

\subsection{Testing the approximation in the PAC method for $z<0.1$}\label{sec:3.6}

When utilizing low redshift data, it becomes necessary to re-evaluate the approximation in Equation~\ref{eq:1}. \citet{2011ApJ...734...88W} has demonstrated an accuracy of 3 per cent even at $r_{\rm{p}}=30h^{-1}\rm{Mpc}$ for various $r$-band absolute magnitude bins at $z=0.07$. We intend to expand the analysis to lower redshifts, encompass a broader range of galaxy populations, and incorporate additional sources of error that were not considered in \citet{2011ApJ...734...88W}.

The first error arises from the different definitions of $\Wp(r_{\rm{p}})$ and $\omega_{12}(\theta)$ that was comprehensively addressed in \citet{2011ApJ...734...88W}. Assuming we have a ${\rm{pop}}_1$ sample at a specific redshift, by definition, we have
\begin{align}
    \omega_{12}(\theta)&=\frac{\int_0^{\infty}\Omega \, \bar{n}_2(r_2) r_2^2\xi(r_{12}){\rm d}r_2}{\Omega \bar{S}_2}\,,
\end{align}
where $\Omega$ is the total sky coverage, $r_2$ denotes the comoving distance to ${\rm{pop}}_2$, $\xi$ is the cross-correlation between ${\rm{pop}}_1$ and ${\rm{pop}}_2$ and
\begin{equation}
    r_{12}=\sqrt{r_1^2+r_2^2-2r_1r_2\cos{\theta}}\,.
\end{equation}
If we consider volume-limited samples with constant mean densities as in this paper, we have $\bar{n}_2(r_2)=\bar{n}_2$ and
\begin{equation}
    \omega_{12}(\theta)=\frac{\bar{n}_2}{\bar{S}_2}\int_0^{\infty} r_2^2\xi(r_{12}){\rm d}r_2\,.
\end{equation}
While, for $\Wp(r_{\rm{p}})$ we have
\begin{equation}
    \Wp(r_{\rm{p}})=\int_0^{\infty}\xi(r_{12}){\rm d}r_2\,,
\end{equation}
where $r_{\rm{p}}=r_1\sin{\theta}$ is used to obtain the best match. We can define the ratio between $\omega_{12}(\theta)$ and $\Wp(r_{\rm{p}})$ as
\begin{equation}
    \eta_{\rm{true}} = \frac{\bar{n}_2\int_0^{\infty} r_2^2\xi(r_{12}){\rm d}r_2}{\bar{S}_2\int_0^{\infty}\xi(r_{12}){\rm d}r_2}\,,
\end{equation}
while the approximated ratio according to Equation~\ref{eq:1} is
\begin{equation}
    \eta_{\rm{approx}}=\frac{\bar{n}_2r_1^2}{\bar{S}_2}\,.
\end{equation}
Therefore, we can test the accuracy of this approximation by calculating $\Delta\eta/\eta_{\rm{true}}=|\eta_{\rm{true}}-\eta_{\rm{approx}}|/\eta_{\rm{true}}$. It is evident that if $r_1$ is significantly larger than the correlation length, we have
\begin{equation}
\int_0^{\infty} r_2^2\xi(r_{12}){\rm d}r_2\approx r_1^2\int_0^{\infty}\xi(r_{12}){\rm d}r_2
\end{equation}
and $\Delta\eta/\eta_{\rm{true}}\approx 0$, since $r_2^2$ changes much more slowly than $\xi(r_{12})$.

However, there is another source of error that has not been considered in \citet{2011ApJ...734...88W}. Suppose we are interested in obtaining measurements for ${\rm{pop}}_2$ within a specific stellar mass bin $[M_{\rm{min}},M_{\rm{max}}]$ and the stellar masses of ${\rm{pop}}_2$ are calculated at the redshift of ${\rm{pop}}_1$ as described in Section~\ref{sec:3.2}. When we compute the angular correlation, ${\rm{pop}}_1$ is correlated with some galaxies with $M_*>M_{\rm{max}}$ at $r_2>r_1$ and $M_*<M_{\rm{min}}$ at $r_2<r_1$ because the masses of correlated galaxies are either underestimated or overestimated due to the inaccurate redshift assumed. This effect can impact the measurements because both $\bar{n}_2$ and $\xi(r_{12})$ depend on stellar mass, which is also negligible when $r_1$ is significantly larger than the correlation length. To investigate this effect, we re-write the $\omega_{12}(\theta)$ we measured as
\begin{equation}
    \omega_{12}(\theta)=\frac{1}{\bar{S}_2}\int_0^{\infty}\int_{M_{\rm{min}}}^{M_{\rm{max}}}\Phi\bigg( M_*\frac{r_{2,\rm{L}}^2}{r_{1,\rm{L}}^2}\bigg)\xi\bigg(r_{12},M_*\frac{r_{2,\rm{L}}^2}{r_{1,\rm{L}}^2}\bigg)r_2^2{\rm d}M_*{\rm d}r_2\,,
\end{equation}
where $r_{1,\rm{L}}$ and $r_{2,\rm{L}}$ are the luminosity distances to ${\rm{pop}}_1$ and ${\rm{pop}}_2$ and $\Phi$ is the stellar mass function. The above equation is derived under the assumption that the stellar mass of ${\rm{pop}}_2$ scales solely with the square of the luminosity distances within the correlation length, and that ${\rm{pop}}_2$ is complete in stellar mass, allowing $\bar{n}_2$ to be replaced by the stellar mass function $\Phi$. Hence, the ratio can be reformulated as
\begin{equation}
    \eta_{\rm{true}} = \frac{\int_0^{\infty}\int_{M_{\rm{min}}}^{M_{\rm{max}}}\Phi\bigg( M_*\frac{r_{2,\rm{L}}^2}{r_{1,\rm{L}}^2}\bigg)\xi\bigg(r_{12},M_*\frac{r_{2,\rm{L}}^2}{r_{1,\rm{L}}^2}\bigg)r_2^2{\rm d}M_*{\rm d}r_2}{\bar{S}_2\int_0^{\infty}\xi(r_{12}){\rm d}r_2}\,. \label{eq:13}
\end{equation}

To estimate the accuracy of the approximation $\Delta\eta/\eta_{\rm{true}}$ for various redshifts and stellar mass bins, we employ the precise stellar-halo mass relation (SHMR) from \citetalias{2023ApJ...944..200X} and the {\tt{CosmicGrowth}} $N$-body simulation \citep{2019SCPMA..6219511J} with a subhalo catalogue constructed using {\tt{HBT+}} \citep{2012MNRAS.427.2437H,2016MNRAS.457.1208H}. The {\tt{CosmicGrowth}} simulation has a mass resolution of $5\times10^{8}\msh$ and a box size of $600 \, h^{-1} \, \text{Mpc}$, which is appropriate for studying the distribution of galaxies with $M_*>10^{8}\ms$ according to \citetalias{2023ApJ...944..200X}. We derive $\xi(r,M_*)$ and $\Phi(M_*)$ using the SHMR and simulation data, and then estimate $\Delta\eta/\eta_{\rm{true}}$ across different redshifts and for varying stellar masses of ${\rm{pop}}_1$ and ${\rm{pop}}_2$. Figure~\ref{fig:test_PAC} presents the results at 2 characteristic scales, $r_{\rm{p}}=1.25h^{-1}\rm{Mpc}$ and $12.5h^{-1}\rm{Mpc}$, across 5 redshifts $z=0.01, 0.025, 0.075$, and $0.1$, and within 18 stellar mass bins for ${\rm{pop}}_1$ and ${\rm{pop}}_2$, spanning the range from $10^{8.0}\ms$ to $10^{11.6}\ms$. We find that $\Delta\eta/\eta_{\rm{true}}$ is smaller for smaller scales, lower ${\rm{pop}}_2$ mass, and higher redshift, with its dependence on the mass of ${\rm{pop}}_1$ being nearly negligible. There are indications that $\Delta\eta/\eta_{\rm{true}}$ begins to increase when $M_{\rm{pop}_2}<10^{9.0}\ms$. This behaviour is attributed to the fact that within these mass bins, ${\rm{pop}}_1$ starts to correlate with some galaxies having $M_{\rm{pop}_2}<10^{8.0}\ms$, whose host halos or subhalos are unresolved by the simulation. In general, we can maintain an accuracy of better than 3 per cent for almost all the stellar mass bins at $z=0.05$ within a scale of up to $12.5h^{-1}\rm{Mpc}$. It is worth noting that the measurements become less accurate in the most massive ${\rm{pop}}_2$ bins at lower redshifts. However, these redshifts are primarily intended for the study of low mass galaxies and have fewer massive galaxies. The PAC method remains highly accurate for $M_{\rm{pop}_2}<10^{9.0}\ms$ even at $z=0.01$, and we anticipate that this level of accuracy extends to lower mass galaxies. Although the test above was conducted for the entire sample, we expect that the conclusion also holds for the red and blue sub-samples. This demonstrates the potential of the PAC method for exploring the unknown region of faint galaxies. 

\begin{figure*}
    \centering
    \includegraphics[width=\textwidth]{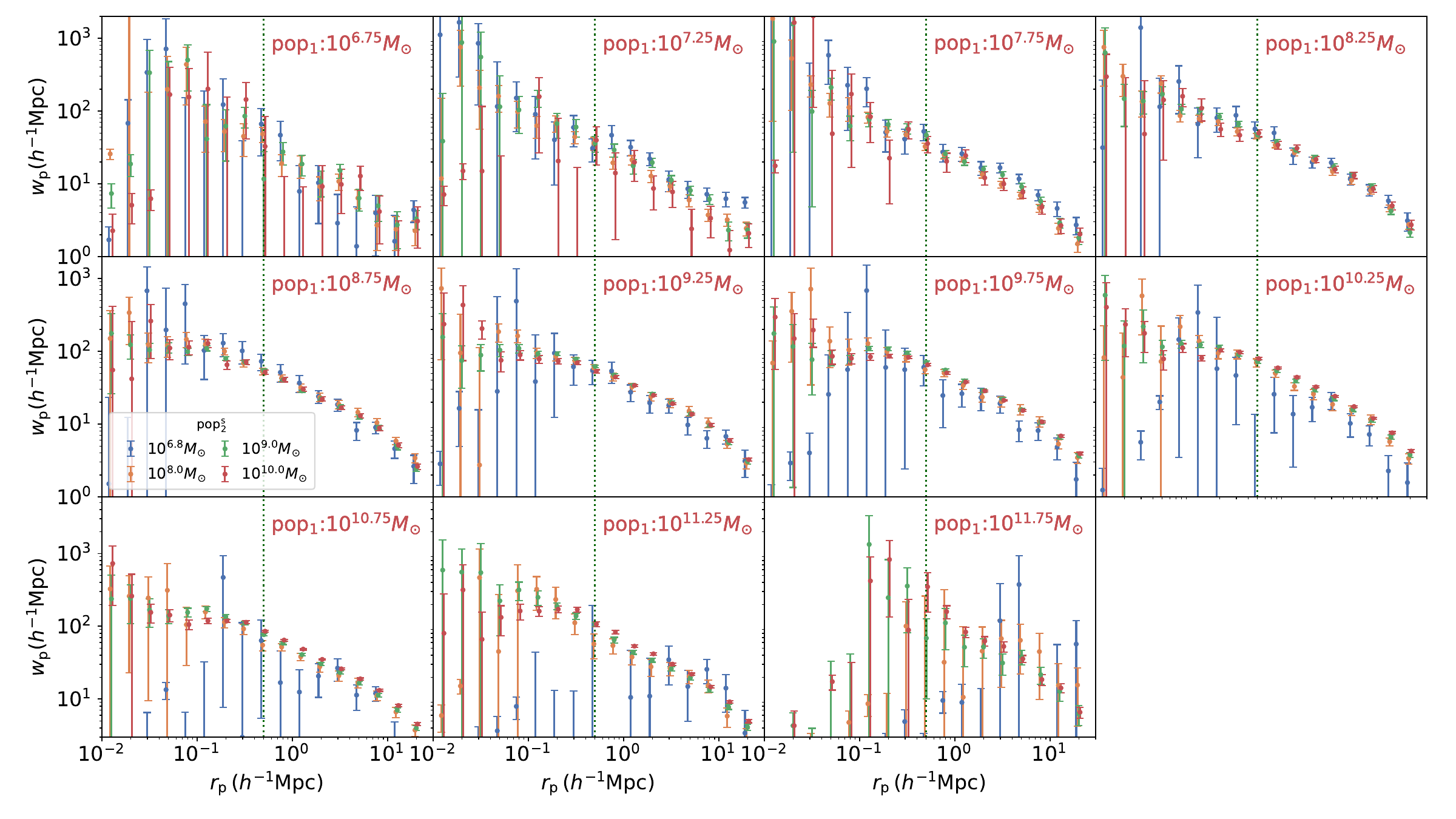}
    \caption{Comparison of $\Wp$ for \textit{blue} ${\rm{pop}_2^{\rm{s}}}$ across various stellar mass bins with $M_* < 10^{10.0}\ms$. Each panel presents the comparison for one of the 11 ${\rm{pop}_1}$ stellar mass bins. The dotted vertical lines indicate a scale of 0.05$^{\circ}$ at $z = 0.2$, where $\Wp$ may be affected by fibre incompleteness. Plots are slightly shifted horizontally for better visualization.}
    \label{fig:wp_blue}
\end{figure*}

\begin{figure*}
    \centering
    \includegraphics[width=\textwidth]{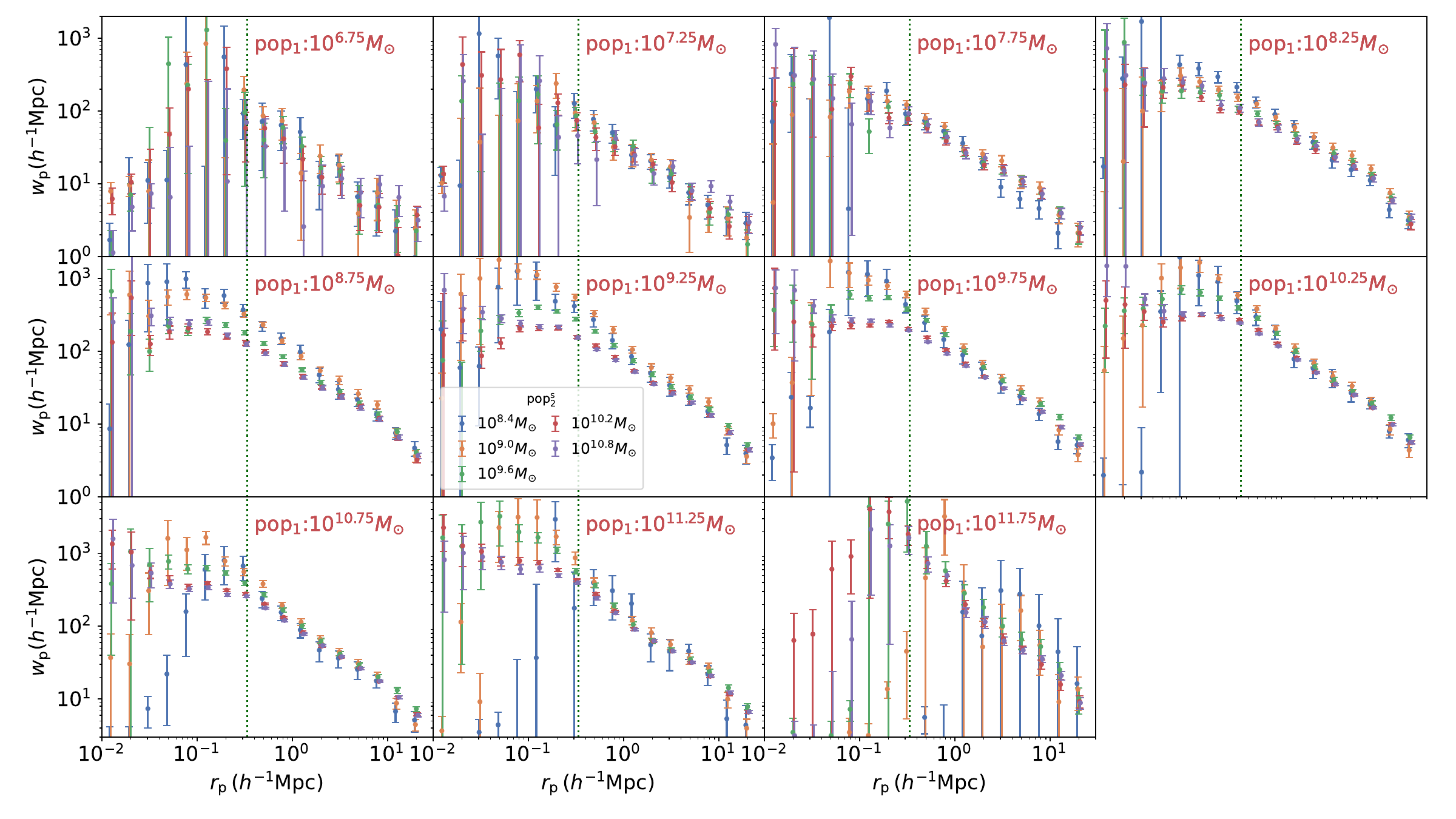}
    \caption{Comparison of $\Wp$ for {\textit{red}} ${\rm{pop}_2^{\rm{s}}}$ across various stellar mass bins at $M_*<10^{10.8}\ms$. Each panel shows the comparison for one of the 11 ${\rm{pop}_1}$ stellar mass bins. The dotted vertical lines indicate a scale of 0.05$^{\circ}$ at $z=0.2$, where $\Wp$ may be impacted by fibre incompleteness. Plots are slightly shifted horizontally for better visualization.}
    \label{fig:wp_red}
\end{figure*}
\begin{figure*}
    \centering
    \includegraphics[width=\textwidth]{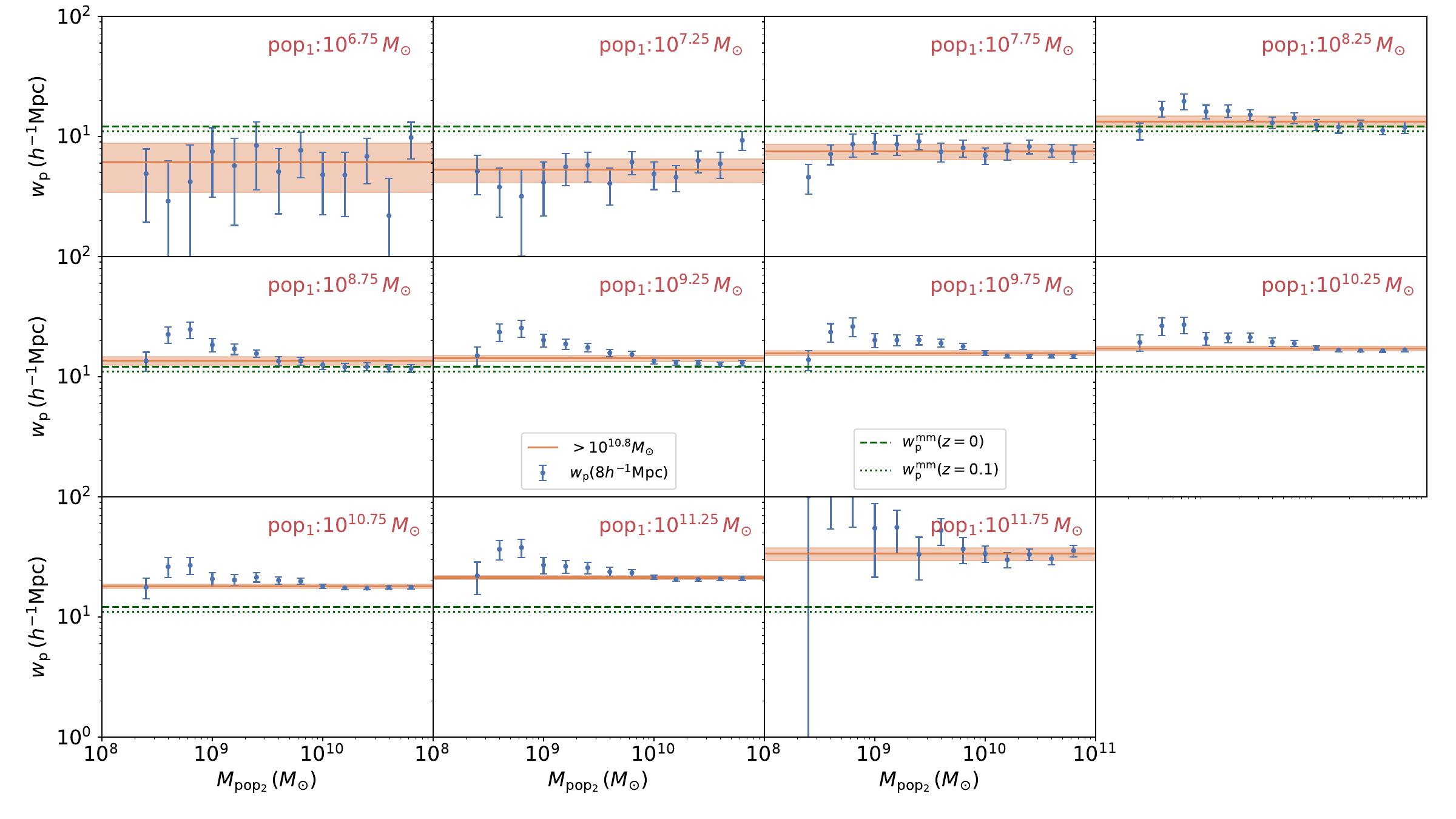}
    \caption{Comparison of $\Wp$ at $8h^{-1}\rm{Mpc}$ for \textit{red} ${\rm{pop}_2^{\rm{s}}}$ across various stellar mass bins. Each panel corresponds to one of the 11 ${\rm{pop}_1}$ stellar mass bins. The orange solid lines with shaded regions represent the results from the $\Wp$ of all ${\rm{pop}_2^{\rm{s}}}$ with $M_*<10^{10.8}\ms$. For comparison, the projected matter correlations $\Wp^{\rm{mm}}$ at redshifts 0 (dashed lines) and 0.1 (dotted lines) calculated using {\texttt{CAMB}} \citep{2000ApJ...538..473L} are also shown.}
    \label{fig:wp_red_8mpc}
\end{figure*}

\section{First Application with DESI Y1 BGS}\label{sec:4}
In this section, we present the first application of the improved PAC method using the DESI Y1 BGS and DECaLS samples. We derive $\nwp$ for blue and red ${\rm{pop}}_2$ galaxies down to stellar masses of $10^{5.3}\ms$ and $10^{6.3}\ms$, respectively. Following the approach of \citetalias{2022ApJ...939..104X}, we obtain measurements of the GSMFs for blue and red galaxies at these mass thresholds by combining $\nwp$ with $\Wp$ from the BGS dataset. We then combine these to obtain the total GSMF.

\subsection{Methodology for measuring GSMF}\label{sec:4.1}
\citetalias{2022ApJ...939..104X} proposed a method to directly derive the galaxy GSMF by combining $\nwp$ from the PAC method with $\Wp$ from spectroscopic surveys, which may not be complete in stellar mass. In this approach, to match ${\rm{pop}}_2$ from the photometric sample, the corresponding ${\rm{pop}}_2^{\rm{s}}$ is selected from the spectroscopic sample within the same stellar mass bin and cross-correlated with ${\rm{pop}}_1$ to obtain $\Wp$. The GSMF is then estimated by comparing $\nwp$ with $\Wp$ within each stellar mass bin. This method is straightforward and relies on the key assumption that galaxy bias is primarily determined by stellar mass, such that ${\rm{pop}}_2^{\rm{s}}$ and ${\rm{pop}}_2$ have similar biases, even if ${\rm{pop}}_2^{\rm{s}}$ is highly incomplete. In this paper, we use a similar approach with several improvements and extensions.

Although $\Wp$ is primarily determined by stellar mass, residual scatter has been observed in numerous studies, which has been shown to depend on secondary properties such as colour and morphology \citep{2006MNRAS.368...21L,2016ApJ...819...58W,2022ApJ...926..130X,2022ApJ...928...10G}. For example, massive red central galaxies are found in more massive halos compared to massive blue ones, resulting in stronger clustering \citep{2016ApJ...819...58W,2022ApJ...926..130X}. Moreover, red galaxies have a higher satellite fraction than blue galaxies \citep{2022ApJ...928...10G}. These red satellites reside in more massive halos, which enhances the clustering of red galaxies at small scales (1-halo term) and slightly increases it at large scales (2-halo term). Consequently, at a fixed stellar mass, red galaxies exhibit higher clustering than blue galaxies, with the difference becoming more pronounced on smaller scales \citep{2006MNRAS.368...21L}. Therefore, the method used in \citetalias{2022ApJ...939..104X} may overestimate the GSMF, as red galaxies, with their higher mass-to-light ratios, are often among the first to be excluded when the sample becomes incomplete. This bias skews the sample toward a higher fraction of blue galaxies, leading to lower observed $\Wp$.

To mitigate this bias, we further divide ${\rm{pop}}_2$ into red and blue samples within each stellar mass bin. Next, we select ${\rm{pop}}_2^{\rm{s}}$ to match both stellar mass and colour in order to estimate the corresponding $\Wp$. The GSMFs for blue and red galaxies are then derived separately. Finally, the total GSMF is obtained by combining these two components. This approach can be considered a more refined treatment of galaxy bias, accounting for the fact that galaxy bias depends on both stellar mass and colour, rather than just stellar mass. In principle, this method could be extended by incorporating additional galaxy properties or subdividing the sample into finer colour bins instead of just blue and red. However, we find that the current approach provides satisfactory results, and thus, we do not pursue further refinements.

Additionally, we also divide ${\rm{pop}}_1$ into different stellar mass bins following \citetalias{2022ApJ...939..104X}. For a given ${\rm{pop}}_2$ sample, the value of $\bar{n}_2$ from each ${\rm{pop}}_1$ sample should be consistent, as the method is independent of the properties of ${\rm{pop}}_1$. Therefore, comparing results from ${\rm{pop}}_1$ across different stellar mass bins serves as a robust test of our method. Additionally, refining ${\rm{pop}}_1$ into smaller stellar mass bins makes the $\nwp$ and $\Wp$ measurements easier to interpret.

\subsection{Measurements of $\nwp$}
Let $\mathcal{A}\equiv\nwp$ for a more concise representation. We assume that $\mathcal{A}$ is calculated in $N_{\rm{s}}$ sky regions, such as DECaLS NGC and DECaLS SGC in this work. Each sky region is further divided into $N_{\rm{sub}}$ sub-regions for error estimation using jackknife resampling. We calculate $\mathcal{A}_{i,j}$ according to Equation~\ref{eq:3} in the $i$th sky region and $j$th jackknife sub-sample. To combine measurements from different sky regions, we weight them by the respective region areas $w_{\rm{s},i}$ as follows:

\begin{equation}
\mathcal{A}_{j} = \frac{\sum_{i=1}^{N_{\rm{s}}}\mathcal{A}_{i,j}w_{\rm{s},i}}{\sum_{i=1}^{N_{\rm{s}}}w_{\rm{s},i}}\,.
\end{equation}

Subsequently, we estimate the mean values and the covariance matrices of the mean values using $N_{\rm{sub}}$ sub-samples:

\begin{equation}
\mathcal{A} = \frac{\sum_{j=1}^{N_{\rm{sub}}}\mathcal{A}_{j}}{N_{\rm{sub}}}\,,
\end{equation}

\begin{align}
    C_{\mathcal{A},ab} = \frac{N_{\rm{sub}}-1}{N_{\rm{sub}}}\sum_{j=1}^{N_{\rm{sub}}}(\mathcal{A}_{j}(r_{\rm{p}}^a)-\mathcal{A}(r_{\rm{p}}^a))
    (\mathcal{A}_{j}(r_{\rm{p}}^b)-\mathcal{A}(r_{\rm{p}}^b))\,,
\end{align}
where $a$ and $b$ denotes the $a$th and $b$th radial bins.

We measure $\nwp$ for ${\rm{pop}_2}$ from DECaLS over a wide stellar mass range of $[10^{6.3}\ms, 10^{11.9}\ms]$ for red galaxies and $[10^{5.3}\ms, 10^{11.9}\ms]$ for blue galaxies, using a bin width of $\Delta \log_{10}(M_*/\ms) = 0.2$, cross-correlated with ${\rm{pop}_1}$ from DESI BGS in 11 stellar mass bins within $[10^{6.5}\ms, 10^{12.0}\ms]$ and with a bin width of $\Delta \log_{10}(M_*/\ms) = 0.5$. The calculation is done across the redshift ranges where ${\rm{pop}_2}$ is complete, according to the results in Figure~\ref{fig:SM_limits}. The completeness of stellar mass for ${\rm{pop}_1}$ is not considered, as the results are independent of the properties of ${\rm{pop}_1}$, as mentioned in Section~\ref{sec:4.1}. Our analysis covers spatial scales ranging from $0.01\,h^{-1}\rm{Mpc}$ to $25\,h^{-1}\rm{Mpc}$, showcasing an additional advantage of the PAC method as it remains unaffected by fibre incompleteness, which is a main challenge for small scale measurements in spectroscopic surveys. 

\subsection{Measurements of $\Wp$}
The cross-correlation functions $\xi(r_{\rm{p}}, \pi)$ are estimated using the Landy-Szalay estimator \citep{1993ApJ...412...64L} within the corresponding $r_{\rm{p}}$ bins and 40 $\pi$ bins, ranging from $0$ to $40\,h^{-1}\rm{Mpc}$ in equal linear intervals, computed with \texttt{Corrfunc} \citep{2020MNRAS.491.3022S}. The projected correlation functions $\Wp$ are then calculated as:
\begin{equation}
    \Wp(r_{\rm{p}}) = 2\int_0^{\pi_{\rm{max}}}\xi(r_{\rm{p}},\pi)d\pi\,,
\end{equation}
where we set $\pi_{\rm{max}} = 40\,h^{-1}\rm{Mpc}$. We also tested \(\pi_{\rm{max}} = 80\,h^{-1} \rm{Mpc}\) and found only a slight impact on the final GSMF results, well within the statistical error. The covariance matrices are estimated through jackknife resampling, employing the same sub-regions as those used for $\nwp$.

For $\nwp$ in each ${\rm{pop}_2}$ and ${\rm{pop}_1}$ mass bin, we select ${\rm{pop}}_2^{\rm{s}}$ from BGS and cross-correlate it with ${\rm{pop}_1}$ to obtain the corresponding $\Wp$. We find that we can measure $\Wp$ for ${\rm{pop}}_2^{\rm{s}}$ down to $10^{6.3}\ms$ for blue galaxies and $10^{8.1}\ms$ for red galaxies. At lower masses, however, there are too few galaxies in BGS to derive $\Wp$. However, as discussed in Section~\ref{sec:4.4}, we have $\nwp$ measurements down to $10^{5.3}\ms$ for blue galaxies and $10^{6.3}\ms$ for red galaxies. To utilize the $\nwp$ measurements and extend GSMF estimates to lower masses, we develop methods to estimate $\Wp$ at these lower masses. 

For blue galaxies, we present the measurements of $\Wp$ for ${\rm{pop}}_2^{\rm{s}}$ across stellar mass bins ranging from $10^{6.8}\ms$ to $10^{10.0}\ms$ in Figure~\ref{fig:wp_blue}. Each panel shows the results for one of the 11 ${\rm{pop}_1}$ stellar mass bins. To illustrate the scale at which $\Wp$ may be impacted by fibre incompleteness, we include a dotted line at $0.05^{\circ}$ for $z=0.2$. As shown in Figure~\ref{fig:bgs_distribution}, low-mass galaxies may be found at lower redshifts, allowing for $\Wp$ measurements at smaller $r_{\rm{p}}$. The angle $0.05^{\circ}$ is selected as the patrol diameter of the fibre positioners extends up to $\sim0.0495^{\circ}$.  This choice is validated by tests in \citet{2019MNRAS.484.1285S}. We find that blue galaxies have nearly identical $\Wp$ values at $M_* < 10^{10.0} \ms$. This can be explained by the fact that blue galaxies are predominantly central galaxies \citep{2022ApJ...928...10G,2023ApJ...954..207G}. Their correlation functions are largely driven by halo-halo correlation, with subhalo contributions being minimal. It is well-established that halo bias remains nearly constant for low-mass halos ($M_{\rm{h}}<10^{12.0}\msh$) \citep{1998ApJ...503L...9J,2010ApJ...724..878T}, which could explain why low-mass blue galaxies show nearly the same correlation functions. To extend the GSMF for blue galaxies to $M_* < 10^{6.3} \ms$, for these mass bins, we measure the $\Wp$ from all the blue ${\rm{pop}}_2^{\rm{s}}$ with $M_* < 10^{9.0} \ms$, and assume that it is the $\Wp$ of ${\rm{pop}}_2$. To validate this approximation, we also apply it to ${\rm{pop}}_2$ with $10^{6.3} \ms < M_* < 10^{9.1} \ms$ and compare the resulting GSMF with the fiducial results.

For red galaxies, we present their $\Wp$ measurements across several stellar mass bins ranging from $10^{8.4} \ms$ to $10^{10.8} \ms$ in Figure~\ref{fig:wp_red}. In contrast to blue galaxies, we observe a clear dependence of $\Wp$ on stellar mass, with a steeper slope at lower masses. This could be attributed to the fact that a significant fraction of low-mass red galaxies are satellites, and their fraction, host halo mass, and radial distribution may vary with stellar mass. As shown in Figure~\ref{fig:wp_red}, lower mass red galaxies appear to have a higher satellite fraction, which contributes to the steeper $\Wp$ slope. Despite this stellar mass dependence, we find that $\Wp$ values for different stellar mass bins are similar at $r_{\rm{p}} \sim 10 \, h^{-1} \rm{Mpc}$. This consistency is expected. At \(8\,h^{-1} \mathrm{Mpc}\), $\Wp$ is dominated by the 2-halo term, meaning its amplitude is primarily determined by the mean host halo bias of the red galaxies. Halo bias changes slightly for \( M_{\rm{h}} < 10^{12} \msh \) (0.6-0.8) and even up to \( M_{\rm{h}} < 10^{13} \msh \) \citep[$\sim 1.0$;][]{1998ApJ...503L...9J,2010ApJ...724..878T}. As long as most low-mass galaxies reside in halos with \( M_{\rm{h}} < 10^{13} \msh \), $\Wp$ at \(8\,h^{-1} \mathrm{Mpc}\) should remain relatively constant across different stellar masses. This holds true for two reasons: (1) If they are centrals, they must be hosted by low-mass halos. (2) For satellites, the number density of halos with \( M_{\rm{h}} > 10^{13} \msh \) decreases exponentially, so most satellites are expected to reside in relatively lower-mass host halos. To estimate $\Wp$ for red galaxies with $M_* < 10^{8.1} \ms$, we assume that their $\Wp$ at $8 \, h^{-1} \rm{Mpc}$ is consistent with the $\Wp$ of all red ${\rm{pop}}_2^{\rm{s}}$ galaxies with $M_* < 10^{10.8} \ms$. Additionally, we assume that the correlation function follows a power-law form, $\xi = (r/r_0)^{-\gamma}$, which implies that $\Wp$ also follows a power law:
\begin{equation}
    \Wp(r_{\rm{p}})=r_{\rm{p}}\left(\frac{r_{\rm{p}}}{r_0}\right)^{-\gamma}\Gamma\left(\frac{1}{2}\right)\Gamma\left(\frac{\gamma-1}{2}\right)/\Gamma\left(\frac{\gamma}{2}\right)\,\,,\label{eq:wp}
\end{equation}
where $\Gamma$ represents the Gamma function. The assumption of a power-law form for $\Wp$ is validated by the $\nwp$ measurements for these stellar mass bins. With these assumptions, there is effectively only one degree of freedom left for $\Wp$. To derive the GSMF for lower mass bins, we simultaneously fit $\nwp$ in each stellar mass bin and $\Wp$ at $8 \, h^{-1} \rm{Mpc}$ from all the red ${\rm{pop}}_2^{\rm{s}}$ galaxies with $M_* < 10^{10.8} \ms$ using Equation~\ref{eq:wp}. This allows us to constrain both $\bar{n}_2$ and $\gamma$. To better illustrate the impact of these assumptions, Figure~\ref{fig:wp_red_8mpc} presents $\Wp$ at $8h^{-1}\rm{Mpc}$ for pop$_2^{\rm{s}}$ across different stellar mass bins. We observe a slight increase in $\Wp$ starting from $10^{9.0}\ms$ toward lower masses, followed by a decrease at $10^{8.4}\ms$. The assumption holds reasonably well for pop$_1$ with $M_*<10^{8.5}\ms$, but shows larger deviations for pop$_1$ with higher masses and for pop$_2$ at $10^{8.6}\ms$ and $10^{8.8}\ms$. To validate this method, we apply it to stellar mass bins up to $10^{10.8} \ms$ and compare the results in mass bins where $\Wp$ can be directly obtained from BGS.

Additionally, in a companion paper, we plan to derive the GSMF for the entire sample by modelling their $\nwp$ using the subhalo abundance matching method within numerical simulations, as developed in \citetalias{2023ApJ...944..200X}. This approach will further validate our results, as it does not rely on assumptions about $\Wp$ and is solely based on the $\nwp$ measurements.

\begin{figure*}
    \centering
    \includegraphics[width=\textwidth]{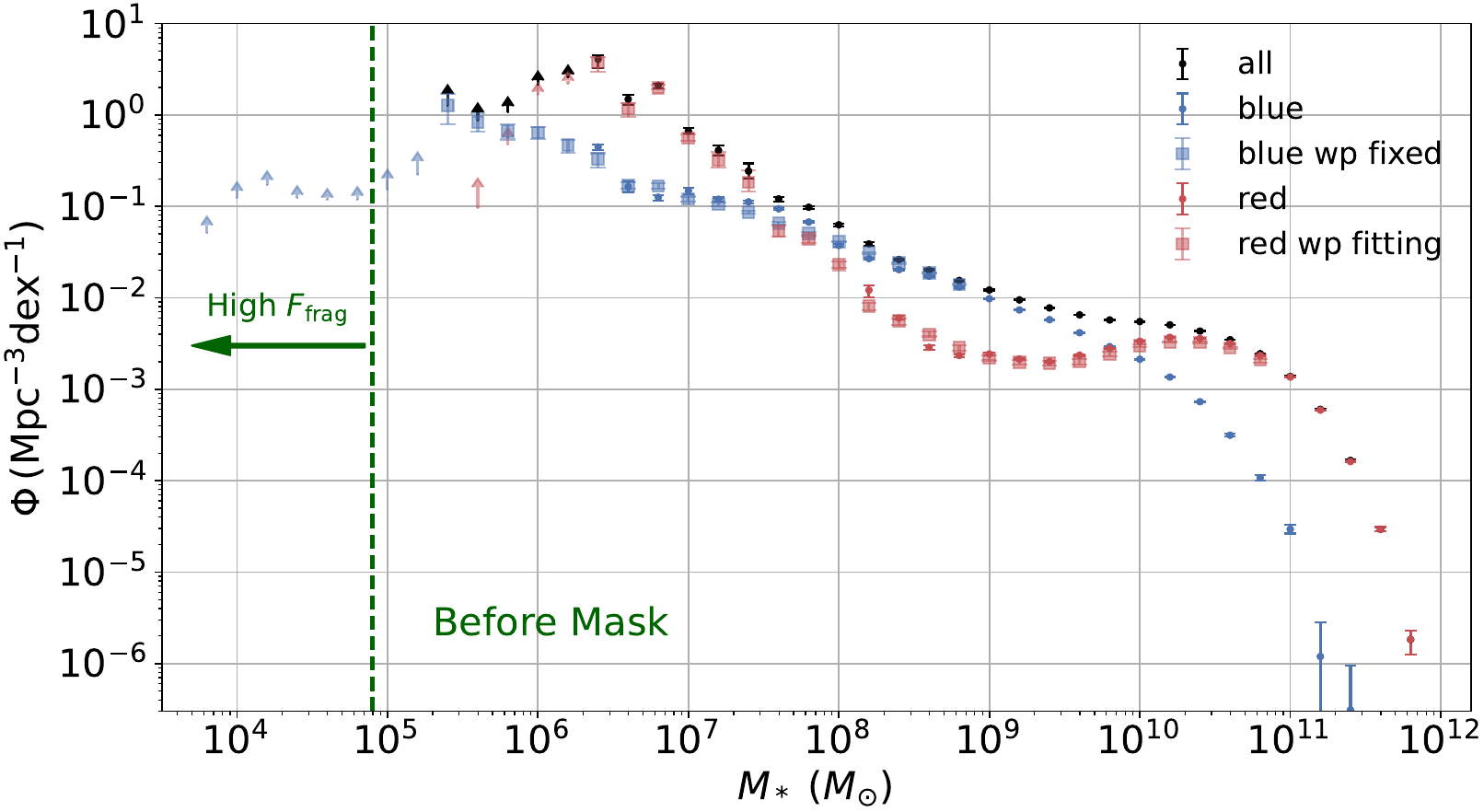}
    \caption{GSMFs for blue, red, and the entire samples before masking regions within 2$R_{e}$ of massive galaxies. Red and blue dots with error bars represent the fiducial results by combining $\nwp$ and $\Wp$ for red and blue galaxies, respectively. Blue squares indicate the results where $\Wp$ is derived from all blue pop$^{\rm{s}}_2$ galaxies with $M_*>10^{9.0}\ms$, while red squares correspond to results where $\Wp$ comes from model fitting. Blue and red arrows denote the lower limits due to incomplete samples, and black dots and arrows represent the GSMF and lower limits for the whole galaxy sample.}
    \label{fig:GSMF_before_mask}
\end{figure*}

\begin{figure*}
    \centering
    \includegraphics[width=\textwidth]{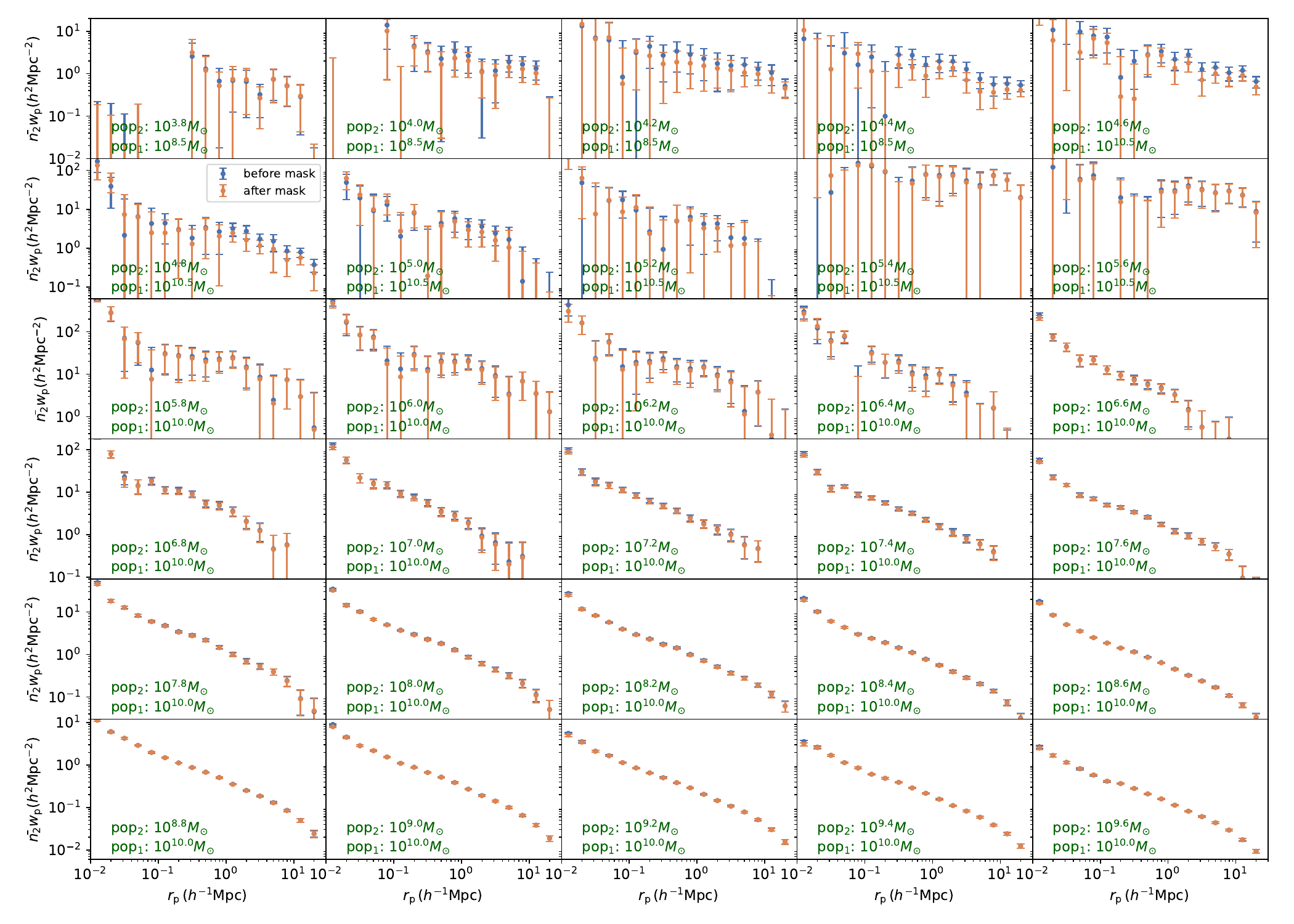}
    \caption{Comparison of \(\nwp\) measurements before (blue) and after (orange) masking regions around massive BGS galaxies for each \textit{blue} pop$_2$ stellar mass bin. Results from one selected pop$_1$ stellar mass bin are shown for each pop$_2$ stellar mass bin. The masking thresholds are set as follows: \(M_*^{\rm{mask}} = 10^{8.0} \, M_\odot\) for pop$_2$ with \(M_* < 10^{8.0} \, M_\odot\), \(M_*^{\rm{mask}} = 10^{9.0} \, M_\odot\) for pop$_2$ with \(10^{8.0} \, M_\odot < M_* < 10^{9.0} \, M_\odot\), and \(M_*^{\rm{mask}} = 10^{10.0} \, M_\odot\) for pop$_2$ with \(10^{9.0} \, M_\odot < M_* < 10^{10.0} \, M_\odot\).}
    \label{fig:blue_compare_mask}
\end{figure*}

\begin{figure*}
    \centering
    \includegraphics[width=\textwidth]{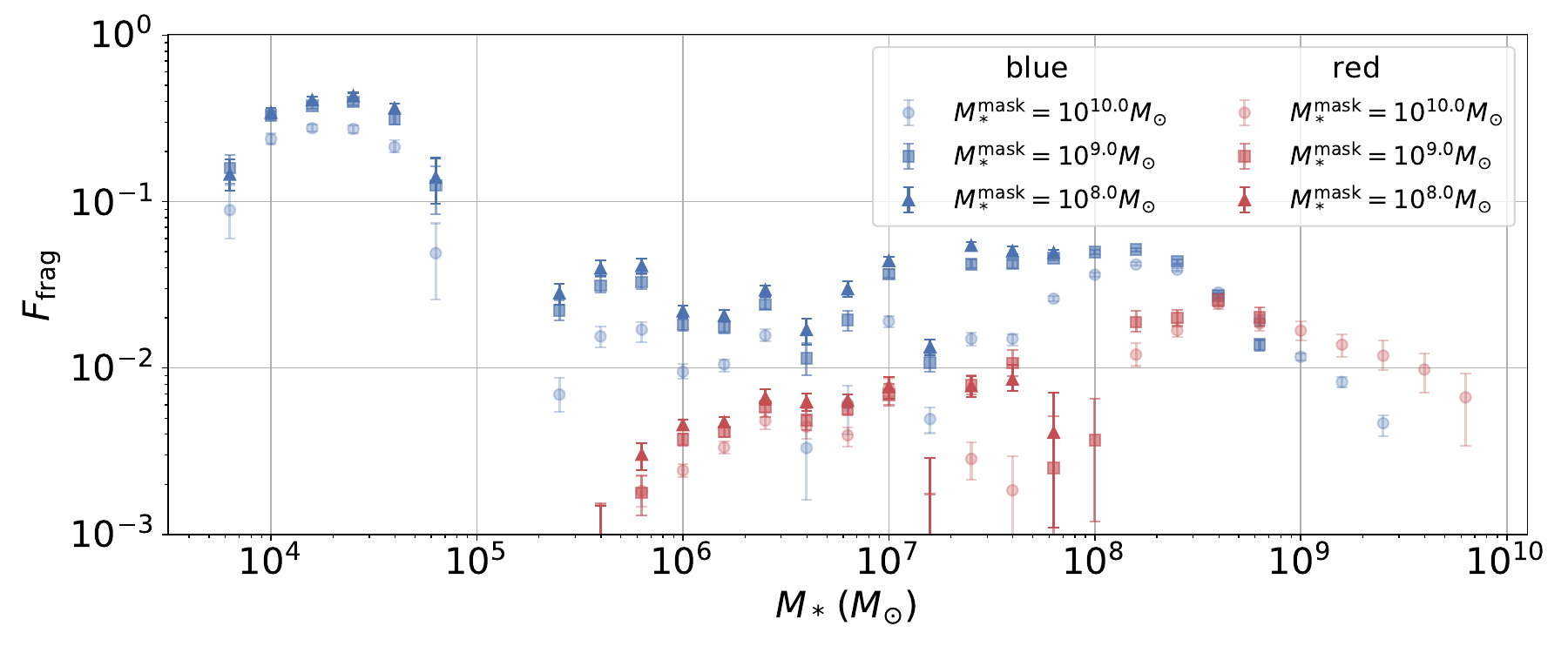}
    \caption{The fraction of objects that are galaxy fragments rather than complete galaxies in DECaLS, plotted as a function of stellar mass for blue and red objects separately. Results are shown for different mass thresholds \(M_{*}^{\rm{mask}}\) of BGS galaxies, around which DECaLS regions are masked.}
    \label{fig:F_fake}
\end{figure*}

\begin{figure*}
    \centering
    \includegraphics[width=\textwidth]{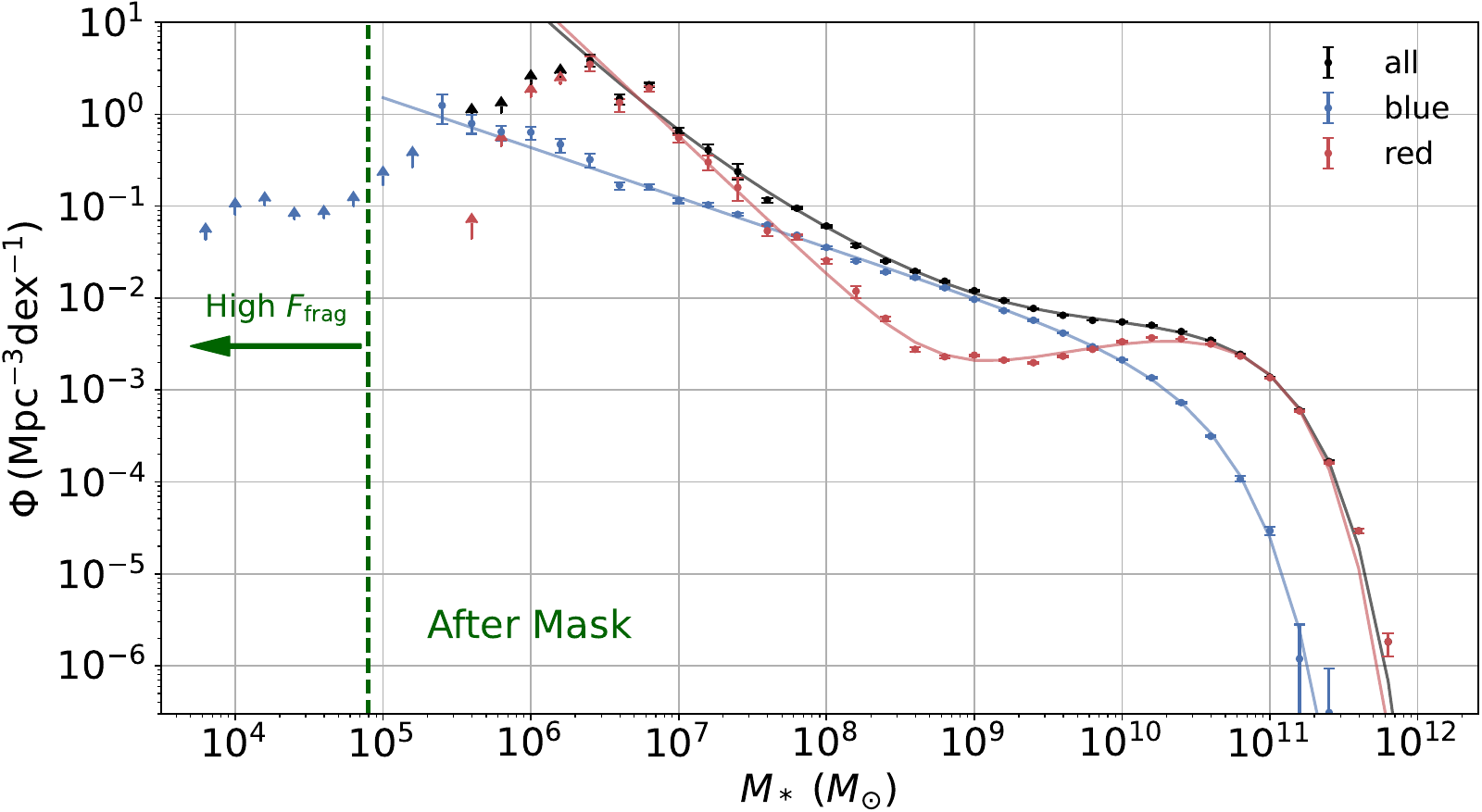}
    \caption{GSMFs for blue, red, and the entire samples, after masking regions within 2$R_{e}$ of massive galaxies, are presented. The best fits for each GSMF, using Schechter functions, are also shown.}
    \label{fig:GSMF_after_mask}
\end{figure*}

\subsection{Obtaining $\bar{n}_2$}\label{sec:4.4}
With these measurements, we can calculate $\bar{n}_2$ by simply dividing $\nwp$ by $\Wp$. However, similar to \citetalias{2022ApJ...939..104X}, instead of employing error propagation, we estimate $\bar{n}_2$ in each sub-region and then combine them to get the final results:
\begin{equation}
    n_{2,j}(r_{\rm{p}}) = \mathcal{A}_j(r_{\rm{p}})/w_{\rm{p},j}(r_{\rm{p}})\,,
\end{equation}
\begin{equation}
    n_{2}(r_{\rm{p}}) = \sum_{j=1}^{N_{\rm{sub}}}n_{2,j}(r_{\rm{p}})/N_{\rm{sub}}\,,
\end{equation}
\begin{equation}
    C_{n_{2},ab} = \frac{N_{\rm{sub}}-1}{N_{\rm{sub}}}\sum_{j=1}^{N_{\rm{sub}}}(n_{2,j}(r_{\rm{p}}^a)-n_{2}(r_{\rm{p}}^a))
    (n_{2,j}(r_{\rm{p}}^b)-n_{2}(r_{\rm{p}}^b))\,.
\end{equation}
To estimate $\bar{n}_2$, we define the $\chi^2$ as
\begin{equation}
    \chi^2=(\mathbf{n}_{2}-\bar{n}_2)^{T}\mathbf{C}_{n_{2}}^{-1}(\mathbf{n}_{2}-\bar{n}_2)\,,\label{eq:20}
\end{equation}
where $\mathbf{C}_{n_{2}}^{-1}$ is the inverse of $\mathbf{C}_{n_{2}}$, and $T$ denotes matrix transposition. Moreover, as mentioned in Section~\ref{sec:4.1}. For each stellar mass bin, pop$_2$ have 11 $\nwp$ measurements from pop$_1$ in different stellar mass bins. They are combined to get the final constraints: 
\begin{equation}
    \chi_{f}^2=\sum_{m=1}^{11}\chi^2_{m}\,\,,
\end{equation}
where $\chi^2_{m}$ is the $\chi^2$ for the $m$th stellar mass bin of pop$_1$. We use the Markov chain Monte Carlo sampler {\tt{emcee}} \citep{2013PASP..125..306F} to perform maximum likelihood analyses of $\{\bar{n}_2\}$ for each stellar mass bin of pop$_2$.

For blue galaxies at lower masses using $\Wp$ from pop$_2^{\rm{s}}$ with $M_* < 10^{9.0}\ms$, the procedure follows similarly to the steps outlined above. However, for red galaxies where $\nwp$ needs to be fitted, we redefine the $\chi^2$ as follows:
\begin{align}
    \chi^2&=(\mathcal{A}^{\rm{PAC}}-\mathcal{A}^{\rm{mod}})^{T}\mathbf{C}_{\mathcal{A}}^{-1}(\mathcal{A}^{\rm{PAC}}-\mathcal{A}^{\rm{mod}})\notag\\
    &+\frac{(w_{\rm{p}}^{\rm{obs}}(r_{\rm{p}}=8h^{-1}{\rm{Mpc}})-w_{\rm{p}}^{\rm{mod}}(r_{\rm{p}}=8h^{-1}\rm{Mpc}))^2}{\sigma_{\Wp}^2(r_{\rm{p}}=8h^{-1}\rm{Mpc})}\,.
\end{align}
Results from different stellar mass bins of pop$_1$ are then combined to compute the total $\chi^2_f$. In this process, we define a distinct parameter $\gamma$ for each pop$_1$ stellar mass bin, while only one $\{\bar{n}_2\}$ is shared among them.

To avoid scales where $\Wp$ is influenced by fibre incompleteness, for each stellar mass bin, we identify two redshifts, $z_{\rm{lim},1}$ and $z_{\rm{lim},2}$, within which BGS contains $95\%$ of the pop$_1$ and pop$_2^{\rm{s}}$ galaxies according to Figure~\ref{fig:bgs_distribution}. The minimum value for $r_{\rm{p}}$ is then set at the scale corresponding to an angular separation of $0.05^\circ$ at max($z_{\rm{lim},1}$, $z_{\rm{lim},2}$). We present the fits in Appendix \ref{sec:A} by comparing $\nwp/\bar{n}_2^{\rm{best}}$ with $\Wp$, where $\bar{n}_2^{\rm{best}}$ represents the best-fit value of $\bar{n}_2$. Results for blue galaxies with corresponding $\Wp$ measurements are shown in Figure~\ref{fig:blue_fitting1} and \ref{fig:blue_fitting2}. Blue galaxies with fixed $\Wp$ derived from $M_* < 10^{9.0} \ms$ are presented in Figure~\ref{fig:blue_fitting_wp1} and \ref{fig:blue_fitting_wp2}. For red galaxies with corresponding $\Wp$ measurements, results are shown in Figure~\ref{fig:red_fitting1} and \ref{fig:red_fitting2}, while red galaxies with modelled $\Wp$ are displayed in Figure~\ref{fig:red_fitting_wp1} and \ref{fig:red_fitting_wp2}. The fits are overall good for all stellar mass bins.

\begin{table*}
    \centering
    \caption{The best-fit parameters and their corresponding \(1\sigma\) uncertainties for the Schechter function fits to the GSMFs.}
    \begin{tabular}{cccccccc}
    \hline
         Sample&  $\log_{10}(M_c/\ms)$&  $\log_{10}(\phi_1/\rm{Mpc}^{-3})$&  $\alpha_1$&  $\log_{10}(\phi_2/\rm{Mpc}^{-3})$&  $\alpha_2$&  $\log_{10}(\phi_3/\rm{Mpc}^{-3})$& $\alpha_3$\\
    \hline
         blue&  $10.45^{+0.04}_{-0.04}$&  $-3.14^{+0.05}_{-0.05}$&  $-1.54^{+0.02}_{-0.02}$&  &  &  & \\
         red&  $10.74^{+0.02}_{-0.02}$&  $-2.51^{+0.02}_{-0.02}$&  $-0.62^{+0.03}_{-0.03}$&  $-6.27^{+0.25}_{-0.26}$&  $-2.50^{+0.08}_{-0.08}$&  & \\
         all&  $10.84^{+0.02}_{-0.03}$&  $-2.59^{+0.03}_{-0.03}$&  $-0.94^{+0.06}_{-0.05}$&  $-4.11^{+0.29}_{-0.27}$&  $-1.85^{+0.08}_{-0.08}$&  $-6.81^{+0.29}_{-0.27}$& $-2.56^{+0.70}_{-0.62}$\\
         \hline
    \end{tabular}
    \label{tab:schechter}
\end{table*}

\subsection{Lower limits for smaller stellar masses}
Although we can only obtain $\bar{n}_2$ for blue and red galaxies at $M_*>10^{5.3}\ms$ and $M_*>10^{6.3}\ms$, respectively, due to the depth of DECaLS, we are still able to acquire incomplete $\nwp$ measurements for stellar masses below these limits and place constraints on the lower limits of $\bar{n}_2$ for these masses.

To achieve this, we first apply an $r$-band magnitude cut of 23.0 to the DECaLS data, a limit reached by over $90\%$ of the regions in DECaLS \citepalias[see][]{2022ApJ...939..104X}. This ensures a uniform sample for calculating $\nwp$ in the incomplete stellar mass bins. We then compute $\nwp$ for blue and red galaxies down to $10^{3.7}\ms$ and $10^{5.5}\ms$, respectively, by cross-correlating DECaLS with the DESI BGS sample within the redshift range $z < 0.02$. Our analysis shows that we successfully obtain $\nwp$ measurements for these stellar mass bins. To determine the lower limits of $\bar{n}_2$ for these bins, we apply fixed $\Wp$ and $\Wp$ fitting methods for blue and red galaxies, respectively. The fits for blue and red galaxies are shown in Figures \ref{fig:blue_fitting_incom} and \ref{fig:red_fitting_incom}.

\begin{figure*}
    \centering
    \includegraphics[width=\textwidth]{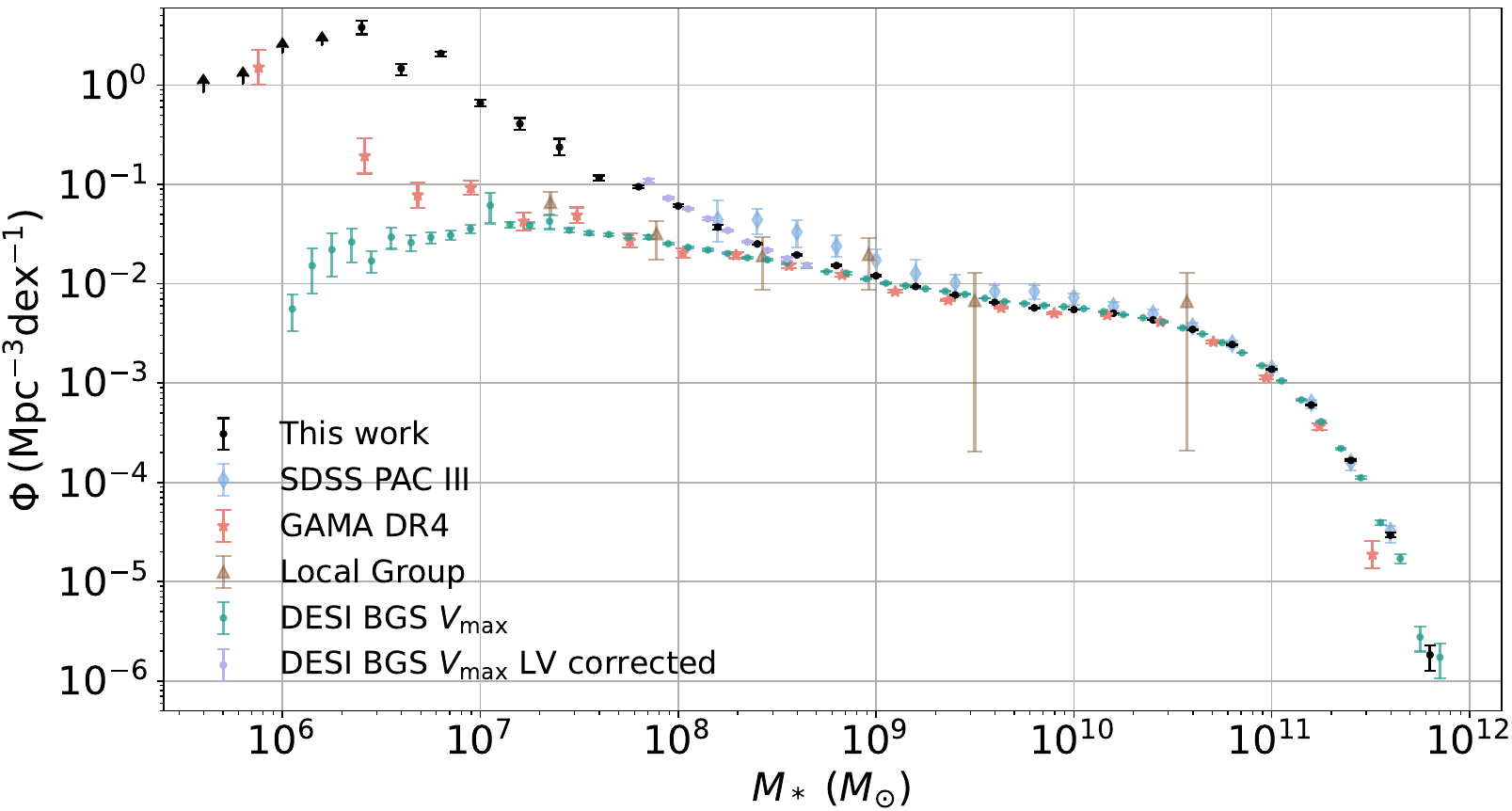}
    \caption{Comparison to previous GSMFs measured in the nearby universe. The GSMF of the full sample (black dots) is compared to results from \citetalias{2022ApJ...939..104X}, obtained using the old PAC method applied to SDSS data (blue diamonds) at \(z < 0.2\), GAMA DR4 results \citep[orange stars;][]{2022MNRAS.513..439D} at \(z < 0.1\), and the DESI BGS Bright sample analyzed with the \(V_{\rm{max}}\) method (green dots) at \(z < 0.2\) (S. Moore 2025, in preparation). Systematic errors in stellar mass for the GAMA results have been corrected using the relationship shown in Figure~\ref{fig:SED_sys}. Additionally, DESI BGS results after correcting for the Local Void bias are shown (purple dots). This correction is based on \citet{2019ApJ...872..180C}, leveraging the ELUCID simulation \citep{2014ApJ...794...94W,2016ApJ...831..164W}, which uses the reconstructed initial density field of the nearby universe. The GSMF from \citep{2022MNRAS.513..439D} for Local Group galaxies \citep[brown triages;][]{2012AJ....144....4M} is also reproduced using the same stellar mass calibration as applied to GAMA.}
    \label{fig:GSMF_compare}
\end{figure*}

\begin{figure*}
    \centering
    \includegraphics[width=\textwidth]{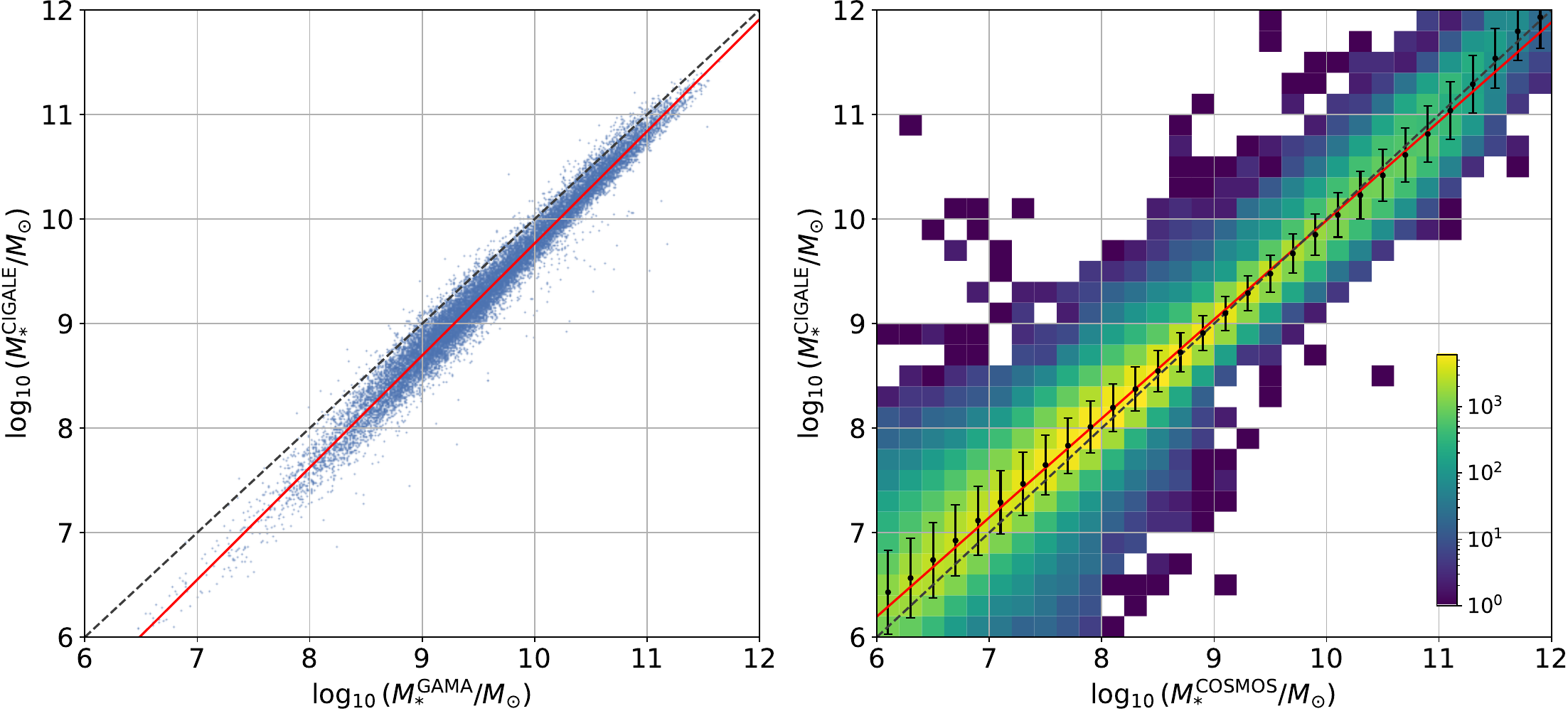}
    \caption{Left: Comparing the stellar mass $M_{*}^{\rm{GAMA}}$ obtained from GAMA DR4 \citep{2022MNRAS.513..439D} with those derived using our \texttt{CIGALE} SED template (described in Section~\ref{sec:2.1}) based on DECaLS $grz$ bands photometry ($M_{*}^{\rm{CIGALE}}$). The relationship between the two can be effectively characterized by a power law: $\log_{10}(M^{\rm{CIGALE}}_{*}/10^{10}\ms)=1.072\log_{10}(M^{\rm{GAMA}}_{*}/10^{10}\ms)-0.237$ (red line). Right: Comparing the stellar mass $M_{*}^{\rm{COSMOS}}$ obtained from COSMOS2020 \citep{2023A&A...677A.184W} with those derived using our \texttt{CIGALE} SED template based on HSC $grz$ bands photometry at $0.2<z_{\rm p}<0.8$. Black dots with errorbars represent the mean relation with the 1 $\sigma$ scatter. The relationship between the two can be effectively characterized by a power law: $\log_{10}(M^{\rm{CIGALE}}_{*}/10^{10}\ms)=0.948\log_{10}(M^{\rm{COSMOS}}_{*}/10^{10}\ms)-0.012$.}
    \label{fig:SED_sys}
\end{figure*}

\begin{figure*}
    \centering
    \includegraphics[width=\textwidth]{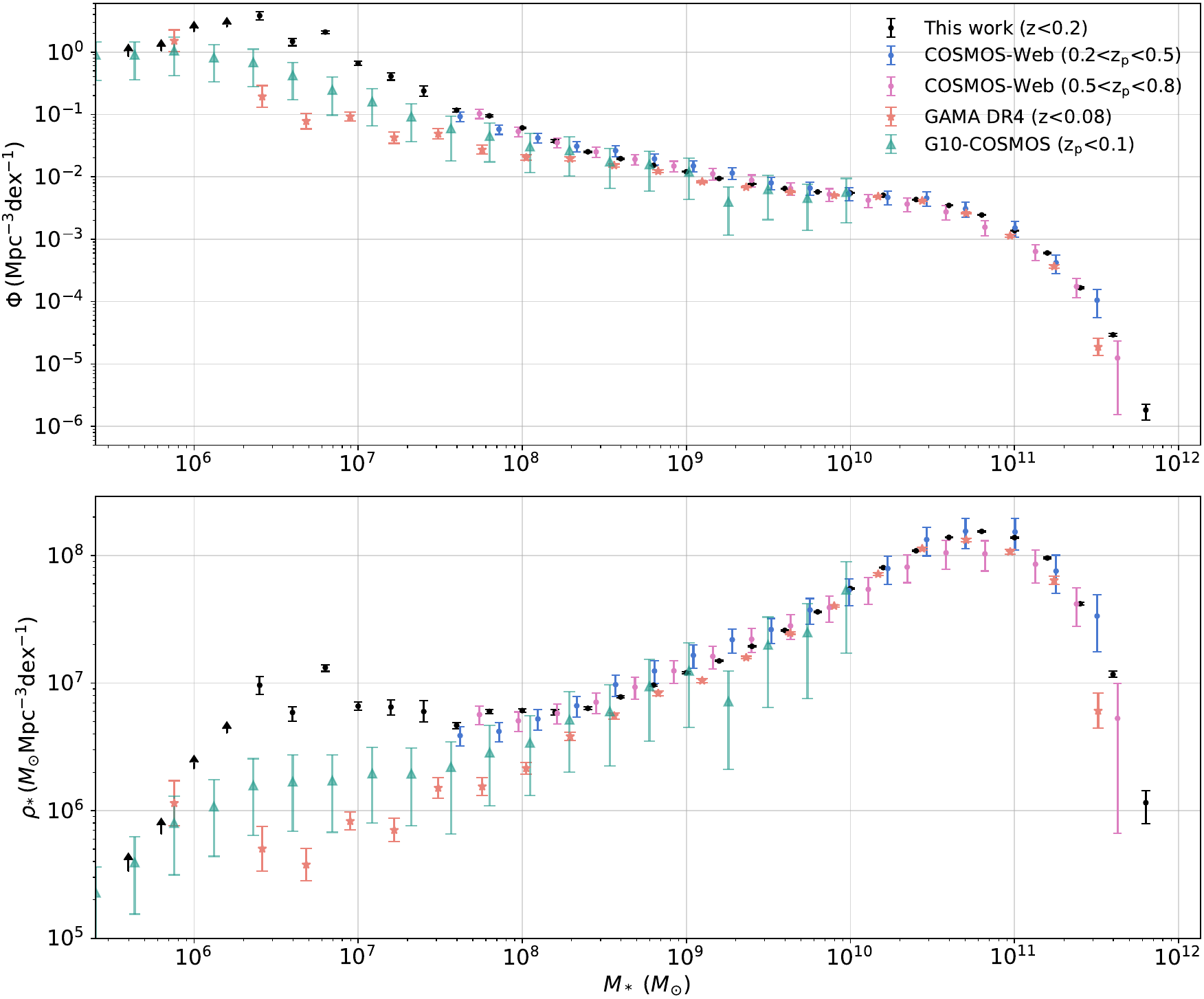}
    \caption{Top: Comparison with galaxy stellar mass functions (GSMFs) from the COSMOS field at higher redshifts. The GSMF of the full sample (black dots) is compared with results from COSMOS-Web \citep{2025A&A...695A..20S} at $0.2 < z_{\rm p} < 0.5$ (blue dots) and $0.5 < z_{\rm p} < 0.8$ (pink dots). Stellar masses from COSMOS-Web have been calibrated to match our scale using the right panel of Figure~\ref{fig:SED_sys}. GSMFs from G10-COSMOS \citep{2017MNRAS.470..283W} at $z_{\rm p} < 1$ are also shown (green triangles) with stellar mass calibrated with the relation for GAMA. Bottom: Stellar mass density within each stellar mass interval.}
    \label{fig:GSMF_COSMOS}
\end{figure*}

\subsection{The GSMF and potential systematics }
After obtaining $\bar{n}_2$ for different stellar mass bins, we derive the GSMFs across a wide range of stellar masses for blue, red, and the whole samples. The GSMFs are shown in Figure~\ref{fig:GSMF_before_mask} with also the tabular form presented in Table \ref{tab:gsmf}. 

Comparing the results for blue galaxies using corresponding $\Wp$ measurements (blue dots) with those using a fixed $\Wp$ derived from $M_* <10^{9.0} \ms$ (blue squares), we find good agreement in the overlapping stellar mass range $10^{6.4}\ms \leq M_* \leq 10^{9.0}\ms$, validating our method for extending the blue GSMF. At $M_* < 10^{8.0}\ms$, we find that the fixed $\Wp$ results are more stable. The reason might be that, although $\Wp$ measurements for blue galaxies at $M_* < 10^{8.0}\ms$ can be measured, they are based on a small number of galaxies in a limited local volume. Therefore, we adopt the fixed $\Wp$ results as the GSMF for blue galaxies in this mass range. For red galaxies, we find good agreement between the results using corresponding $\Wp$ measurements (red dots) and those from fitting $\Wp$ in the overlapping stellar mass range $10^{8.2}\ms \leq M_* \leq 10^{10.8}\ms$ (red squares), validating our method for extending the red GSMF. For $M_* < 10^{8.2}\ms$, we adopt the results from fitting $\Wp$ for red galaxies. As a result, leveraging the deep DECaLS data, we measure the GSMF down to $10^{5.2}\ms$ for blue galaxies and $10^{6.4}\ms$ for red galaxies, while setting lower limits down to $10^{3.8}\ms$ and $10^{5.6}\ms$, respectively. 

While it is exciting to obtain measurements down to \(10^{5.4}\ms\) and lower limits of \(10^{3.8}\ms\) for blue galaxies, caution is warranted, as extended objects in DECaLS are not guaranteed to be galaxies. For example, star-forming regions and spiral arms may be deblended into separate pieces. It is challenge to differentiate between multiple overlapping galaxies and a single structured galaxy \citep{1990MNRAS.247..311B,2016ApJ...816...11D,2018PASJ...70S...8A}. If the density of fragmented sources is significant, it could lead to an overestimation of the GSMF at the low mass end, particularly for blue galaxies. To quantify and correct for these spurious sources, we mask regions in DECaLS within \(2R_e\) of massive BGS galaxies and subsequently recalculate \(\nwp\), where \(R_e\) is the effective radius of the galaxy. Since \(2R_e\) encompasses most of the galaxy's light, this masking should effectively eliminate fake structures while minimizing impact on genuine neighboring galaxies. We test three different mass thresholds: \(M_*^{\rm{mask}} = 10^{10.0}, 10^{9.0}, 10^{8.0} \, \ms\). Regions around BGS galaxies with \(M_* > M_*^{\rm{mask}}\) are masked in DECaLS. Using \(M_*^{\rm{mask}} = 10^{8.0} \, \ms\) results in the masking of regions around most BGS galaxies. We recalculate \(\nwp\) only for \({\rm pop}_2\) with \(M_* < M_*^{\rm{mask}}\), as the host galaxies are expected to be significantly more massive than any deblended substructures.

In Figure~\ref{fig:blue_compare_mask}, we compare the \(\nwp\) measurements for blue galaxies before and after masking regions around massive BGS galaxies. For each stellar mass bin of pop$_2$, we select one pop$_1$ mass bin and show the results from the lowest \(M_{*}^{\rm{mask}}\). We observe that \(\nwp\) consistently decreases after masking, suggesting that substructures within blue galaxies are indeed being fragmented into independent objects. The changes in \(\nwp\) are minor for \(M_* > 10^{5.0} \ms\) but become more pronounced at lower masses. To quantify the fraction of fragmented galaxies, \(F_{\rm{frag}}\), in the sample, we compute the ratio of \(\nwp\) after masking to before masking, following the same method used to obtain \(\bar{n}_2\) in Section~\ref{sec:4.4}. However, since DESI Y1 observes only a fraction \(f_{\rm{com}}=0.656\) of BGS within the Y1 footprint, we mask only \(f_{\rm{com}}\) of the substructures in DECaLS. This partial masking is then corrected to determine the true \(F_{\rm{frag}}\):
\begin{equation}
    F_{\rm{frag}} = \left[1-\frac{(\nwp)_{\rm{after}}}{(\nwp)_{\rm{before}}}\right]/f_{\rm{com}}\,.\label{eq:f_fake}
\end{equation}
The \(F_{\rm{frag}}\) for blue and red galaxies across various stellar mass bins and using different \(M_{*}^{\rm{mask}}\) thresholds is shown in Figure~\ref{fig:F_fake}. We observe that a lower \(M_{*}^{\rm{mask}}\) leads to a higher \(F_{\rm{frag}}\), which is expected since more galaxies are masked. Red galaxies generally have a lower \(F_{\rm{frag}}\) than blue galaxies, peaking at only $2-3\%$ in the highest cases and remaining below $1\%$ in most instances. This is likely because most red galaxies are regular elliptical galaxies with fewer substructures, while blue galaxies typically exhibit spiral arms and clumpy star formation regions.  Although blue galaxies show a higher \(F_{\rm{frag}}\), it remains under $5\%$ for \(M_* > 10^{5.0} \ms\), suggesting that most objects are still actual galaxies. However, for \(M_* < 10^{5.0} \ms\), we find a significant rise in \(F_{\rm{frag}}\), reaching $40\%$, with a peak at \(10^{4.4} \ms\). This likely indicates that many star-forming regions fall within the stellar mass range of \(10^{4.0} \ms\) to \(10^{5.0} \ms\). 

It is important to note that we may underestimate \( F_{\rm{frag}} \) due to insufficient masking. As shown in Figure~\ref{fig:SM_limits}, we can still obtain \(\nwp\) measurements for pop$_2$ galaxies around \(10^{7.0}\ms\) at redshifts beyond \(z=0.05\). However, the BGS sample becomes significantly incomplete for galaxies with \(10^{8.0}\ms\) at these redshifts. This suggests that some structures measured by PAC may not be fully masked due to the limitations of BGS coverage. Although the deblending issue may be less significant at higher redshifts as the substructures become unresolved, the exact impact is difficult to quantify. Additionally, blue galaxies smaller than \(10^{8.0}\ms\) could still contain substructures down to \(10^{4.0}\ms\), which cannot be adequately studied with BGS. While we observe relatively small changes in \(F_{\rm{frag}}\) when reducing \(M_{*}^{\rm{mask}}\) from \(10^{9.0}\ms\) to \(10^{8.0}\ms\), compared to the more substantial changes between \(10^{10.0}\ms\) and \(10^{9.0}\ms\), the contamination from even smaller galaxies remains unquantified. Improved data or methods will be necessary to address the remaining issues effectively. With the constraints on \(F_{\rm{frag}}\), we can adjust for the contribution of spurious sources in the GSMF. For each stellar mass bin of pop\(_2\), we apply the \(F_{\rm{frag}}\) value derived from the lowest \(M_{*}^{\rm{mask}}\) to correct the GSMF accordingly. 

The corrected GSMFs are presented in Figure~\ref{fig:GSMF_after_mask}. We find that the blue, red, and overall GSMFs can be described by single, double, and triple Schechter functions, respectively:
\begin{equation}
    \phi_{\rm{blue}}(M_*)dM_* = e^{-M_*/M_{c}} \phi_1\left(\frac{M_*}{M_c}\right)^{\alpha_1}\frac{dM_*}{M_c}\,,
\end{equation}

\begin{equation}
    \phi_{\rm{red}}(M_*)dM_* = e^{-M_*/M_{c}} \left[ \phi_1\left(\frac{M_*}{M_c}\right)^{\alpha_1} + \phi_2\left(\frac{M_*}{M_c}\right)^{\alpha_2} \right] \frac{dM_*}{M_c}\,,
\end{equation}

\begin{align}
    \phi_{\rm{all}}(M_*)dM_* = e^{-M_*/M_c} & \left[ \phi_1\left(\frac{M_*}{M_c}\right)^{\alpha_1} + \phi_2\left(\frac{M_*}{M_c}\right)^{\alpha_2} \right. \notag \\
    & \left. + \phi_3\left(\frac{M_*}{M_c}\right)^{\alpha_3} \right] \frac{dM_*}{M_c}\,.
\end{align}
Here, \(\phi(M_*)dM_*\) represents the number density of galaxies with masses between \(M_*\) and \(M_* + dM_*\). The number density \(\Phi\) that we typically measure is expressed in equal logarithmic intervals, which relates to \(\phi\) as follows:
\begin{equation}
    \Phi(M_*) = \ln{(10)} M_* \phi(M_*)\,.
\end{equation}
The constraints of the parameters are shown in Table \ref{tab:schechter}.

From Figure~\ref{fig:GSMF_after_mask}, we observe that red galaxies dominate the high-mass end at \(M_* > 10^{10.6} \ms\), while blue galaxies become increasingly prominent at lower masses, overtaking red galaxies at around \(10^{9.8} \ms\). However, the density of red galaxies rises sharply below \(10^{8.6} \ms\), once again dominating at \(M_* < 10^{7.6} \ms\). This increasing trend persists down to the stellar mass limit of \(10^{6.4} \ms\) for red galaxies. If this trend holds beyond our observational limits, dwarf galaxies would predominantly be red. \citet{2010ApJ...721..193P} provided a physical explanation for the red GSMF, where galaxies with \(M_* > M_c\) are primarily influenced by "mass quenching" in central galaxies, and those with \(M_* < M_c\) are driven by "environmental quenching" in satellite galaxies. However, while their model predicted a slight upturn in the red GSMF around \(10^{9.0}\ms\), it is not nearly as steep as what we observe. This discrepancy suggests that the quenching mechanism for red dwarf galaxies might differ from our current understanding. Investigating this further is an interesting and important problem for future studies.

\subsection{Comparison with previous nearby GSMFs}
In Figure~\ref{fig:GSMF_compare}, we compare the GSMF of the entire sample with the results from \citetalias{2022ApJ...939..104X} (blue diamonds), which used the old PAC method on SDSS Main sample \citep{2009ApJS..182..543A}. The results in \citetalias{2022ApJ...939..104X} were obtained by controlling only for stellar mass when selecting pop\(_2^{\rm{s}}\) to derive \(\Wp\), as described in Section~\ref{sec:4.1}. We observe that their GSMF is higher than ours for \(M_* < 10^{10.6} \ms\). This discrepancy, as explained in Section~\ref{sec:4.1}, arises because the SDSS main sample used in \citetalias{2022ApJ...939..104X} is incomplete for stellar masses below \(10^{10.6} \ms\), resulting in a higher fraction of blue galaxies and consequently lower \(\Wp\). This highlights the importance of controlling for both stellar mass and colour, as developed in this study.

We also compare our results with those from GAMA DR4 at \(z < 0.1\) \citep{2022MNRAS.513..439D}. To ensure a meaningful comparison, it is crucial to account for systematic errors in stellar mass, as we utilize different SED templates and photometry compared to GAMA. Therefore, we match GAMA galaxies used in \citep{2022MNRAS.513..439D} to DECaLS and recompute their stellar masses using our SED template and DECaLS photometry. A comparison between the two stellar mass estimates is shown in the left panel of Figure~\ref{fig:SED_sys}. We find that our stellar mass calculations are slightly lower than those from GAMA, with the difference increasing toward lower stellar masses. This trend follows a power law: \(\log_{10}(M^{\rm{CIGALE}}_{*}/10^{10}\ms) = 1.072\log_{10}(M^{\rm{GAMA}}_{*}/10^{10}\ms) - 0.237\). The increasing discrepancy at the low mass end is expected, as most SED templates are calibrated at high masses. Recently, \citet{2024arXiv240903959D} highlighted that stellar mass estimates for low mass galaxies are sensitive to the assumed star formation history in the template, which can lead to differences of up to 0.4 dex. Despite this stellar mass discrepancy, we can still make a reasonable comparison after calibrating the GAMA stellar mass using the relation in Figure~\ref{fig:SED_sys}. We adjust the GSMF from \citep{2022MNRAS.513..439D} accordingly and present the comparison with our results in Figure~\ref{fig:GSMF_compare}. Additionally, since many studies utilize the \citet{2022MNRAS.513..439D} stellar mass function, we also present our GSMFs calibrated to the \citet{2022MNRAS.513..439D} stellar mass in Appendix \ref{sec:B}, allowing for easier comparison with other studies.

We find that our results are in good agreement with \citet{2022MNRAS.513..439D} for stellar masses in the range \(10^{9.0} \ms < M_* < 10^{11.0} \ms\). However, we observe slight discrepancies for \(M_* > 10^{11.0} \ms\) and an increasing divergence below \(10^{9.0} \ms\). Both discrepancies are likely attributed to cosmic variance. At the high mass end, the limited survey area of GAMA, which covers only about 250 deg\(^2\), results in a scarcity of massive galaxies at \(z < 0.1\) available for calculating the GSMF, leading to significant uncertainties. In contrast, as shown as green dots in Figure \ref{fig:GSMF_compare}, when we compare our GSMF with that from the much larger DESI Y1 BGS Bright sample at \(z < 0.2\) using the same SED template and $V_{\rm{max}}$ method (S. Moore 2025, in preparation), we find strong consistency up to \(10^{11.8} \ms\). This indicates that the discrepancies with GAMA for massive galaxies might be primarily due to its small sample size. Another possible factor could be differences in scatter in the estimated stellar masses, which could affect the GSMF at the steep, high-mass end. At the low mass end, the depth of GAMA is only \(r_{\rm{KiDS}} \sim 19.65\), which is much shallower than DECaLS with \(r \sim 23.0\). As a result, most low mass galaxies in GAMA are found in the local volume surrounding the Milky Way. It is well known that the Milky Way resides in a Local Void (LV) \citep{2008ApJ...676..184T,2010Natur.465..565P,2020A&A...633A..19B}. Consequently, using the local volume to study the GSMF is likely to yield underestimated results. This discrepancy at $M_{*}<10^{9.0}M_{\odot}$ is also found when comparing to the BGS $V_{\rm{max}}$ results, which has the depth of $r=19.5$. Moreover, \citet{2022MNRAS.513..439D} found that the GSMF from GAMA is broadly consistent with those from the Local Volume \citep{2019Ap.....62..293K} and the Local Group \citep{2012AJ....144....4M} in the stellar mass range $10^{8.0}\ms < M_* < 10^{10.0}\ms$. We reproduce their measurements in the Local Group in Figure~\ref{fig:GSMF_compare} (brown triangles), applying the same stellar mass calibration in Figure~\ref{fig:SED_sys} used for GAMA, under the assumption that \citet{2022MNRAS.513..439D} made a consistent comparison.

To quantify the bias from the LV, \citet{2019ApJ...872..180C} provide a comprehensive correction approach using the ELUCID simulation \citep{2014ApJ...794...94W,2016ApJ...831..164W}. This simulation was conducted using the reconstructed initial density field of the nearby universe within a box size of \(500h^{-1}{\rm{Mpc}}\). They observed a rapid decrease in halo and galaxy densities toward lower redshifts, beginning from
$z=0.03$. They derived the correction for the SDSS GSMF (\(r < 17.6\)) by comparing the GSMF from the simulated SDSS region to those across the entire simulation box. Additionally, they examined corrections for samples with \(r\)-band magnitude limits ranging from \(16\) to \(20\). Recently, \citet{2024ApJ...971..119W} found that the corrected low mass end of the SDSS GSMF aligns well with results from their deeper \(r = 19.5\) sample, which combines DESI Y1 BGS and DESI Legacy Imaging Surveys samples. This indicates that the cosmic variance correction from \citet{2019ApJ...872..180C} is highly effective. Additionally, \citet{2024ApJ...971..119W} applied the correction for $r=19.5$ from \citet{2019ApJ...872..180C} to their GSMF, resulting in a much steeper increase at lower masses. We adopt the correction \(f_2\) from \citet{2024ApJ...971..119W} and apply it to the BGS $V_{\rm{max}}$ results. The LV bias corrected BGS $V_{\rm{max}}$ GSMF is also shown in Figure~\ref{fig:GSMF_compare}. We find that it now aligns well with our GSMF down to \(10^{7.6} \ms\).We note that the correction from \citet{2019ApJ...872..180C} was performed within the SDSS footprint, which does not exactly match the DESI Y1 BGS sample. Additionally, as the GSMF from GAMA shows good agreement with the BGS \(V_{\rm{max}}\) results, despite covering different sky regions, this suggests that the bias related to the LV may depend primarily on redshift rather than specific sky areas. This factor was not fully explored by \citet{2019ApJ...872..180C} and warrants further investigation. Moreover, \citet{2019ApJ...872..180C} did not provide corrections of the LV bias for lower masses due to the resolution limits of the ELUCID simulation. Thus, whether the discrepancies observed at lower masses can be fully attributed to cosmic variance requires further investigation. 

We acknowledge that GAMA has applied a LSS correction using the DFTOOLS method described in \citet{2018MNRAS.474.5500O}. However, this approach is based on a simplified assumption. Specifically, it uses more massive galaxies—which are observable over a larger volume—to estimate density variance. It then assumes that this variance applies equally to low-mass galaxies, adjusting their number densities accordingly. This assumption overlooks key complexities. It relies on the notion of linear galaxy bias, where galaxy overdensity is assumed to respond linearly to the matter overdensity. This is only valid when comparing galaxy samples in very large volumes with little mass variation $\delta_g\ll1$. The local volume is not large enough and involves numerous non-linear processes. For instance, Figure 9 of \citet{2019ApJ...872..180C} shows, using a constrained simulation, the halo and galaxy number density change below $z<0.03$ differs significantly across mass bins. We would expect an even more complex behavior when considering how galaxy colour responds to environmental density. Furthermore, the DFTOOLS LSS correction requires normalization to determine the absolute correction factor. In GAMA's case, this was done by fixing the total stellar mass within the survey volume at z<0.08 \citep{2022MNRAS.513..439D}. However, since the region at z<0.03 is underdense, this approach likely underestimates the true correction.

If the GSMF from BGS in the local volume and our PAC method for the cosmic average in Figure \ref{fig:GSMF_compare} are both accurate, the dominance of blue galaxies in the local volume and red galaxies in the cosmic average at $M_*<10^{7.6}\ms$ may suggest a significant dependence of low-mass galaxy colour on the environment. In other words, the discrepancy between the local volume and the cosmic average may primarily be due to the absence of low-mass red galaxies in the local volume. This is supported by our finding that the GSMF for blue galaxies in the cosmic average in Figure \ref{fig:GSMF_after_mask} closely resembles the total GSMF from BGS in the local volume in Figure \ref{fig:GSMF_compare}. We plan to further investigate this using constrained simulations with a semi-analytical model in the future.

The comparison of our results with GAMA and BGS highlights the significant advantages of the PAC method in obtaining robust measurements of the GSMF, particularly at lower masses.

\subsection{Comparison with COSMOS GSMFs at higher redshifts}

Since the LV correction from \citet{2019ApJ...872..180C} is highly model-dependent, we further validate our results by comparing our GSMF at $z < 0.2$ with measurements from COSMOS-Web \citep{2025A&A...695A..20S} at photometric redshifts $0.2 < z_{\rm p} < 0.5$ and $0.5 < z_{\rm p} < 0.8$. COSMOS-Web \citep{2023ApJ...954...31C} combines deep ground- and space-based photometric data across 33 bands in the COSMOS field \citep{2007ApJS..172...38S}, including recent observations from JWST \citep{2006SSRv..123..485G}, and provides photometric redshifts and physical properties for galaxies over an area of approximately 0.431 deg$^2$ \citep{2025A&A...695A..20S}. This region currently offers the only reliable constraints on the GSMF at $M_* < 10^{9.0}\ms$ beyond the local universe. To ensure a fair comparison, we calibrate the COSMOS-Web stellar masses to our scale. As the COSMOS-Web catalogue has not yet been published, we instead use the COSMOS2020 catalogue \citep{2022ApJS..258...11W}, given the consistency of the COSMOS data reduction pipelines. We select all galaxies in COSMOS2020 with photometric redshifts $0.2 < z_{\rm p} < 0.8$ and compute their stellar masses using our \texttt{GIGALE} SED templates (Section~\ref{sec:2.1}), based solely on HSC $grz$ photometry for optimal consistency. These results are then compared to the COSMOS2020 stellar masses derived with \texttt{Le Phare}, as shown in the right panel of Figure~\ref{fig:SED_sys}. We find that our stellar masses are quite consistent with those from COSMOS2020, and their relation can be described by $\log_{10}(M^{\rm{CIGALE}}_{*}/10^{10}\,M_\odot) = 0.948\,\log_{10}(M^{\rm{COSMOS}}_{*}/10^{10}\,M_\odot) - 0.012$. The comparison of GSMFs is presented in Figure~\ref{fig:GSMF_COSMOS}. We find that our GSMF agrees well with the COSMOS-Web measurements down to $10^{7.6}\ms$ at the two higher redshifts, within the uncertainties. This further indicate a potential higher number density of dwarf galaxies than reported in the local volume. \citet{2025A&A...695A..20S} did not provide separate GSMFs for blue and red galaxies, but the sharp rise in the red GSMF at $M_* < 10^{8.5}\ms$ is evident ($\alpha_2\sim -2$) in the COSMOS2020 results \citep{2023A&A...677A.184W} for $z<0.8$. We plan to explore this further once the full COSMOS-Web catalogues are released. 

In Figure~\ref{fig:GSMF_COSMOS}, we also reproduce the GSMF from G10-COSMOS by \citet{2017MNRAS.470..283W} (see also \citealt{2018MNRAS.475.2891D, 2018MNRAS.480.3491W}), based on a combination of spectroscopic and photometric redshifts at $z_{\rm p} < 0.1$ (green triangles). The stellar masses are calibrated using the relation for GAMA, under the assumption that \citet{2017MNRAS.470..283W} applied a consistent methodology to G10-COSMOS as was used for GAMA, as detailed in \citet{2018MNRAS.475.2891D}. The results from G10-COSMOS appear to lie between those from GAMA and COSMOS-Web/our work. However, caution is warranted when using photometric redshifts at such low redshifts.

In the bottom panel of Figure~\ref{fig:GSMF_COSMOS}, we also show the galaxy stellar mass density (GSMD) in each stellar mass bin. Our results suggest that the upturn in the GSMF leads to a relatively constant or slightly increasing GSMD at $M_* < 10^{8.0}\ms$, in contrast to previous findings where the GSMD continues to decline toward lower stellar masses.

\section{Conclusions and prospects}\label{sec:5}
In this paper, we improve the PAC method originally developed in \citetalias{2022ApJ...925...31X} by introducing several enhancements to make it more rigorous and to eliminate the need for redshift bins.

As a first application, we use the DECaLS photometric sample and the BGS Bright spectroscopic sample from DESI Y1 to measure $\nwp$ across various stellar mass bins, ranging from \(10^{5.3} \ms\) to \(10^{11.5} \ms\) for blue galaxies and \(10^{6.3} \ms\) to \(10^{11.9} \ms\) for red galaxies, all of which are complete in stellar mass. Additionally, we measure several incomplete bins down to \(10^{3.7} \ms\) for blue galaxies and \(10^{5.5} \ms\) for red galaxies. Following \citetalias{2022ApJ...939..104X} with some extensions, we combine these $\nwp$ results with the corresponding $\Wp$ measurements from the BGS sample, which may not be fully complete in stellar mass. Assuming that galaxy bias is primarily determined by stellar mass and colour, we then calculate $\bar{n}_2$ for these mass bins. This approach enables us to derive the GSMF down to \(10^{5.3} \ms\) for blue galaxies and \(10^{6.3} \ms\) for red galaxies, with lower limits extending down to \(10^{3.7} \ms\) and \(10^{5.5} \ms\), respectively. We also examine the influence of fragmented sources in the DECaLS samples on our GSMF measurements, which could be misidentified features like fragmented spiral arms or star-forming regions. The impact on red galaxies is negligible. However, for blue galaxies, the fake fraction \( F_{\rm{frag}} \) is notably higher at \( M_* <10^{9.0} \ms \), though it remains below $5\%$ for \( M_* > 10^{5.0} \ms \). For \( M_* < 10^{5.0} \ms \), \( F_{\rm{frag}} \) becomes significant, reaching $40\%$ at \( 10^{4.6} \ms \). These fake sources have been corrected for in our GSMF.

We find that the GSMFs for blue, red, and the entire galaxy populations can be well described by single, double, and triple Schechter functions. The power-law indices at the low-mass ends are \( -1.54^{+0.02}_{-0.02} \) for blue galaxies and \( -2.50^{+0.02}_{-0.02} \) for red galaxies, resulting in an index ranging between \( -1.85 \) and \( -2.56 \) for the entire GSMF. The slope of our GSMF at the low-mass end is significantly steeper than that of GAMA \citep{2022MNRAS.513..439D} and BGS Bright , whose index is closer to that of our blue GSMF. This discrepancy is reconciled after applying the correction for the local volume, as provided by \citet{2019ApJ...872..180C}, to the BGS $V_{\rm{max}}$ GSMF. Additionally, the steep rise in the GSMF at lower masses has not been successfully reproduced by current galaxy formation models, suggesting an incomplete understanding of the physics governing galaxy formation.

Our measurements of the GSMF, extending to the \(10^{6.0} \ms\) range, hold the potential to provide valuable constraints on various aspects, including the nature of dark matter \citep{2000ApJ...542..622C,2001ApJ...556...93B,2005PhRvD..71f3534V,2012MNRAS.420.2318L,2013MNRAS.428..882M,2016MNRAS.456.4346H,2017MNRAS.464.4520B}, the epoch of reionization \citep{1992MNRAS.256P..43E,2001ARA&A..39...19L,2008MNRAS.390..920O,2017MNRAS.465.3913B,2018ApJ...863..123B}, and the physics of galaxy formation. For instance, warm dark matter models predict a cutoff at high \(k\) in the power spectrum, leading to the suppression of both the halo and stellar mass functions at low masses \citep{2017MNRAS.464.4520B}. The steep rise we observe in our GSMF at the low-mass end may place strong constraints on the mass of dark matter particles. Furthermore, upcoming cosmological photometric surveys, such as LSST, Euclid, CSS-OS, and Roman, are expected to probe depths 2-3 magnitudes fainter than DECaLS. By applying the PAC method, we have the potential to thoroughly explore the full range of the galaxy population in the local universe and address these fundamental questions.

In addition to its relevance for dwarf galaxy studies in the local universe, the PAC method offers many other applications. By leveraging the diverse spectroscopic samples from DESI, including BGS, LRG, ELG, and QSO, which together map the universe up to \(z \sim 2\), we can consistently investigate the evolution of various galaxy properties. Moreover, we can establish highly accurate galaxy-halo connections for spectroscopic tracers, thanks to the extensive information available on their spatial distribution in relation to different types of galaxies, as developed in \citetalias{2023ApJ...944..200X}. This enables precise modeling of small-scale measurements based on these tracers, such as galaxy-galaxy lensing and redshift space distortions \citep{2012ApJ...758...50L,2022MNRAS.515..871Y,2023ApJ...948...99Z,2024ApJ...973..102X}, leading to stringent constraints on the structure formation of the universe.

With this paper, we begin applying the PAC method to Stage-IV cosmological surveys, aiming to obtain new measurements that we hope will significantly advance our understanding of the Universe.

\section*{Acknowledgements}
We thank Zheng Zheng, Jeremy Tinker and Jiaxi Yu for their valuable comments and great help during the DESI Collaboration Wide Review. We thank the anonymous referee for the valuable comments that helped improve the manuscript.

K.X. is supported by the funding from the Center for Particle Cosmology at U Penn. Y.P.J. is supported by NSFC (12133006, 11890691, 11621303), by grant No. CMS-CSST-2021-A03, and by 111 project No. B20019. Y.P.J. gratefully acknowledges the support of the Key Laboratory for Particle Physics, Astrophysics and Cosmology, Ministry of Education. S.B. is supported by the UK Research and Innovation (UKRI) Future Leaders Fellowship [grant number MR/V023381/1]. W.W. is supported by NSFC (12273021) and the National Key R\&D Program of
China (2023YFA1605600, 2023YFA1605601). This work made use of the Gravity Supercomputer at the Department of Astronomy, Shanghai Jiao Tong University. This work used the DiRAC@Durham facility managed by the Institute for Computational Cosmology on behalf of the STFC DiRAC HPC Facility (www.dirac.ac.uk). The equipment was funded by BEIS capital funding via STFC capital grants ST/K00042X/1, ST/P002293/1, ST/R002371/1 and ST/S002502/1, Durham University and STFC operations grant ST/R000832/1. DiRAC is part of the National e-Infrastructure. 

This material is based upon work supported by the U.S. Department of Energy (DOE), Office of Science, Office of High-Energy Physics, under Contract No. DE–AC02–05CH11231, and by the National Energy Research Scientific Computing Center, a DOE Office of Science User Facility under the same contract. Additional support for DESI was provided by the U.S. National Science Foundation (NSF), Division of Astronomical Sciences under Contract No. AST-0950945 to the NSF’s National Optical-Infrared Astronomy Research Laboratory; the Science and Technologies Facilities Council of the United Kingdom; the Gordon and Betty Moore Foundation; the Heising-Simons Foundation; the French Alternative Energies and Atomic Energy Commission (CEA); the National Council of Science and Technology of Mexico (CONACYT); the Ministry of Science, Innovation and Universities of Spain (MICIU/AEI/10.13039/501100011033), and by the DESI Member Institutions: \url{https://www.desi.lbl.gov/collaborating-institutions}. 

The DESI Legacy Imaging Surveys consist of three individual and complementary projects: the Dark Energy Camera Legacy Survey (DECaLS), the Beijing-Arizona Sky Survey (BASS), and the Mayall z-band Legacy Survey (MzLS). DECaLS, BASS and MzLS together include data obtained, respectively, at the Blanco telescope, Cerro Tololo Inter-American Observatory, NSF’s NOIRLab; the Bok telescope, Steward Observatory, University of Arizona; and the Mayall telescope, Kitt Peak National Observatory, NOIRLab. NOIRLab is operated by the Association of Universities for Research in Astronomy (AURA) under a cooperative agreement with the National Science Foundation. Pipeline processing and analyses of the data were supported by NOIRLab and the Lawrence Berkeley National Laboratory. Legacy Surveys also uses data products from the Near-Earth Object Wide-field Infrared Survey Explorer (NEOWISE), a project of the Jet Propulsion Laboratory/California Institute of Technology, funded by the National Aeronautics and Space Administration. Legacy Surveys was supported by: the Director, Office of Science, Office of High Energy Physics of the U.S. Department of Energy; the National Energy Research Scientific Computing Center, a DOE Office of Science User Facility; the U.S. National Science Foundation, Division of Astronomical Sciences; the National Astronomical Observatories of China, the Chinese Academy of Sciences and the Chinese National Natural Science Foundation. LBNL is managed by the Regents of the University of California under contract to the U.S. Department of Energy. The complete acknowledgments can be found at \url{https://www.legacysurvey.org/}. 

Any opinions, findings, and conclusions or recommendations expressed in this material are those of the author(s) and do not necessarily reflect the views of the U. S. National Science Foundation, the U. S. Department of Energy, or any of the listed funding agencies. 

The authors are honored to be permitted to conduct scientific research on Iolkam Du'ag (Kitt Peak), a mountain with particular significance to the Tohono O'odham Nation.

\section*{Data Availability}
Data points for each figure are available in a machine readable form: \url{https://doi.org/10.5281/zenodo.14879599}. Other data will be shared upon reasonable request to the corresponding author.



\bibliographystyle{mnras}
\bibliography{example} 




\appendix
\section{Measurements and fittings}\label{sec:A}
\begin{figure*}
    \centering    
    \includegraphics[width=0.98\textwidth]{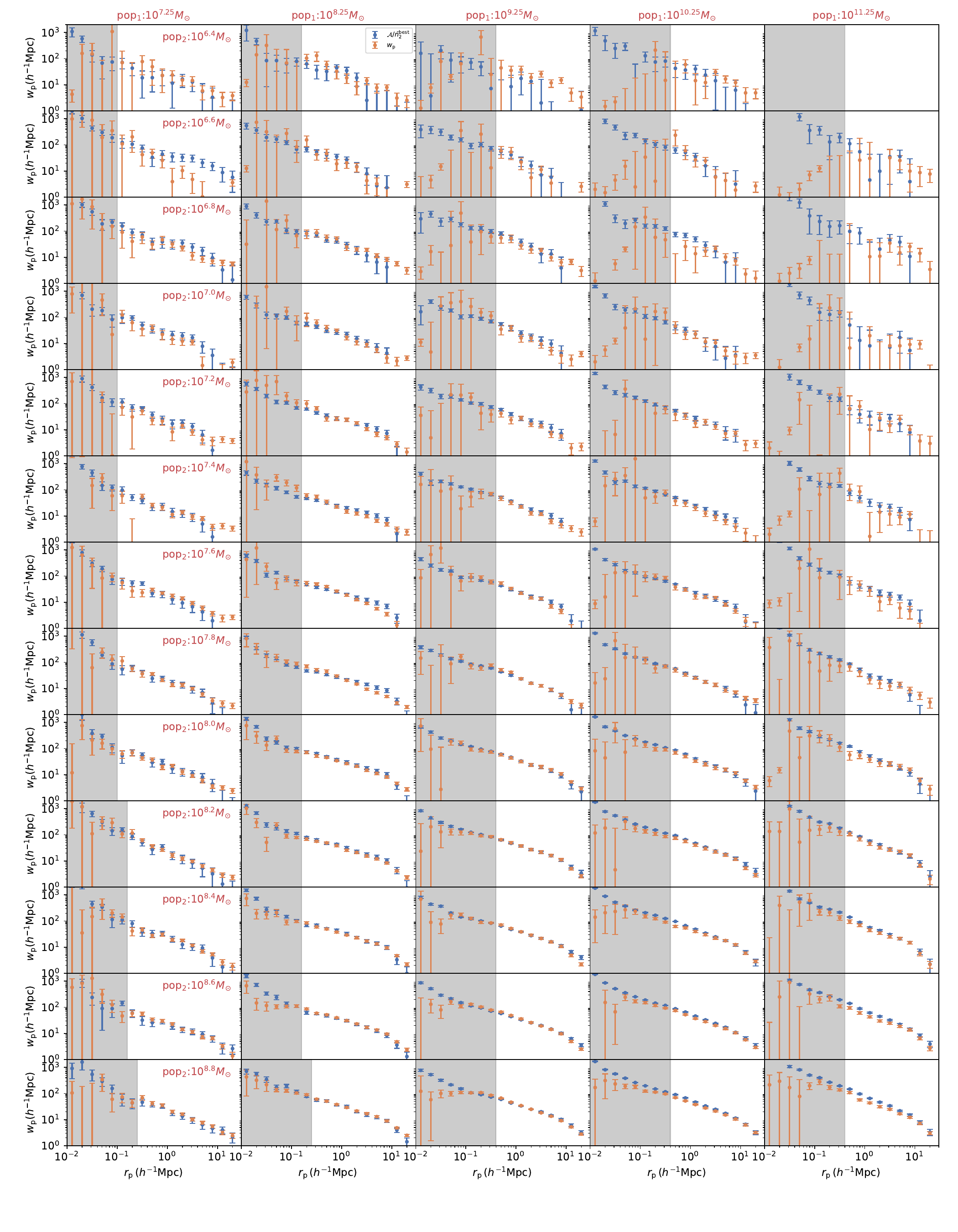}
    \caption{Comparison of $\mathcal{A}/\bar{n}_2^{\rm{best}}$ (blue) with $\Wp$ (orange) for {\textit{blue}} galaxies. Each row presents the results for one pop$_2$ stellar mass bin, and each column corresponds to one pop$_1$ stellar mass bin. For clarity, only 5 out of the 11 pop$_1$ bins are shown. Shaded regions indicate areas where $\Wp$ is affected by fibre incompleteness. The continuation of this figure is provided in Figure~\ref{fig:blue_fitting2}.}  
    \label{fig:blue_fitting1}
\end{figure*}

\begin{figure*}
    \centering    
    \includegraphics[width=0.98\textwidth]{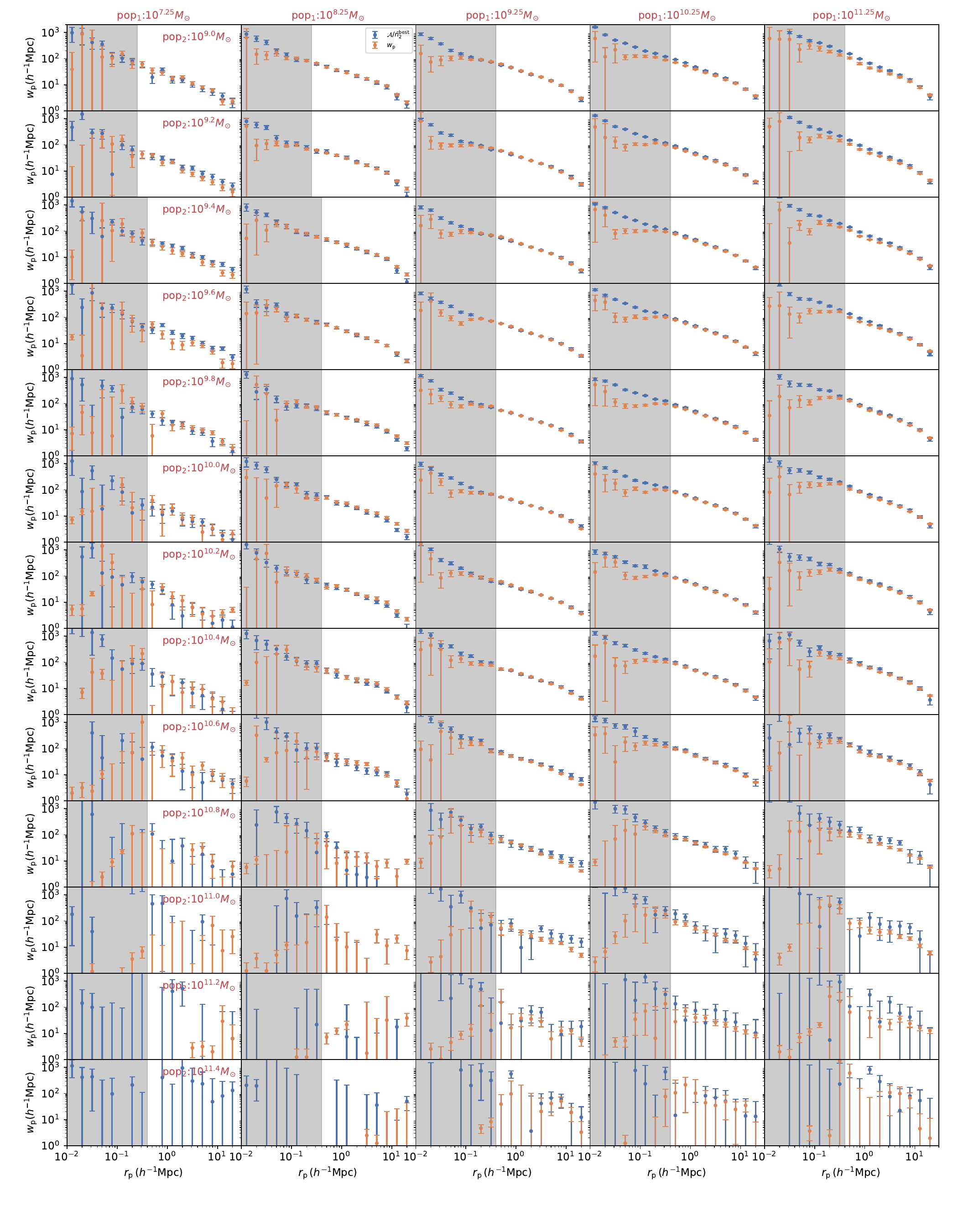}
    \caption{The continuation of Figure~\ref{fig:blue_fitting1}.}  
    \label{fig:blue_fitting2}
\end{figure*}

\begin{figure*}
    \centering    
    \includegraphics[width=0.98\textwidth]{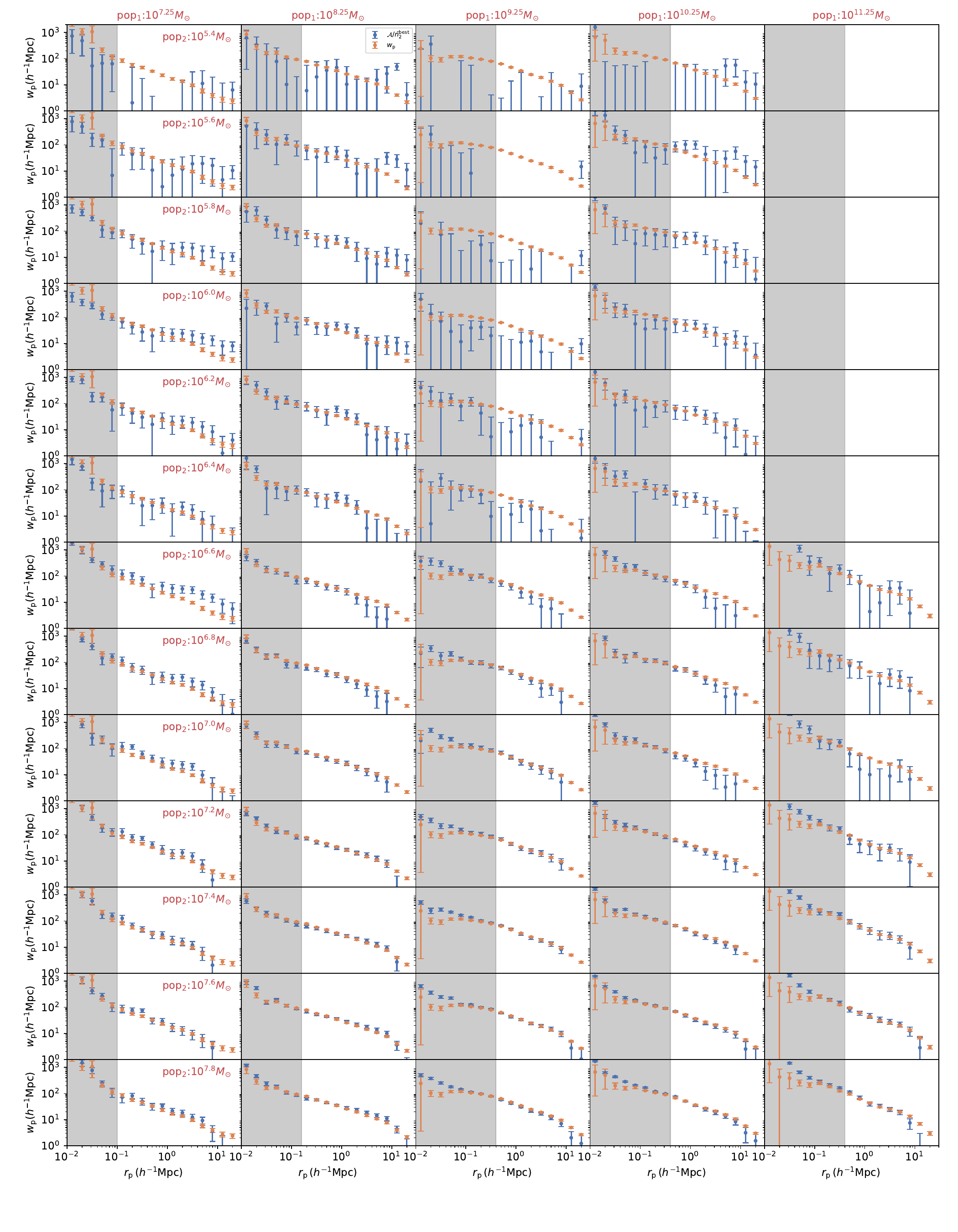}
    \caption{Similar to Figure~\ref{fig:blue_fitting1} for {\textit{blue}} galaxies, but here $\Wp$ is derived from all blue pop$_2^{\rm{s}}$ galaxies with $M_*<10^{9.0}\ms$. The continuation of this figure is provided in Figure~\ref{fig:blue_fitting_wp2}.}  
    \label{fig:blue_fitting_wp1}
\end{figure*}

\begin{figure*}
    \centering    
    \includegraphics[width=0.98\textwidth]{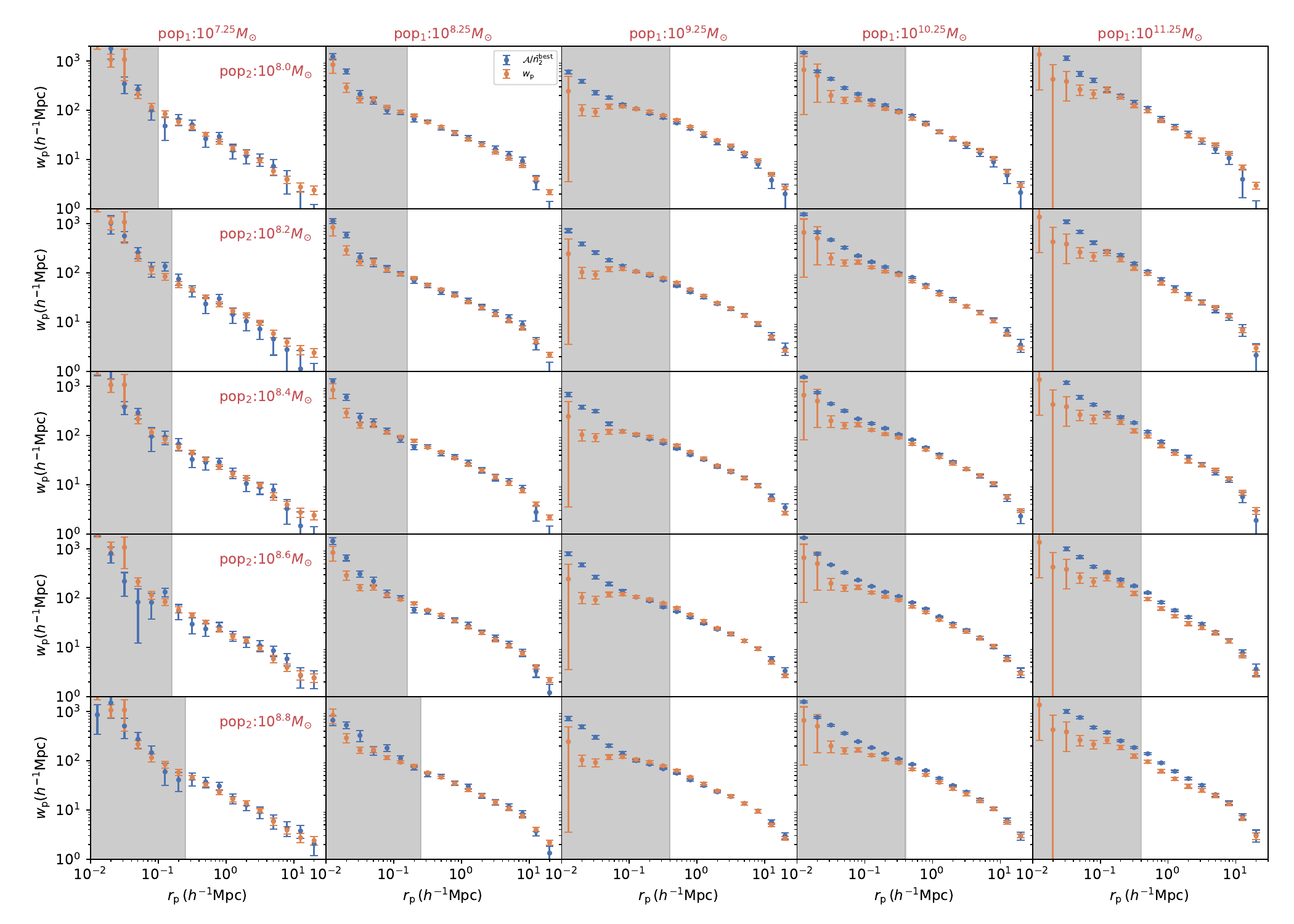}
    \caption{The continuation of Figure~\ref{fig:blue_fitting_wp1}.}  
    \label{fig:blue_fitting_wp2}
\end{figure*}

\begin{figure*}
    \centering    
    \includegraphics[width=0.98\textwidth]{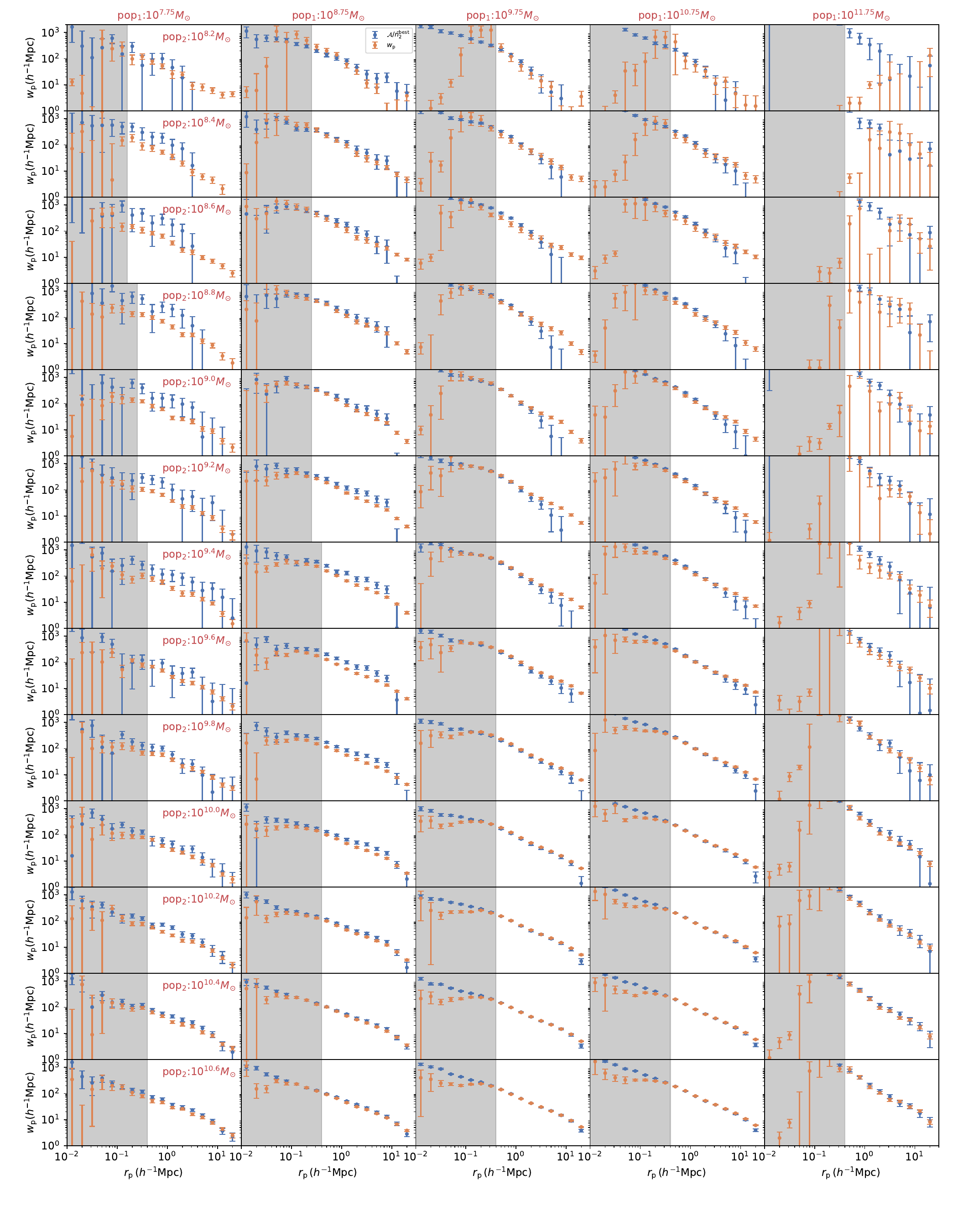}
    \caption{Comparison of $\mathcal{A}/\bar{n}_2^{\rm{best}}$ (blue) with $\Wp$ (orange) for {\textit{red}} galaxies. Each row presents the results for one pop$_2$ stellar mass bin, and each column corresponds to one pop$_1$ stellar mass bin. Similar to the blue galaxy case, only 5 out of the 11 pop$_1$ bins are shown, but we select different mass bins than those used for blue galaxies. Shaded regions indicate areas where $\Wp$ is affected by fibre incompleteness. The continuation of this figure is provided in Figure~\ref{fig:red_fitting2}.}  
    \label{fig:red_fitting1}
\end{figure*}

\begin{figure*}
    \centering    
    \includegraphics[width=0.98\textwidth]{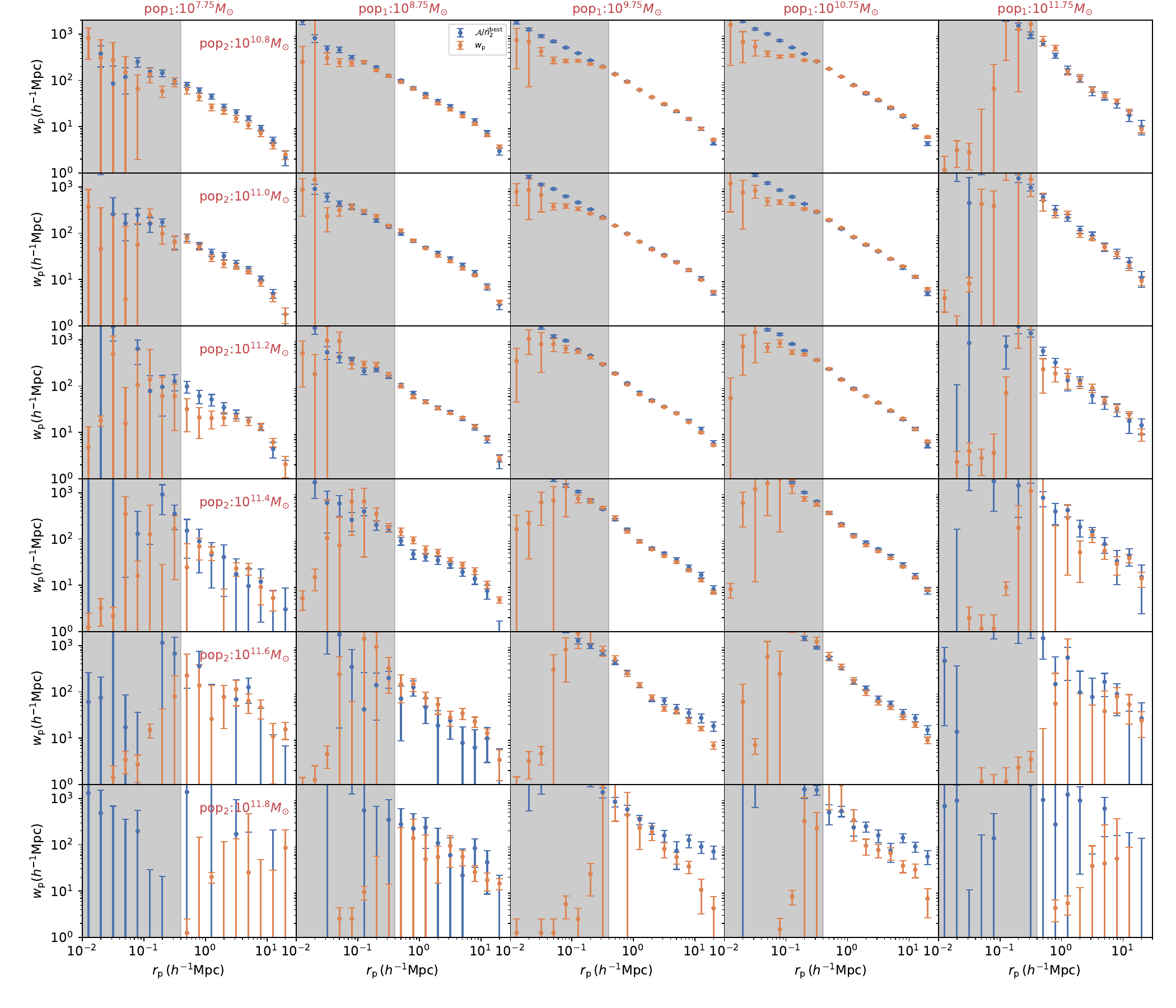}
    \caption{The continuation of Figure~\ref{fig:red_fitting1}.}  
    \label{fig:red_fitting2}
\end{figure*}

\begin{figure*}
    \centering    
    \includegraphics[width=0.98\textwidth]{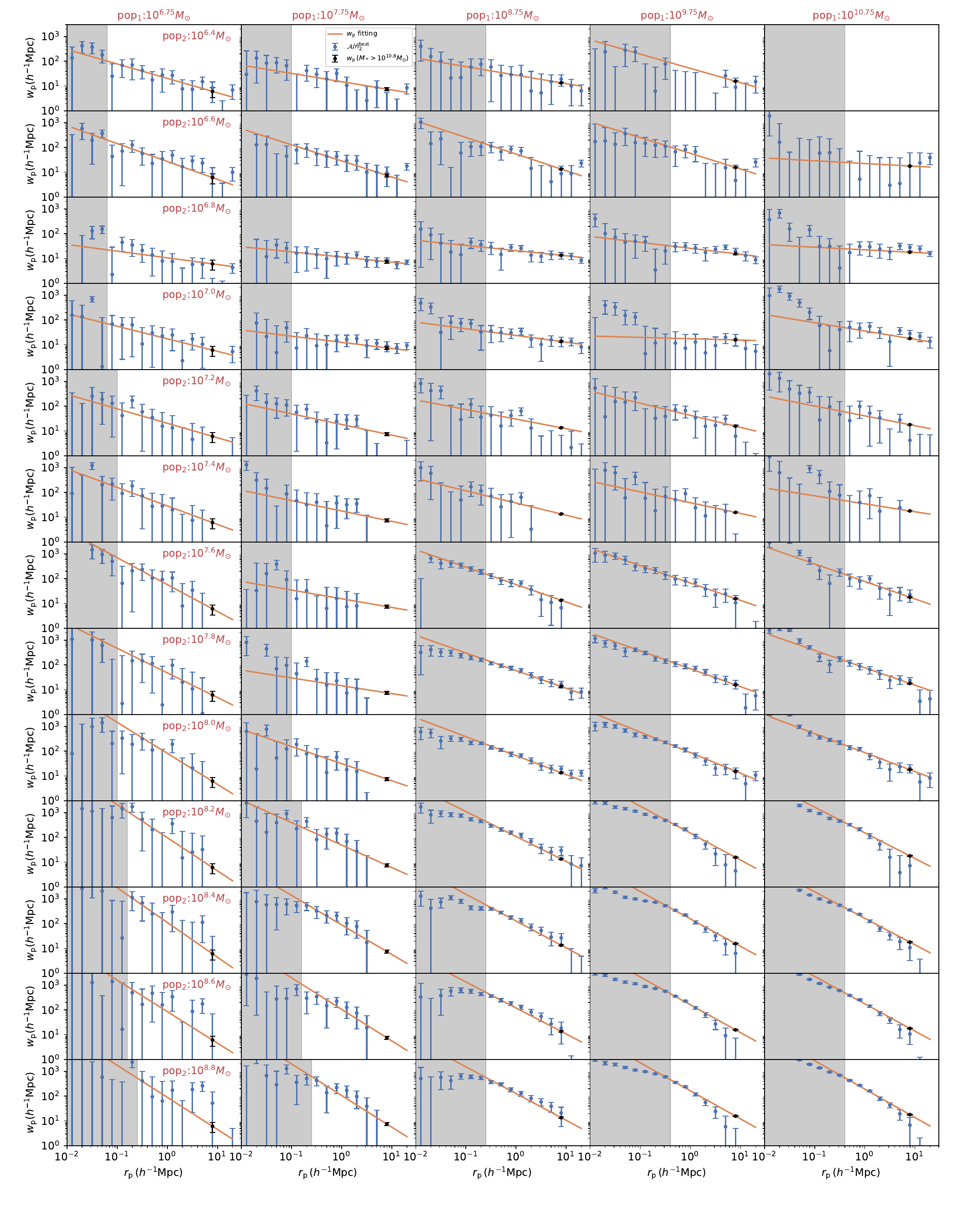}
    \caption{Similar to Figure~\ref{fig:red_fitting1} for \textit{red} galaxies, but here $\Wp$ represents the best-fit model, assuming a power-law with a constraint at $8h^{-1}$ Mpc from all red pop$_2^{\rm{s}}$ galaxies with $M_*<10^{10.8}\ms$. The continuation of this figure is provided in Figure~\ref{fig:red_fitting_wp2}.}  
    \label{fig:red_fitting_wp1}
\end{figure*}

\begin{figure*}
    \centering    
    \includegraphics[width=0.98\textwidth]{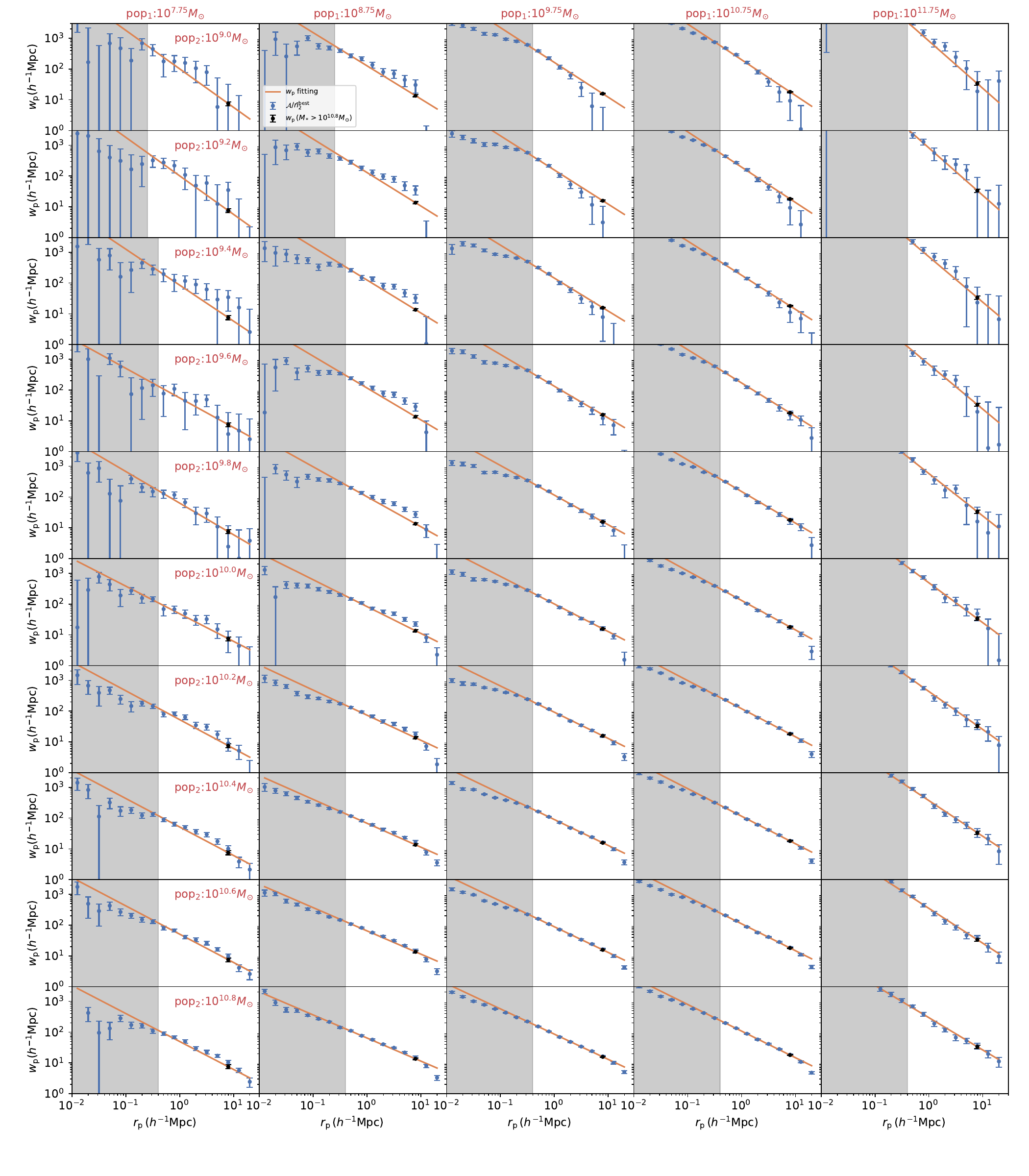}
    \caption{The continuation of Figure~\ref{fig:red_fitting_wp1}.}  
    \label{fig:red_fitting_wp2}
\end{figure*}

\begin{figure*}
    \centering    
    \includegraphics[width=0.98\textwidth]{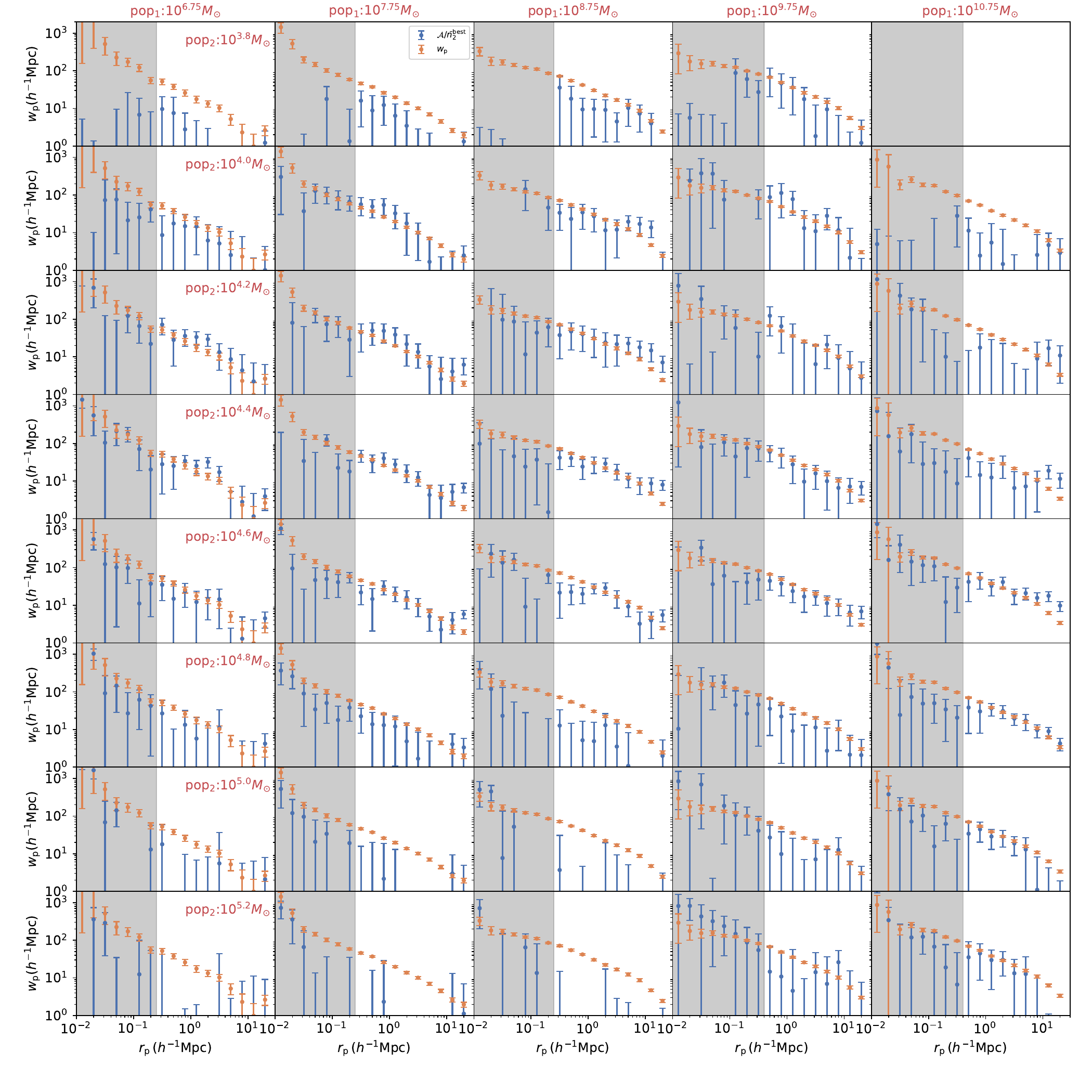}
    \caption{Similar to Figure~\ref{fig:blue_fitting_wp1} for {\textit{blue}} galaxies with fixed $\Wp$, but for incomplete stellar mass bins ranging from $10^{3.8} \ms$ to $10^{5.2} \ms$.}  
    \label{fig:blue_fitting_incom}
\end{figure*}

\begin{figure*}
    \centering    
    \includegraphics[width=0.98\textwidth]{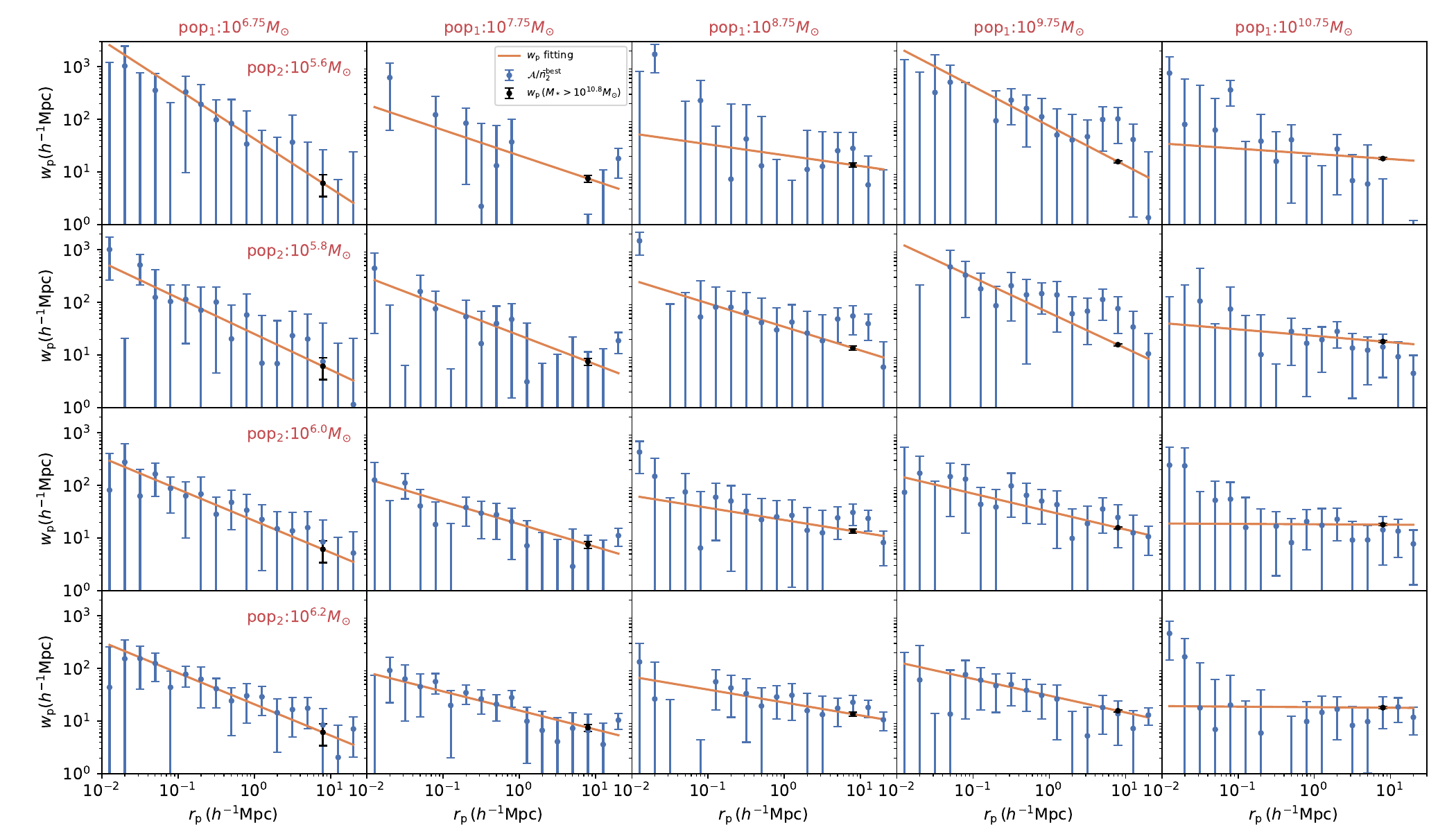}
    \caption{Similar to Figure~\ref{fig:red_fitting_incom} for {\textit{red}} galaxies with $\Wp$ from model fitting, but for incomplete stellar mass bins ranging from $10^{5.6} \ms$ to $10^{6.2} \ms$.}  
    \label{fig:red_fitting_incom}
\end{figure*}

In this section, we present the fits used to determine \(\bar{n}_2\) by comparing \(\nwp / \bar{n}_2^{\rm{best}}\) with \(\Wp\), where \(\bar{n}_2^{\rm{best}}\) represents the best-fit value of \(\bar{n}_2\). Results for blue galaxies with corresponding \(\Wp\) measurements are shown in Figures \ref{fig:blue_fitting1} and \ref{fig:blue_fitting2}. For blue galaxies with \(\Wp\) derived from \(M_* < 10^{9.0} \ms\), the results are presented in Figures \ref{fig:blue_fitting_wp1} and \ref{fig:blue_fitting_wp2}. Red galaxy results with corresponding \(\Wp\) measurements are shown in Figures \ref{fig:red_fitting1} and \ref{fig:red_fitting2}, while red galaxies with modeled \(\Wp\) are displayed in Figures \ref{fig:red_fitting_wp1} and \ref{fig:red_fitting_wp2}. The fits used to establish lower limits for incomplete stellar mass bins are presented in Figures \ref{fig:blue_fitting_incom} and \ref{fig:red_fitting_incom}. Overall, the fits are satisfactory across all stellar mass bins.

\section{The GSMF in tabular form and calibration to GAMA stellar mass}\label{sec:B}
In this Section, we show the tabular form of the GSMFs presented in Figure~\ref{fig:GSMF_after_mask} in Table \ref{tab:gsmf}. Additionally, we also show the GSMFs calibrated to GAMA stellar mass function in Table \ref{tab:gsmf}.

\begin{table*}
    \centering
    \caption{Tabular form of the GSMF \(\log_{10}(\Phi/{\rm{Mpc^{-3}dex^{-1})}}\) for each stellar mass bin $\log_{10}(M_*/\ms)$ as presented in Figure~\ref{fig:GSMF_after_mask}. Results calibrated to the GAMA stellar mass are also included.}
    \renewcommand{\arraystretch}{1.5}
    \begin{tabular}{cccccccc}
    \hline
       $\log_{10}(M_{*})$  & $\log_{10}(\Phi_{\rm{blue}})$ & $\log_{10}(\Phi_{\rm{red}})$ & $\log_{10}(\Phi_{\rm{all}})$ & $\log_{10}(M^{\rm{GAMA}}_{*})$ & $\log_{10}(\Phi_{\rm{blue}}^{\rm{GAMA}})$ & $\log_{10}(\Phi_{\rm{red}}^{\rm{GAMA}})$ & $\log_{10}(\Phi_{\rm{all}}^{\rm{GAMA}})$\\
    \hline
        3.8 & $-1.366_{-0.104}^{+0.079}$ &   &   & 4.44 & $-1.465_{-0.112}^{+0.085}$ &   &   \\
        4.0 & $-1.092_{-0.089}^{+0.079}$ &   &   & 4.62 & $-1.171_{-0.096}^{+0.085}$ &   &   \\
        4.2 & $-0.996_{-0.058}^{+0.048}$ &   &   & 4.81 & $-1.068_{-0.062}^{+0.051}$ &   &   \\
        4.4 & $-1.148_{-0.042}^{+0.033}$ &   &   & 5.00 & $-1.231_{-0.045}^{+0.036}$ &   &   \\
        4.6 & $-1.135_{-0.035}^{+0.040}$ &   &   & 5.18 & $-1.218_{-0.037}^{+0.043}$ &   &   \\
        4.8 & $-1.005_{-0.072}^{+0.062}$ &   &   & 5.37 & $-1.078_{-0.077}^{+0.066}$ &   &   \\
        5.0 & $-0.772_{-0.208}^{+0.104}$ &   &   & 5.55 & $-0.828_{-0.223}^{+0.112}$ &   &   \\
        5.2 & $-0.584_{-0.298}^{+0.129}$ &   &   & 5.74 & $-0.626_{-0.319}^{+0.138}$ &   &   \\
        5.4 & $0.095_{-0.199}^{+0.121}$ &   &   & 5.93 & $0.102_{-0.213}^{+0.129}$ &   &   \\
        5.6 & $-0.101_{-0.112}^{+0.094}$ & $-1.355_{-0.273}^{+0.173}$ & $-0.074_{-0.109}^{+0.087}$ & 6.11 & $-0.108_{-0.120}^{+0.101}$ & $-1.453_{-0.258}^{+0.186}$ & $-0.079_{-0.117}^{+0.094}$ \\
        5.8 & $-0.192_{-0.096}^{+0.068}$ & $-0.348_{-0.167}^{+0.059}$ & $0.015_{-0.060}^{+0.072}$ & 6.30 & $-0.206_{-0.103}^{+0.072}$ & $-0.373_{-0.180}^{+0.063}$ & $0.016_{-0.064}^{+0.077}$ \\
        6.0 & $-0.197_{-0.081}^{+0.056}$ & $0.187_{-0.121}^{+0.045}$ & $0.328_{-0.061}^{+0.050}$ & 6.49 & $-0.211_{-0.087}^{+0.060}$ & $0.201_{-0.129}^{+0.049}$ & $0.352_{-0.066}^{+0.054}$ \\
        6.2 & $-0.328_{-0.091}^{+0.055}$ & $0.328_{-0.080}^{+0.032}$ & $0.408_{-0.057}^{+0.034}$ & 6.67 & $-0.352_{-0.098}^{+0.059}$ & $0.352_{-0.086}^{+0.034}$ & $0.438_{-0.061}^{+0.036}$ \\
        6.4 & $-0.494_{-0.087}^{+0.067}$ & $0.543_{-0.078}^{+0.084}$ & $0.583_{-0.070}^{+0.069}$ & 6.86 & $-0.529_{-0.094}^{+0.072}$ & $0.582_{-0.084}^{+0.090}$ & $0.625_{-0.075}^{+0.074}$ \\
        6.6 & $-0.771_{-0.050}^{+0.028}$ & $0.126_{-0.101}^{+0.039}$ & $0.169_{-0.068}^{+0.045}$ & 7.05 & $-0.827_{-0.054}^{+0.030}$ & $0.135_{-0.108}^{+0.042}$ & $0.181_{-0.073}^{+0.049}$ \\
        6.8 & $-0.791_{-0.033}^{+0.027}$ & $0.282_{-0.033}^{+0.024}$ & $0.320_{-0.029}^{+0.021}$ & 7.23 & $-0.848_{-0.036}^{+0.029}$ & $0.303_{-0.036}^{+0.026}$ & $0.343_{-0.031}^{+0.022}$ \\
        7.0 & $-0.938_{-0.030}^{+0.024}$ & $-0.254_{-0.056}^{+0.023}$ & $-0.181_{-0.031}^{+0.031}$ & 7.42 & $-1.006_{-0.032}^{+0.026}$ & $-0.273_{-0.060}^{+0.025}$ & $-0.194_{-0.033}^{+0.033}$ \\
        7.2 & $-0.986_{-0.020}^{+0.019}$ & $-0.521_{-0.086}^{+0.072}$ & $-0.389_{-0.062}^{+0.058}$ & 7.61 & $-1.058_{-0.021}^{+0.021}$ & $-0.559_{-0.093}^{+0.077}$ & $-0.417_{-0.066}^{+0.062}$ \\
        7.4 & $-1.090_{-0.014}^{+0.014}$ & $-0.797_{-0.142}^{+0.107}$ & $-0.624_{-0.083}^{+0.085}$ & 7.79 & $-1.169_{-0.015}^{+0.015}$ & $-0.855_{-0.153}^{+0.115}$ & $-0.669_{-0.089}^{+0.091}$ \\
        7.6 & $-1.204_{-0.012}^{+0.011}$ & $-1.269_{-0.057}^{+0.050}$ & $-0.932_{-0.027}^{+0.024}$ & 7.98 & $-1.291_{-0.013}^{+0.012}$ & $-1.360_{-0.061}^{+0.054}$ & $-1.000_{-0.029}^{+0.025}$ \\
        7.8 & $-1.316_{-0.011}^{+0.011}$ & $-1.336_{-0.035}^{+0.025}$ & $-1.023_{-0.015}^{+0.014}$ & 8.17 & $-1.411_{-0.012}^{+0.012}$ & $-1.432_{-0.037}^{+0.027}$ & $-1.097_{-0.017}^{+0.015}$ \\
        8.0 & $-1.449_{-0.016}^{+0.016}$ & $-1.591_{-0.037}^{+0.016}$ & $-1.216_{-0.014}^{+0.015}$ & 8.35 & $-1.554_{-0.017}^{+0.018}$ & $-1.706_{-0.040}^{+0.017}$ & $-1.304_{-0.015}^{+0.016}$ \\
        8.2 & $-1.597_{-0.013}^{+0.015}$ & $-1.926_{-0.079}^{+0.053}$ & $-1.429_{-0.022}^{+0.022}$ & 8.54 & $-1.713_{-0.014}^{+0.016}$ & $-2.065_{-0.085}^{+0.057}$ & $-1.532_{-0.023}^{+0.024}$ \\
        8.4 & $-1.716_{-0.014}^{+0.010}$ & $-2.221_{-0.029}^{+0.021}$ & $-1.598_{-0.011}^{+0.010}$ & 8.73 & $-1.840_{-0.015}^{+0.011}$ & $-2.381_{-0.031}^{+0.023}$ & $-1.714_{-0.012}^{+0.010}$ \\
        8.6 & $-1.775_{-0.009}^{+0.008}$ & $-2.557_{-0.025}^{+0.024}$ & $-1.709_{-0.008}^{+0.009}$ & 8.91 & $-1.903_{-0.010}^{+0.009}$ & $-2.742_{-0.027}^{+0.026}$ & $-1.833_{-0.008}^{+0.009}$ \\
        8.8 & $-1.887_{-0.009}^{+0.008}$ & $-2.638_{-0.021}^{+0.016}$ & $-1.816_{-0.009}^{+0.007}$ & 9.10 & $-2.024_{-0.010}^{+0.009}$ & $-2.829_{-0.023}^{+0.017}$ & $-1.948_{-0.009}^{+0.007}$ \\
        9.0 & $-2.014_{-0.009}^{+0.005}$ & $-2.621_{-0.017}^{+0.009}$ & $-1.920_{-0.006}^{+0.006}$ & 9.28 & $-2.159_{-0.010}^{+0.006}$ & $-2.811_{-0.018}^{+0.010}$ & $-2.059_{-0.006}^{+0.007}$ \\
        9.2 & $-2.137_{-0.006}^{+0.007}$ & $-2.675_{-0.013}^{+0.007}$ & $-2.027_{-0.005}^{+0.007}$ & 9.47 & $-2.291_{-0.007}^{+0.007}$ & $-2.869_{-0.014}^{+0.008}$ & $-2.174_{-0.005}^{+0.007}$ \\
        9.4 & $-2.241_{-0.006}^{+0.005}$ & $-2.703_{-0.009}^{+0.009}$ & $-2.113_{-0.005}^{+0.005}$ & 9.66 & $-2.403_{-0.007}^{+0.006}$ & $-2.899_{-0.010}^{+0.010}$ & $-2.266_{-0.005}^{+0.006}$ \\
        9.6 & $-2.382_{-0.005}^{+0.007}$ & $-2.632_{-0.009}^{+0.005}$ & $-2.187_{-0.005}^{+0.005}$ & 9.84 & $-2.554_{-0.005}^{+0.007}$ & $-2.823_{-0.010}^{+0.005}$ & $-2.346_{-0.005}^{+0.005}$ \\
        9.8 & $-2.529_{-0.005}^{+0.004}$ & $-2.557_{-0.007}^{+0.005}$ & $-2.242_{-0.004}^{+0.004}$ & 10.03 & $-2.712_{-0.006}^{+0.005}$ & $-2.742_{-0.007}^{+0.005}$ & $-2.404_{-0.004}^{+0.004}$ \\
        10.0 & $-2.672_{-0.006}^{+0.004}$ & $-2.475_{-0.006}^{+0.007}$ & $-2.260_{-0.003}^{+0.003}$ & 10.22 & $-2.866_{-0.006}^{+0.005}$ & $-2.655_{-0.006}^{+0.007}$ & $-2.423_{-0.004}^{+0.003}$ \\
        10.2 & $-2.868_{-0.006}^{+0.006}$ & $-2.432_{-0.004}^{+0.005}$ & $-2.296_{-0.004}^{+0.003}$ & 10.40 & $-3.075_{-0.006}^{+0.006}$ & $-2.608_{-0.005}^{+0.005}$ & $-2.462_{-0.005}^{+0.003}$ \\
        10.4 & $-3.137_{-0.008}^{+0.008}$ & $-2.444_{-0.004}^{+0.004}$ & $-2.364_{-0.003}^{+0.003}$ & 10.59 & $-3.364_{-0.009}^{+0.008}$ & $-2.621_{-0.004}^{+0.004}$ & $-2.535_{-0.004}^{+0.003}$ \\
        10.6 & $-3.503_{-0.012}^{+0.014}$ & $-2.500_{-0.005}^{+0.003}$ & $-2.459_{-0.004}^{+0.003}$ & 10.78 & $-3.756_{-0.013}^{+0.015}$ & $-2.681_{-0.005}^{+0.003}$ & $-2.637_{-0.004}^{+0.003}$ \\
        10.8 & $-3.968_{-0.028}^{+0.031}$ & $-2.632_{-0.004}^{+0.003}$ & $-2.612_{-0.004}^{+0.003}$ & 10.96 & $-4.256_{-0.030}^{+0.033}$ & $-2.822_{-0.004}^{+0.004}$ & $-2.801_{-0.004}^{+0.003}$ \\
        11.0 & $-4.531_{-0.046}^{+0.048}$ & $-2.867_{-0.005}^{+0.003}$ & $-2.860_{-0.004}^{+0.005}$ & 11.15 & $-4.859_{-0.050}^{+0.051}$ & $-3.074_{-0.006}^{+0.004}$ & $-3.067_{-0.004}^{+0.005}$ \\
        11.2 & $-5.924_{-1.000}^{+0.373}$ & $-3.228_{-0.007}^{+0.004}$ & $-3.221_{-0.007}^{+0.006}$ & 11.34 & $-6.352_{-1.000}^{+0.400}$ & $-3.462_{-0.007}^{+0.005}$ & $-3.454_{-0.008}^{+0.006}$ \\
        11.4 & $-6.509_{-1.000}^{+0.482}$ & $-3.791_{-0.008}^{+0.010}$ & $-3.779_{-0.010}^{+0.012}$ & 11.52 & $-6.980_{-1.000}^{+0.517}$ & $-4.065_{-0.009}^{+0.010}$ & $-4.052_{-0.010}^{+0.012}$ \\
        11.6 &   & $-4.532_{-0.024}^{+0.025}$ & $-4.532_{-0.024}^{+0.025}$ & 11.71 &   & $-4.860_{-0.026}^{+0.026}$ & $-4.860_{-0.026}^{+0.026}$ \\
        11.8 &   & $-5.738_{-0.164}^{+0.094}$ & $-5.738_{-0.164}^{+0.094}$ & 11.90 &   & $-6.153_{-0.176}^{+0.101}$ & $-6.153_{-0.176}^{+0.101}$ \\
    \hline
    \end{tabular}
    \label{tab:gsmf}
\end{table*}

\section{Author Affiliations}\label{sec:author}
\scriptsize
$^{1}$ Center for Particle Cosmology, Department of Physics and Astronomy, University of Pennsylvania, Philadelphia, PA 19104, USA\\
$^{2}$ Institute for Computational Cosmology, Department of Physics, Durham University, South Road, Durham DH1 3LE, UK\\
$^{3}$ State Key Laboratory of Dark Matter Physics, Tsung-Dao Lee Institute \& School of Physics and Astronomy, Shanghai Jiao Tong University, Shanghai 201210, P.R. China\\
$^{4}$ Department of Astronomy, School of Physics and Astronomy, Shanghai Jiao Tong University, Shanghai 200240, China\\
$^{5}$ Lawrence Berkeley National Laboratory, 1 Cyclotron Road, Berkeley, CA 94720, USA\\
$^{6}$ Physics Dept., Boston University, 590 Commonwealth Avenue, Boston, MA 02215, USA\\
$^{7}$ Dipartimento di Fisica ``Aldo Pontremoli'', Universit\`a degli Studi di Milano, Via Celoria 16, I-20133 Milano, Italy\\
$^{8}$ INAF-Osservatorio Astronomico di Brera, Via Brera 28, 20122 Milano, Italy\\
$^{9}$ Department of Physics \& Astronomy, University College London, Gower Street, London, WC1E 6BT, UK\\
$^{10}$ Instituto de F\'{\i}sica, Universidad Nacional Aut\'{o}noma de M\'{e}xico,  Circuito de la Investigaci\'{o}n Cient\'{\i}fica, Ciudad Universitaria, Cd. de M\'{e}xico  C.~P.~04510,  M\'{e}xico\\
$^{11}$ NSF NOIRLab, 950 N. Cherry Ave., Tucson, AZ 85719, USA\\
$^{12}$ Departamento de F\'isica, Universidad de los Andes, Cra. 1 No. 18A-10, Edificio Ip, CP 111711, Bogot\'a, Colombia\\
$^{13}$ Observatorio Astron\'omico, Universidad de los Andes, Cra. 1 No. 18A-10, Edificio H, CP 111711 Bogot\'a, Colombia\\
$^{14}$ Institut d'Estudis Espacials de Catalunya (IEEC), c/ Esteve Terradas 1, Edifici RDIT, Campus PMT-UPC, 08860 Castelldefels, Spain\\
$^{15}$ Institute of Cosmology and Gravitation, University of Portsmouth, Dennis Sciama Building, Portsmouth, PO1 3FX, UK\\
$^{16}$ Institute of Space Sciences, ICE-CSIC, Campus UAB, Carrer de Can Magrans s/n, 08913 Bellaterra, Barcelona, Spain\\
$^{17}$ Fermi National Accelerator Laboratory, PO Box 500, Batavia, IL 60510, USA\\
$^{18}$ Center for Cosmology and AstroParticle Physics, The Ohio State University, 191 West Woodruff Avenue, Columbus, OH 43210, USA\\
$^{19}$ Department of Physics, The Ohio State University, 191 West Woodruff Avenue, Columbus, OH 43210, USA\\
$^{20}$ The Ohio State University, Columbus, 43210 OH, USA\\
$^{21}$ Department of Physics, The University of Texas at Dallas, 800 W. Campbell Rd., Richardson, TX 75080, USA\\
$^{22}$ Institute for Astronomy, University of Edinburgh, Royal Observatory, Blackford Hill, Edinburgh EH9 3HJ, UK\\
$^{23}$ Institute of Astronomy, University of Cambridge, Madingley Road, Cambridge CB3 0HA, UK\\
$^{24}$ Sorbonne Universit\'{e}, CNRS/IN2P3, Laboratoire de Physique Nucl\'{e}aire et de Hautes Energies (LPNHE), FR-75005 Paris, France\\
$^{25}$ Instituci\'{o} Catalana de Recerca i Estudis Avan\c{c}ats, Passeig de Llu\'{\i}s Companys, 23, 08010 Barcelona, Spain\\
$^{26}$ Institut de F\'{i}sica d’Altes Energies (IFAE), The Barcelona Institute of Science and Technology, Edifici Cn, Campus UAB, 08193, Bellaterra (Barcelona), Spain\\
$^{27}$ Department of Physics and Astronomy, Siena College, 515 Loudon Road, Loudonville, NY 12211, USA\\
$^{28}$ Space Sciences Laboratory, University of California, Berkeley, 7 Gauss Way, Berkeley, CA  94720, USA\\
$^{29}$ University of California, Berkeley, 110 Sproul Hall \#5800 Berkeley, CA 94720, USA\\
$^{30}$ Instituto de Astrof\'{i}sica de Andaluc\'{i}a (CSIC), Glorieta de la Astronom\'{i}a, s/n, E-18008 Granada, Spain\\
$^{31}$ Departament de F\'isica, EEBE, Universitat Polit\`ecnica de Catalunya, c/Eduard Maristany 10, 08930 Barcelona, Spain\\
$^{32}$ Department of Physics and Astronomy, Sejong University, 209 Neungdong-ro, Gwangjin-gu, Seoul 05006, Republic of Korea\\
$^{33}$ CIEMAT, Avenida Complutense 40, E-28040 Madrid, Spain\\
$^{34}$ University of Michigan, 500 S. State Street, Ann Arbor, MI 48109, USA\\
$^{35}$ National Astronomical Observatories, Chinese Academy of Sciences, A20 Datun Rd., Chaoyang District, Beijing, 100012, P.R. China\\

\bsp	
\label{lastpage}
\end{document}